\documentclass[a4paper,11pt]{article}
\pdfoutput=1 % if your are submitting a pdflatex (i.e. if you have
\usepackage{jheppub} % for details on the use of the package, please
\usepackage[T1]{fontenc} % if needed
\usepackage{amsmath}
\usepackage{amssymb}
\usepackage{amsfonts}
\usepackage{mathtools}
\usepackage{microtype}
\usepackage{tikz}
\usepackage{bm}
\usepackage[utf8]{inputenc}
\usepackage{blkarray}
\usepackage{bigstrut}
\usepackage{nccmath}
\usepackage{enumitem}
\usepackage{braket}
\usepackage{dsfont}
\usepackage{lipsum}
\usepackage{array,multirow}
\usepackage{physics}
\usepackage{xcolor}
\usepackage{color}
\usepackage{float}
\usepackage{subfig}
\usepackage[export]{adjustbox}

\newcommand{\id}{\mathrm{d}}

\definecolor{hjaltegreen}{rgb}{0.0,0.5,0.0}

\def \be {\begin{equation}}
\def \ee {\end{equation}}
\newcommand{\bea}{\begin{eqnarray}}
\newcommand{\eea}{\end{eqnarray}}
\newcommand{\bei}{\begin{itemize}}
\newcommand{\eei}{\end{itemize}}

\setlist[itemize,1]{leftmargin=1.5em}
\newcommand{\setmuskip}[2]{#1=#2\relax}

\title{\boldmath Decomposition of Feynman Integrals by Multivariate Intersection Numbers}

\author[a,b]{Hjalte Frellesvig,}
\author[a,b]{\ Federico Gasparotto,}
\author[a,b]{\ Stefano Laporta,}
\author[b,a]{\ Manoj K. Mandal,} 
\author[a,b]{\ Pierpaolo Mastrolia,}
\author[b,a]{\ Luca Mattiazzi,}
\author[c]{\ Sebastian Mizera}

\affiliation[a]{Dipartimento di Fisica e Astronomia, Universit\`a di Padova, Via Marzolo 8, 35131 Padova, Italy}

\affiliation[b]{INFN, Sezione di Padova, Via Marzolo 8, 35131 Padova, Italy}

\affiliation[c]{Institute for Advanced Study, Einstein Drive, Princeton, NJ 08540, USA}

\emailAdd{\{hjalte.frellesvig,
federico.gasparotto,
stefano.laporta,
manojkumar.mandal,
pierpaolo.mastrolia,
luca.mattiazzi\}@pd.infn.it}
\emailAdd{smizera@ias.edu}

\abstract{
We present a detailed description of the recent idea for a direct decomposition of Feynman integrals onto a basis of master integrals by projections, as well as a direct derivation of the differential equations satisfied by the master integrals, employing multivariate intersection numbers. We discuss a recursive algorithm for the computation of multivariate intersection numbers, and provide three different approaches for a direct decomposition of Feynman integrals, which we dub the {\it straight decomposition}, the {\it bottom-up decomposition}, and the {\it top-down decomposition}. These algorithms 
exploit the unitarity structure of Feynman integrals by computing intersection numbers supported on cuts, in various orders, thus showing the synthesis of the intersection-theory concepts with unitarity-based methods and integrand decomposition. We perform explicit computations to exemplify all of these approaches applied to Feynman integrals, paving a way towards potential applications to generic multi-loop integrals.
}

\begin{document} 
\maketitle
\flushbottom

%\newpage

\section{Introduction}
\label{sec:introduction}
Feynman integrals in dimensional regularization admit parametric
integral representations which expose their nature as Aomoto-Gel'fand
integrals, thereby enabling a novel form of investigation of their
algebraic structure by means of {\it intersection theory of twisted de
Rham (co)homology} for general hypergeometric
functions~\cite{Mastrolia:2018uzb,Frellesvig:2019kgj,Frellesvig:2019uqt}. 
Accordingly, {\it intersection numbers} of differential forms \cite{cho1995} can
be employed to define a {\it scalar product} on a {\it vector space}
of Feynman integrals \cite{Mastrolia:2018uzb}, such that projecting any 
multi-loop integral onto a basis of master integrals (MIs) becomes
conceptually identical to decomposing a generic vector into a
basis of a vector space. 

Univariate intersection numbers, as shown in the original studies \cite{Mastrolia:2018uzb,Frellesvig:2019kgj}, 
were sufficient to validate a novel method based on intersection
theory for deriving integral
relations, which was used for the direct
derivation of contiguity relations for Lauricella $F_{D}$ functions, 
as well as for Feynman integrals on maximal cuts, {\it i.e.} with on-shell internal lines,
that admit a one-fold integral representations. 
As proposed in \cite{Frellesvig:2019kgj}, applications of
this novel method to the decomposition of full Feynman integrals
in terms of a complete set of MIs, including the ones corresponding
to subdiagrams, as well as deriving contiguity relations for special
functions admitting multi-fold integral representation,
required the use of {\it multivariate} intersection numbers \cite{matsumoto1994,matsumoto1998,OST2003,doi:10.1142/S0129167X13500948,goto2015,goto2015b,Yoshiaki-GOTO2015203,Mizera:2017rqa,matsubaraheo2019algorithm}.

A recursive algorithm for computing multivariate intersection numbers
was proposed in \cite{Mizera:2019gea} and 
later refined and applied to a few paradigmatic cases of Feynman
integral decomposition \cite{Frellesvig:2019uqt}.
This recursive algorithm was developed in order to compute intersection numbers for twisted
cohomologies associated to $n$-forms, which in the general case may contain poles that are not necessarily simple. 
In the case of logarithmic (dlog) differential forms, owing to the
presence of simple poles only, the computation of
the intersection numbers is known to be simpler~\cite{matsumoto1998,Mizera:2017rqa}.

Recent complementary work
\cite{Abreu:2019wzk,Abreu:2019xep} shows that intersection numbers play a fundamental role in the definition of a diagrammatic coaction for MIs, which combined with the master integral decomposition studied in this paper, 
as well as in Refs.~\cite{Mastrolia:2018uzb,Frellesvig:2019kgj,Frellesvig:2019uqt}
paves a way towards comprehensive computations of scattering amplitudes using the tools of intersection theory.

The intersection theory-based decomposition has also been recently applied to the study
of Feynman integrals in $d = 4 \pm 2\epsilon$ space-time dimensions, from which an
unexpected relation between the behaviors around $\epsilon \to 0$ and $\epsilon \to \infty$ emerged \cite{Mizera:2019vvs} and 
was used to investigate the properties of canonical systems of differential
equations~\cite{Henn:2013pwa}. A further interesting step for the construction of canonical integrals with intersection theory has been reported in~\cite{Chen:2020uyk}. Moreover, it was observed that using recursion relations for computing intersection numbers
can be further refined by relating them to dlog forms at each step of the recursive algorithm~\cite{Weinzierl:2020xyy}. Other recent intersection-theory approaches include \cite{Kaderli:2019dny,Kalyanapuram:2020vil,Weinzierl:2020nhw}.

This work can be considered as an extension of ~\cite{Frellesvig:2019uqt}, which contained the essential mathematical details that brought us to the formulation of a decomposition algorithm based on multivariate intersection numbers.
In the current work, we address more extensively the problem of evaluating intersection numbers for multivariate forms, providing an explicit description of the 2-form case, and showing their application to the complete decomposition of Feynman integrals in terms of MIs.
We show how intersection numbers can be used to establish linear and quadratic relations for Feynman integrals, and, more generally, for Aomoto-Gel'fand generalized hypergeometric functions. The former set of relations yields results that are equivalent to the known integration-by-parts identities (IBPs) \cite{Chetyrkin:1981qh}, while the
latter allow for a systematic classification of relations which, for certain type of integrals were originally detected within the application of number-theoretic methods to Feynman integrals, giving rise to interesting  conjectures \cite{Broadhurst:2016hbq, Broadhurst:2016myo,Broadhurst:2018tey,Zhou:2017vhw,Zhou:2017jnm}, proven to be true quite recently \cite{fresn2020quadratic,fresn2020quadratic2}. 
A special set of quadratic relations have been presented in \cite{Lee:2018jsw}, and it would be interesting to investigate if they can be classified as Twisted Riemann Period Relations \cite{cho1995}.

In particular, we focus on different ways of using 
intersection theory in order to derive linear relations for Feynman integrals, as well as the systems of differential equations and the finite difference equations they obey.
Moreover, we present here for the first time, a novel algorithm for Feynman integral decomposition, which we will refer to as {\it top-down decomposition}, 
showing that the coefficients of MIs can be suitably extracted by projections via
intersection numbers within an iterative strategy,
starting from the integrals that correspond to graphs with the highest number of internal lines, and ending with those 
corresponding to graphs with the lowest possible number of internal lines (given, in the general case, by the product of as many tadpoles as the number of loops).

Following \cite{Frellesvig:2019kgj,Frellesvig:2019uqt}, we also make use of two other algorithms: the \emph{bottom-up decomposition} and \emph{straight decomposition} to similar aim.
All these strategies combine the advantages of 
the integrand decomposition techniques~\cite{Ossola:2006us,Ellis:2007br,Ellis:2008ir,Mastrolia:2012bu,Zhang:2012ce,Mastrolia:2012an,Mastrolia:2011pr,Badger:2013gxa},  
the unitarity-based methods~\cite{Bern:1994cg,Bern:1994zx,Britto:2004nc,Britto:2005ha,Britto:2006sj,Anastasiou:2006jv,Mastrolia:2006ki,BjerrumBohr:2007vu,Forde:2007mi,Badger:2008cm,Mastrolia:2009dr,NigelGlover:2008ur,Britto:2009wz,Britto:2010um,Kosower:2011ty}, 
and the intersections theory-based decomposition.

This work constitutes an important step forward towards the development of a complete algorithm for the decomposition of Feynman integral by means of intersection theory based concepts. In particular, the use of intersection-numbers within the top-down decomposition enhances the effectiveness of the unitarity-based decomposition and of the integrand-decomposition. In fact, on the one side, the applicability of generalised-unitarity for the direct extraction of integral coefficients is known to be limited by two factors: the lack of (complex) integration techniques for evaluating phase-space integrals corresponding to generic generalised cuts (multiple-cut techniques are indeed available only for a limited set of cuts, and mainly at one-loop); the lack of systematic criteria for disentangling the coefficients of master integrals that belong to the same sector on the maximal-cut, {\it i.e.} sharing a common set of cut-denominators (usually occurring in dealing with diagrams with more than one loop). On the other side, the integrand decomposition techniques, which implement generalised unitarity at the integrand level, provide a decomposition in terms of a set of integrals which is not minimal: in fact, it is known that these integrals can be further reduced to a minimal basis of master integrals by means of IBPs. 
Using intersection numbers those problems are bypassed, and integrals can be expressed in terms of a minimal set of master integrals.

As originally defined \cite{cho1995}, intersection theory for twisted cohomologies and the evaluation of multivariate intersection numbers are applicable to differential forms obeying certain genericity conditions, whose purpose is to regulate boundaries of integration and ensure that they integrate to analytic functions. In the physics language, this corresponds to the analytic regularization of Feynman integrals \cite{speer1969theory}. To simplify computations, we employ this regularization whenever necessary. It has the additional benefit of resolving the ambiguities that arise when there is a non-trivial overlap between critical points and singularities \cite{Lee:2013hzt,Frellesvig:2019kgj,Weinzierl:2020xyy}. Recent mathematical developments, employing the notion of intersection numbers for the {\it relative twisted cohomology}~\cite{matsumoto2018relative}, seem to offer the possibility of studying the vector space properties of Aomoto-Gel'fand hypergeometric integrals in absence of analytic regulators. This creates a natural path for further investigations of the connections between intersection theory and Feynman integrals, which are left for the future.

The paper is organized as follows: In Sec.~\ref{sec:derahm} we
begin by recalling the basics of the
Feynman integrals in terms of twisted de Rham (co)homologies and their intersection theory. We show the representation of both the integral and its dual, together with the master decomposition formula needed for their direct decomposition. We discuss different ways to compute the dimension of the cohomology group. The differential equations satisfied by the forms and the dual forms are also provided.
Then follows Sec.~\ref{sec:multivariate} in which we discuss multivariate intersection numbers. We start with an explicit construction of the 2-variable intersection numbers, which is expressed in terms of the univariate ones recursively. This procedure is generalized, resulting in the final formula for the $n$-variable intersection numbers. We also present an explicit example showing all the
steps of the computation of a specific $2$-variable intersection number, and discuss a few properties satisfied by the intersection numbers, as well as the simplified formula valid in the case of dlog forms.
In Sec.~\ref{sec:feynintdeco}, we discuss strategies for the decomposition of an arbitrary Feynman integral. Specifically, we show three different approaches, namely the {\it straight decomposition}, the {\it bottom-up decomposition}, and the {\it top-down decomposition}. Sec.~\ref{sec:examples} is dedicated to examples. We first consider the one-loop massless box and perform the decomposition with all these three approaches to show the steps involved explicitly. Moreover, we show the decomposition for the QED triangle as well as the differential equation for the QED sunrise. After that, we provide a few tables with all the key ingredients necessary for the computation of the multivariate intersection numbers needed to obtain the direct decomposition, as well as their differential equations, for the cases of the $1$-loop box with 4 different masses, the $2$-loop sunrise with 3 different masses, the $2$-loop planar and non-planar massless triangle-boxes, as well as 2-loop massless double-box on a triple cut. Finally, Sec.~\ref{sec:conclusions} contains our conclusions and discussion. The paper ends with Appendix~\ref{appendix} containing the explicit forms of the multivariate intersection numbers used for the $1$-loop massless box, the QED triangle, and the QED sunrise.

\section{Feynman integrals and differential forms}
\label{sec:derahm}

We consider Aomoto-Gel'fand generalized hypergeometric integrals of the form
\begin{equation}
I = \int_{\mathcal{C}_R}u(\mathbf{z}) \, \varphi_L (\mathbf{z}),
\label{eq:integral_definition}
\end{equation}
where $u(\mathbf{z})$ is a multivalued function, $u(\mathbf{z})=\mathcal{B}(\mathbf{z})^{\gamma}$ (or $u(\mathbf{z})=\prod_{i} \mathcal{B}_{i}(\mathbf{z})^{\gamma_i}$). In the context of the Feynman integrals addressed in this manuscript $\mathcal{B}$ is the Baikov (graph) polynomial, which has the property that it vanishes on the boundary of the integration domain in~(\ref{eq:integral_definition})
\begin{equation}
\mathcal{B}(\partial \mathcal{C}_R)=0,
\label{eq:B_vanishing_on_parital_C}
\end{equation}
while $\gamma$ depends on the space-time dimensionality $d$, and on the number of loops and external legs. We assume $\gamma$ to not be an integer, $\gamma \notin \mathbb{Z}$,
which follows from dimensional regularization.\\
On the other hand, $\varphi(\mathbf{z})$ is a single valued differential form
\begin{equation}
\varphi_L(\mathbf{z}) = \hat{\varphi}_L(\mathbf{z}) \, d^n \mathbf{z}\,, \qquad
\hat{\varphi}_L(\mathbf{z}) = \frac{f(\mathbf{z})}{z_1^{a_1} \dots z_n^{a_n}}\,,
\end{equation}
where $\hat{\varphi}_L(\mathbf{z})$ denotes its differential-stripped version, $f(\mathbf{z})$ is a \emph{rational function} and $a_i$ are integer exponents, $a_i \in \mathbb{Z}$.\\
One of the key assumptions is that all the poles present in $\varphi_L$ must be regulated by $u(\mathbf{z})$. In genuine Feynman integrals this assumption is often violated; in this work we present two different strategies for overcoming this apparent obstacle.

It is possible to identify equivalence classes of differential $n$-forms entering the integral \eqref{eq:integral_definition}. Forms in the same class are those that differ by a covariant derivative and give the same result upon integration, as will be explained below.
\\
\subsection{The cohomology group and its dual}

Consider an $(n{-}1)$-differential form $\xi_L$. In the absence of boundary terms due to (\ref{eq:B_vanishing_on_parital_C}) we have:
\begin{equation}
0=\int_{\mathcal{C}_R}d(u \, \xi_L)=\int_{\mathcal{C}_R} \left( du \wedge \xi_L+ u \, d \xi_L \right)= \int_{\mathcal{C}_R}u \left(\frac{du}{u} \wedge+ d \right) \xi_L= \int_{\mathcal{C}_R}u \, \nabla_{\omega} \, \xi_L,
\end{equation}
where
\begin{equation}
\nabla_{\omega}= d + \omega \wedge, \qquad \omega= d \log u.
\label{eq:nablasmallomegaplus}
\end{equation}
Thus we can write
\begin{equation}
\int_{\mathcal{C}_R}u \, \varphi_L= \int_{\mathcal{C}_R} u \, \left(\varphi_L+\nabla_{\omega} \xi_L \right)    
\end{equation}
The forms $\varphi_L$ and $\varphi_L+\nabla_{\omega} \xi_L$, which give the same result upon integration, are in the same equivalence class
\begin{equation}
\varphi_L \sim \varphi_L+\nabla_{\omega} \xi_L.
\label{eq:equivalence_relation}
\end{equation}
Differential $n$-forms modulo the equivalence relation ($\ref{eq:equivalence_relation}$) belong to a vector space, the \emph{twisted cohomology group} $H_{\omega}^{n}$, and elements in this vector space are denoted by $\langle \varphi_L|$.\\
In a similar way one can define an equivalence relation among integration contours which give the same result upon integration. Integration contours modulo the equivalence relation, are denoted by $| \mathcal{C}_R]$ and belong to the vector space $H_{n}^{\omega}$, referred to as the \emph{twisted homology group}.\\ 
The integral of eq.~(\ref{eq:integral_definition}) can be regarded as a paring between $\langle \varphi_L|$ and the function $u(\mathbf{z})$, integrated over the contour $| \mathcal{C}_R]$
\begin{equation}
I= \int_{\mathcal{C}_R}u(\mathbf{z}) \, \varphi_L(\mathbf{z})= \langle \varphi_L| \mathcal{C}_R].    
\label{}
\end{equation}
Given this terminology, we may now
define a \emph{dual} integral, given by
\begin{equation}
\Tilde{I}=\int_{\mathcal{C}_L} u(\mathbf{z})^{-1} \, \varphi_R(\mathbf{z})
\label{eq:dual_integral_definition}   \end{equation}
and consider the covariant derivative
\begin{equation}
\nabla_{-\omega}=d-\omega \wedge, \qquad \omega=d \log u.
\label{eq:nablasmallomegaminus}
\end{equation}
In analogy to \eqref{eq:equivalence_relation} we can derive the  equivalence relation
\begin{equation}
\varphi_R \;\sim\; \varphi_R+ \nabla_{- \omega} \xi_R
\label{eq:dual_equivalence_relation}
\end{equation}
such that differential $n$-forms modulo the equivalence relation eq.~(\ref{eq:dual_equivalence_relation}) belong to the \emph{dual} vector space $(H^{n}_{\omega})^{\ast}=H^n_{- \omega}$; the elements of this space are denoted by $| \varphi_R \rangle$. As done above, one can also consider an equivalence relation among integration contours, which leads to the vector space $(H_{n}^{\omega})^{\ast}=H_{n}^{- \omega}$ whose elements are denoted by $[ \mathcal{C}_{L}|$.\\ 
The dual integral of eq.~(\ref{eq:dual_integral_definition}) is interpreted as paring between $| \varphi_R \rangle$ and the function $u(\mathbf{z})^{-1}$, integrated over the contour $[\mathcal{C}_L|$
\begin{equation}
\Tilde{I}
%= \int_{\mathcal{C}_L} u(\mathbf{z})^{-1} \, \varphi_R(\mathbf{z})
= [\mathcal{C}_L| \varphi_R \rangle.
\end{equation}

Aomoto-Gel'fand (AG) integrals are known to obey {\it Gauss contiguity relations}. Similarly Feynman integrals obey linear relations, dubbed {\it integration by parts identities} \cite{Chetyrkin:1981qh}.
Those identities can be used to identify a minimal set of functions which constitute a basis that generates a vector space \cite{Frellesvig:2019uqt}, which -- by borrowing the terminology from Feynman multi-loop calculus -- we will refer to as {\it master integrals} (MIs).
Linear relations among integrals can therefore be used to {\it decompose} any AG/Feynman integral in terms of MIs, as well as to derive (systems of first order) difference and differential equations for MIs.

Let us observe that dual integrals are AG integrals where $u^{-1}$ appears in the integrand (instead of $u$). 
In the case of dimensionally regulated Feynman integrals, $u$ consists of graph polynomials raised to a non-integer power that depends on the space-time dimensions $d$. Therefore, dual integrals represent integrals in a different dimension (for which the exponents of the graph polynomials becomes the opposite of the one contained in $u$). \\ 

Linear relations for Feynman integrals can be derived by projections using intersection numbers, purely
algebraically, in the same way as any vector admits a decomposition in terms of a basis, within a vector space.
Intersection theory for twisted de Rham (co)-homology provides the mathematical framework of a vector space structure, characterized by its {\it dimension}, its {\it bases} and its {\it scalar product}, which we present in the following.

\subsection{Dimension of twisted cohomology groups}
\label{subsec:Number_of_Master_Integrals}

In ref.~\cite{Lee:2013hzt},  
the number of MIs within the IBP-decomposition was related to the number of independent contours of integration, generating no surface terms. 
The condition in eq.~(\ref{eq:B_vanishing_on_parital_C}) relates the {\it geometric} properties of the multivariate polynomial ${\cal B}$ to the {\it analytic} properties derived from the integration domain.
In particular, using a correspondence between the basis cycles and the critical points of the graph-polynomial of the considered integral parametrization,
the number of MIs was related to the rank of the homology groups $H_{n}^{\pm \omega}$. 

In refs.~\cite{Mastrolia:2018uzb,Frellesvig:2019kgj,Frellesvig:2019uqt}, we considered a dual, equivalent description of the same problem, in terms of independent differential forms. 
Accordingly, we define $\nu$ as the dimension of the twisted cohomology group, respectively, $H^{n}_{\pm \omega}$, here considered as a vector space,
\begin{equation}
\nu= \text{dim} \,  H^{n}_{\pm \omega} \ .
\end{equation}
The complex Morse (Picard-Lefschetz) theory allows us to determine $\nu$ as the \emph{number of critical points} of the function $\log u(\mathbf{z})$~\cite{Lee:2013hzt}. We define
\begin{equation}
\omega=d \, \log u(\mathbf{z})= \sum_{i=1}^{n} \hat{\omega}_i \, dz_i
\label{eq:small_omega_definition}
\end{equation}
and the number of critical points is given by the number of solutions of the (zero dimensional) system
\begin{equation}
\hat{\omega}_i \equiv \partial_{z_{i}} \log u(\mathbf{z})=0\,, \qquad i=1, \dots, n.
\label{eq:leepom}
\end{equation}
The number of solutions of~(\ref{eq:leepom}) can be determined without computing explicitly its zeros~\cite{Lee:2013hzt}. 
In our applications the function $u({\bf z})$ always takes the form $u({\bf z}) = \prod_j {\cal B}_j^{\gamma_j}({\bf z})$, which gives the equations:
\begin{eqnarray}
\hat{\omega}_{i}=
\sum_j \gamma_j \frac{\partial_{z_i} \mathcal{B}_j}{\mathcal{B}_j}, \quad i=1, \dots, n.
\end{eqnarray}

In the absence of critical points at infinity, the number of solutions of~($\ref{eq:leepom}$) equals to the dimension of the quotient space for the ideal\footnote{We introduce an extra variable $z_0$ in order to prevent the case when $\mathcal{B}_j=0$ for either $j$.}
\begin{eqnarray}
{\cal I} &=& \Big<
\beta_1,\;
\ldots,\;
\beta_n,\;
z_0 \prod_j {\cal B}_j -1
\Big> \  \qquad{}
{\rm with \ } \quad
\beta_k \equiv \textstyle\sum_i \gamma_i \, (\partial_{z_k} {\cal B}_i)
\prod_{j\neq i} {\cal B}_j \ .
\end{eqnarray}
In the special case when $u({\bf z}) = {\cal B}^\gamma({\bf z})$, it becomes simply \cite{Lee:2013hzt}
\be
\mathcal{I}=\langle \partial_{z_1} \mathcal{B},\, \dots,\, \partial_{z_n} \mathcal{B},\, z_0 \, \mathcal{B}-1\rangle \, .
\ee
Considering a Gr{\"o}bner basis $\mathcal{G}$ generating $\mathcal{I}$, the Shape Lemma (see, e.g. \cite{kreuzer2008computational}, and \cite{Mastrolia:2012an} for an application to physics)
ensures that the number $\nu$ of zeros of $\mathcal{I}$, and hence the number of the solutions of the system ($\ref{eq:leepom}$), is the dimension of the quotient ring,
\begin{equation}
    \nu = {\rm dim} ({\mathbb C}[ {\bf z} ] / \langle {\cal G} \rangle ) \ ,
\end{equation}
where ${\mathbb C}[ {\bf z} ]$ is the set of all polynomials that vanish on the zeroes of $\mathcal{I}$ (they identify a discrete variety, $V \subset {\mathbb C}^\nu$).
In particular, the lemma ensures that
the degree of the remainder of the polynomial division modulo $\mathcal{G}$ is $\nu+1$.

In ref.~\cite{Frellesvig:2019uqt}, we recalled that $\nu$
can be computed using one of the many ways of evaluating the topological Euler characteristic $\chi(X)$: $X = \mathbb{CP}^n {-} \mathcal{P}_\omega$, where $\mathcal{P}_\omega \equiv \{ \text{set of poles of }\omega\}$ in projective space. This relation can be written as 
\begin{equation}\label{Euler-characteristic-complement}
\nu = |\chi(X)| = (-1)^n\left( n{+}1  - \chi(\mathcal{P}_\omega)\right),
\end{equation}
where we used $\chi(\mathbb{CP}^n) = n{+}1$ together with the inclusion-exclusion principle for Euler characteristics. In other words, to compute  
$\nu$, it is sufficient to evaluate  $\chi(\mathcal{P}_\omega)$ of the projective variety $\mathcal{P}_\omega$ (see also refs.~\cite{Aluffi:2008sy,Marcolli:2008vr,Bitoun:2017nre}). \\
\\
In the following, we will compute the dimension of the cohomology groups to determine the size of the basis of differential forms for different choices of 
$H^{n}_{\pm \omega}$, each characterized by $\omega$, or correspondingly by $u$.

\subsection{Intersection numbers for twisted (co)homology classes} 

Within twisted de Rham theory, $\langle \varphi_L |$ and $| \varphi_R \rangle$ are elements of the twisted cohomology class $H_{\omega}^{n}$  and the dual 
cohomology class $H_{-\omega}^{n}$ respectively.
Because of a duality between twisted cycles and co-cycles \cite{aomoto2011theory}, 
$[ \mathcal{C}_L |$ and $| \mathcal{C}_R ]$ can be considered as elements of the  homology class $H^{\omega}_{n}$  and the dual homology class $H^{-\omega}_{n}$. 
Beside the two type of pairings that defined the integrals and the dual integrals, respectively $\langle \varphi_L|{\cal C}_R]$ and 
$[{\cal C}_L|\varphi_R\rangle$, defined above, one can consider:
\begin{itemize}
    \item {\it intersection numbers of twisted cycles} 
     $[ \mathcal{C}_L | \mathcal{C}_R ]$, as introduced in \cite{MANA:MANA19941660122};
    \item {\it intersection numbers of twisted co-cycles} 
    $\langle \varphi_L | \varphi_R \rangle$, which were first considered in \cite{cho1995}.
\end{itemize}
While we refer the interested reader to consult the original publications on the topics, 
we will briefly review some properties of intersection numbers for twisted co-cycles (here refereed to also as twisted forms, or simply $n$-forms), which are relevant to our later discussion. 

Given the integrals $I = \langle \varphi_L | {\cal C}_R]$ and 
${\tilde I} = [ {\cal C}_L | \varphi_R \rangle$, 
we define the intersection number between the corresponding 
$n$-forms, $\varphi_L = {\hat \varphi}_L \, dz_1 \wedge \ldots \wedge dz_n$ 
and $\varphi_R = {\hat \varphi}_R \, dz_1 \wedge \ldots \wedge dz_n$ as
\be
\langle \varphi_L | \varphi_R \rangle = \frac{1}{(2\pi i)^n} \int_{X} \varphi_L \wedge \varphi_R \ .
\ee
In the general case, the integral over $X$ can be performed by iteration, and applying Stokes' theorem 
one variable at a time (namely, by splitting it into one-dimensional fibers), it reduces to a nested sequence of contour integrations, 
performed by Cauchy's residue theorem, as it will be shown later (see also sec. 3.2 of \cite{Mizera:2020wdt}).
As required by the proper mathematical definition of the intersection number for twisted cohomology, we assume that $\varphi_L$ and $\varphi_R$ have compact support near the boundary of $X$ and that, until differently specified, they have poles which are regulated by the multi-valued function $u$.

%\subsubsection{Properties}
Two interesting properties of intersection numbers play a role in devising the decomposition algorithm we propose:
\begin{itemize}
    \item Intersection numbers are invariant under a change of differential forms within the same equivalence classes, namely
    \begin{equation}
        \langle \varphi_{L}| \varphi_{R} \rangle= \langle \varphi_{L}^{\prime}| \varphi_{R} \rangle = \langle \varphi_{L}| \varphi_{R}^{\prime} \rangle = \langle \varphi_{L}^{\prime}| \varphi_{R}^{\prime} \rangle \,,
    \end{equation}
    where
    \begin{align}
        \varphi_{L}^{\prime} &= \varphi_{L}+ \nabla_{\omega} \xi_{L} \, , \\
        \varphi_{R}^{\prime} &= \varphi_{R}+ \nabla_{{-}\omega} \xi_{R} \, ,
    \end{align}
    and the covariant derivatives $\nabla_{\pm\omega}$ defined in eqs.~($\ref{eq:nablasmallomegaplus}$) and ($\ref{eq:nablasmallomegaminus}$), explicitly read
    \begin{equation}
        \nabla_{\pm\omega} = \sum_{i=1}^{n} \,  dz_i \left( \partial_{z_i} \pm \hat{\omega}_i \right) \wedge\, ,
    \end{equation}
    while $\xi_{L}$ and $\xi_{R}$ are {\it arbitrary} $(n{-}1)$-forms with poles regulated by $u$.

   The invariance of intersection numbers under the replacement of forms belonging to the same equivalence class can be useful: 
    {\it i)} for substituting differential forms having higher poles with (equivalent) forms that have simple poles \cite{Weinzierl:2020nhw}, as it will be recalled in Sec. \ref{sec:equivalentsimplepoles};
   {\it ii)} for substituting differential forms having poles that are not regulated with (equivalent) forms that are fully regulated, as it will be shown in 
   Sec. \ref{subsubsec:masslessbox_topdown}.

\item Intersection numbers obey 
    the symmetry relation
    \begin{eqnarray}
        \langle \varphi_{L}| \varphi_{R} \rangle_{\omega}=(-1)^{n} \, \langle \varphi_{R}| \varphi_{L} \rangle_{{-}\omega} \, ,
    \end{eqnarray}
    which follows directly from the definition and the fact that commuting $\varphi_L$ with $\varphi_R$ yields a sign change of $(-1)^n$.
    We stress that the right-hand side is evaluated with respect to ${-}\omega$ rather than $\omega$.
\end{itemize}

Before providing the details for the evaluation of intersection numbers, which are going to be presented in Sec.~\ref{sec:multivariate}, 
let us recall their main applications: the derivation of {\it linear and quadratic relations} and of {\it systems of differential equations} for
AG/Feynman Integrals. They can be presented in full generality, purely algebraically, without any specific reference to the number of integration variable and to the explicit computation of intersection numbers.

\subsection{Linear and quadratic relations}

The reduction of a given integral, $I=\langle \varphi_L| \mathcal{C}_R]$, in terms of a set of $\nu$ MIs, $J_i=\langle e_i | \mathcal{C}_R]$
\begin{equation}
I= \sum_{i=1}^{\nu} c_i \, J_i
\label{eq:decomposition_MIs}
\end{equation}
can be interpreted in terms of differential forms, as
\begin{equation}
\langle \varphi_L|= \sum_{i=1}^{\nu} c_i \, \langle e_i|\,,
\label{eq:decomposition_master_forms}
\end{equation}
since the integration cycle is the same for all the integrals of eq.~\eqref{eq:decomposition_MIs}.
Likewise, the decomposition of a \emph{dual} integral $\Tilde{I}=[ \mathcal{C}_L| \varphi_R \rangle$ in terms of a set of $\nu$ \emph{dual} MIs $\Tilde{J}_i=[ \mathcal{C}_L| h_i \rangle$
\begin{equation}
\Tilde{I}= \sum_{i=1}^{\nu} \Tilde{c}_i \, \Tilde{J}_i
\end{equation}
becomes
\begin{equation}
| \varphi_R \rangle= \sum_{i=1}^{\nu} \Tilde{c}_i \, | h_i \rangle.
\label{eq:decomposition_dual_master_forms}
\end{equation}

The coefficients $c_i$, and $\Tilde{c}_i$ in eqs.~(\ref{eq:decomposition_master_forms}),~(\ref{eq:decomposition_dual_master_forms}) are determined by the \emph{master decomposition formulas} \cite{Mastrolia:2018uzb,Frellesvig:2019kgj}
\begin{align}
c_i &= \sum_{j=1}^{\nu} \, \langle \varphi_L | h_j \rangle \, \left( \mathbf{C}^{-1} \right)_{ji} \,, \label{eq:masterdeco} \\
\Tilde{c}_i & =\sum_{j=1}^{\nu} \, \left( \mathbf{C}^{-1} \right)_{ij} \, \langle e_j | \varphi_R \rangle \,, \label{eq:masterdecodual}
\end{align}
where we introduced the (inverse of the) \emph{metric matrix}
\begin{equation}\label{eq:metric_matrix}
\mathbf{C}_{ij}= \langle e_i | h_j \rangle.
\end{equation}
In the above formula $\mathbf{C}$ is a $(\nu \times \nu)$-matrices of intersection numbers of basic forms $\langle e_i|$ and dual-forms $|h_i\rangle$, which, in general, differs from the identity matrix, but, for suitably chosen bases can reduce to it, hence simplifying eqs.~(\ref{eq:masterdeco},\ref{eq:masterdecodual}).
The formal derivation of the latter two equations are given in Appendix \ref{sec:App:masterdecoproofs}.

By substituting eq.~(\ref{eq:masterdeco}) in eq.~(\ref{eq:decomposition_master_forms}) (or eq.~\eqref{eq:masterdecodual} in eq.~(\ref{eq:decomposition_dual_master_forms})), we obtain a representation of the identity operator in the cohomology space
\begin{eqnarray}
\sum_{i,j=1}^{\nu} | h_i \rangle \left( \mathbf{C}^{-1} \right)_{ij} \langle e_j | = \mathbb{I}_c 
\label{eq:identityexp}
\end{eqnarray}
Similarly, in the homology space, the resolution of the identity is 
\begin{eqnarray}
\sum_{i,j=1}^{\nu} | \mathcal{C}_{R,i}]  \left( \mathbf{H}^{-1} \right)_{ij} [ \mathcal{C}_{L,j} | = \mathbb{I}_h \, ,
\end{eqnarray}
where $\mathbf{H}_{ij} = [\mathcal{C}_{L,i} | \mathcal{C}_{R,j}] $ is the metric matrix for the twisted cycles. 
The operators $\mathbb{I}_c$
and $\mathbb{I}_h$ can be inserted either in the bilinear pairing between the twisted cocyles or the twisted cycles,
to obtain the quadratic identities
\begin{eqnarray}
\langle \varphi_L | \varphi_R \rangle &=& \sum_{i,j=1}^{\nu} \langle \varphi_L | \mathcal{C}_{R,i}]  \left( \mathbf{H}^{-1} \right)_{ij} [ \mathcal{C}_{L,j} | \varphi_R \rangle \\
[ \mathcal{C}_{L} | \mathcal{C}_{R}] &=& \sum_{i,j=1}^{\nu} [ \mathcal{C}_{L} | h_i \rangle \left( \mathbf{C}^{-1} \right)_{ij} \langle e_j | \mathcal{C}_{R}] \, ,
\end{eqnarray}
which are known as {\it{Twisted Riemann's Period Relations}} (TRPR) \cite{cho1995}. 
TRPR relates intersection numbers for (co)-homologies to products of integrals and dual integrals.

Let us emphasize that the coefficients $c_i$ in eq.(\ref{eq:masterdeco}) are independent of the choice of the 
dual basis $|h_j \rangle$. 
A suitable choice of the dual basis may simplify the intermediate steps of the evaluation, which requires the separate calculations of the intersection numbers 
$\langle \varphi_L | h_j \rangle$ 
and $\langle e_i | h_j \rangle$. Similar considerations hold for $ \Tilde{c}_i$ in eq.(\ref{eq:masterdecodual}), which are independent of $\langle e_i |$).
Since the master decomposition formula eq.(\ref{eq:masterdeco}) involves the inverse of the matrix $\mathbf{C}$, further simplifications arise when it is close to a diagonal matrix, hence implying that $|h_j \rangle$ and $\langle e_i |$ are as 
{\it othogonal} as possible.
The construction of orthonormal bases of forms can be achieved by the Gram-Schmidt algorithm, using the intersection numbers as scalar products. In the case of 1-form, orthonormal bases can be built directly, simply using the expression of $\omega$ \cite{Frellesvig:2019kgj}.

%For illustration purposes, we set ${\hat e}_i = {\hat h}_i$ throughout, namely implying 
%\begin{equation}
%    \mathbf{C}_{ij}= \langle e_i | e_j \rangle \ ,
%\end{equation}
%although this choice, for the comments given above, might turn out not to be the most convenient choice, %computationally.

Recent mathematical literature on intersection numbers of twisted cycles and co-cycles include application to Gel'fand-Kapranov-Zelevinski systems \cite{matsubaraheo2019euler,matsubaraheo2019algorithm,goto2020homology} and to quadratic relations \cite{Broadhurst:2016hbq, Broadhurst:2016myo,Broadhurst:2018tey,Zhou:2017vhw,Zhou:2017jnm,fresn2020quadratic,fresn2020quadratic2}.

\subsection{Differential equation for forms and dual forms}
\label{sec:DEofFormsAndDualForms}

The decomposition of differential forms in terms of master forms,
implemented by the use of eqs.(\ref{eq:masterdeco},\ref{eq:masterdecodual}),
yields the direct derivation of the systems of differential equations \cite{Mizera:2019gea,Frellesvig:2019uqt}.
In particular, the basis $\langle e_i |$ and the dual basis $| h_i \rangle$ obey a system of first order differential equations, with respect to any external variable, say $x$, respectively reading as,
\begin{eqnarray}
\label{eq:sysdiffeq}
\partial_x \langle e_i | &=& {\bf \Omega }_{ij} \,
\langle e_j | \ ,
\\
\partial_x | h_i \rangle &=& -| h_j \rangle\, \widetilde{{\bf \Omega }}_{ji} \,
 \ .
\label{eq:sysdiffeq:Dual}
\end{eqnarray} 
The matrices $\bf{\Omega}$ and $\widetilde{\bf{\Omega}}$ arise from the decompositions,

\begin{eqnarray}
\partial_x \langle e_i |  &=& 
\langle (\partial_x + \sigma ) e_i |  
=
\underbrace{\langle (\partial_x + \sigma \wedge) e_i | h_k \rangle \,
\left( {\bf C}^{-1} \right)_{kj}}_{{\bf \Omega}_{ij}} \,
\langle e_j |
 \ ,
\\
\partial_x | h_i \rangle &=& 
| (\partial_x - \sigma ) h_i \rangle  
=
| h_j \rangle \underbrace{\left( {\bf C}^{-1} \right)_{jk} \,
\langle e_k| (\partial_x - \sigma \wedge) h_i \rangle}_{- \widetilde{\bf \Omega}_{ji}} 
\ ,
\label{eq:DEQ:sigma}
\end{eqnarray}
where $ \sigma \equiv \partial_x \log u$.
Let us observe that the combinations 
$\partial_x \pm \sigma\wedge \equiv \nabla_{x, \pm \sigma}$ 
may be also interpreted as covariant derivatives. The systems of differential equations for forms directly translates into systems of differential equations for MIs, as follows,
\begin{eqnarray}
\partial_x \, J_i = {\bf \Omega}_{ij} \, J_j \ , \qquad 
\partial_x \, {\tilde J}_i = - {\bf {\tilde \Omega}}_{ij} \, {\tilde J}_j \ , 
\quad i,j=1,\ldots, \nu \ .
\end{eqnarray}
We observe that although the integration domain of $J_i$ and ${\tilde J}_i$, respectively ${\cal C}_R$ and ${\cal C}_L$, may depend on the $x$ variable, the condition that $u$ vanishes at the integration boundaries, $u(\partial{\cal C}_{R,L} ) = 0$, preserves the commutation between the $x$-differentiation and integration.
For the case of Feynman integrals,
$\bf{\Omega}$ and $\widetilde{\bf{\Omega}}$ on the space-time dimension $d$ and on kinematic variables, including $x$.
Quite generally, for Aomoto-Gel'fand integrals, these matrices depend on the external variables and on the parameters appearing in the definition of $u$.

Using the above formulas, one can relate the matrices $\mathbf{\Omega}$ and $\widetilde{\mathbf{\Omega}}$ through the identity
\begin{equation}
\partial_x \langle e_i | h_j \rangle = \big( \partial_x \langle e_i| \big) | h_j \rangle + \langle e_i | \big( \partial_x | h_j \rangle \big) = {\bf \Omega }_{ik} \,
\langle e_k | h_j \rangle - \langle e_i | h_k \rangle \, \widetilde{\bf \Omega}_{kj} \ ,
\end{equation}
which, in matrix notation, reads as,
\begin{align}
\partial_x \mathbf{C}
    = \mathbf{\Omega}\, \mathbf{C}
    -
    \mathbf{C}\, \widetilde{\mathbf{\Omega}}\, .
\end{align}
In particular, for orthonormal bases, $\mathbf{C}=\mathbb{I}$, 
therefore $\mathbf{\Omega} = \widetilde{\mathbf{\Omega}}$. 

Intersection theory has been recently used to identify special bases of Feynman integrals admitting canonical systems of differential equations \cite{Henn:2013pwa} (see also \cite{Argeri:2014qva}), according to the structure of $u$ \cite{Chen:2020uyk}.

\section{Multivariate intersection numbers}
\label{sec:multivariate}
Multivariate intersection numbers constitute the key operation 
for generating linear and quadratic relations among integrals and dual integrals. 
In particular, they enter the decomposition of differential forms in terms of a set of master forms, therefore of the corresponding integrals in terms of master integrals, according to the master decomposition formulas eq.(\ref{eq:masterdeco}, \ref{eq:masterdecodual}). 
It is important to observe that these decomposition formulas hold for generic $n$-forms.
Therefore, algorithms for the evaluation of intersection numbers play an important role in the development of novel strategies for computing scattering amplitudes in Physics as well as for deriving relations among transcendental functions in Mathematics.

\subsection{Intersection numbers of logarithmic forms}
\label{sec:mlogforms}
Intersection numbers for multivariate logarithmic forms were considered in \cite{matsumoto1998,Mizera:2017rqa,Mizera:2019gea}. In particular, if $\varphi_L$ and $\varphi_R$ are both logarithmic differential forms (dlog forms), and $\omega_i$ have simple poles, the intersection numbers 
can be evaluated as: 
 \begin{align}
 \langle \varphi_L | \varphi_R \rangle
 =
({-}1)^n
 \!\!\sum_{(z_1^\ast, \dots,  z_n^\ast)} 
 \!\! {\det}^{-1} 
  \begin{bmatrix} 
  \partial_{z_1} \hat{\omega}_1 &
  \dots &
  \partial_{z_n} \hat{\omega}_1 \\ 
  \vdots & \ddots & \vdots \\
  \partial_{z_1} \hat{\omega}_n & 
  \dots &
  \partial_{z_n} \hat{\omega}_n
  \end{bmatrix} 
  \widehat{\varphi}_L\, \widehat{\varphi}_R \,\Bigg|_{(z_1, \dots, z_n) = (z_1^\ast, \dots  z_n^\ast)}
  \label{eq:IntersectionNumberdLogForm}
 \end{align}
where the sum goes over all the $\nu$ \emph{critical points}, identified with the $n$-ples ${(z_1^\ast, \dots,  z_n^\ast)}$ that solve the system of equations
\begin{equation}
    \hat{\omega}_{i}\equiv \partial_{z_{i}} \log u(\mathbf{z})=0, \quad i=1, \dots, n,
\end{equation}
as in eq. \eqref{eq:leepom}. When at least one of the forms is non-logarithmic, the formula \eqref{eq:IntersectionNumberdLogForm} is only valid asymptotically in the limit $\gamma \to \infty$. In those cases, one can still calculate intersection numbers making use of the above formula within a series expansion in $1/\gamma$, as it was successfully applied to the computation of differential equations for certain Feynman integrals in \cite{Mizera:2019vvs}.

\subsection{Intersection numbers of general forms}

Logarithmic differential forms have been subject of intense mathematical developments. Nonetheless, generic Feynman integrals may correspond to pairing of forms that are not necessarily logarithmic, and therefore 
it becomes necessary to devise algorithms for computation of intersection numbers for general rational forms.

The evaluation of intersection numbers of multivariate differential forms has been introduced in \cite{Mizera:2019gea} and systematized in \cite{Frellesvig:2019uqt} for the derivation of linear relations of Feynman integrals as well as of hypergeometric functions, by adopting an iterative procedure. According to this approach, the calculation of the intersection number of two $n$-forms proceeds recursively, 
in terms of the intersection numbers of $(n-1)$-forms, until reaching the terminating condition, given by the univariate intersection numbers \cite{Mastrolia:2018uzb, Frellesvig:2019kgj}.

One of the goals of this work is to provide a pedagogical introduction to the evaluation of multivariate intersection numbers (for twisted de Rham cohomology) by means of the recursive algorithm.
To this aim, let us consider integrals with $n$ integration variables $\{z_{i_1}, \ldots, z_{i_n}\}$, which can be seen as iterative integrals, with a nested structure that follows from the chosen ordering $\{i_1, \dots, i_k\}$ of the integers $\{1, \dots, n\}$.
In order to compute multivariate intersection numbers for differential $k$-forms, we need to compute the dimension of the cohomology groups for all differential $k$-forms, from $k=1$ to $k=n$. 
They can be obtained, for instance, by counting the number $\nu_{\mathbf{k}}$ of solutions of the system of equations given by eq.~\eqref{eq:leepom},
\begin{eqnarray}
\hat{\omega}_j \equiv \partial_{z_{j}} \log u(\mathbf{z})=0 \,, 
\qquad j=i_1, \dots, i_k \ ,
\end{eqnarray}
where ${\bf k} = \{i_1, \dots, i_k\}$ is a subset of $\{1, \dots, n\}$ with $k$ distinct elements. 
In this way, one obtains a list of dimensions 
\begin{eqnarray}
&& \nu_{\bf 1} \ , \nu_{\bf 2} \ , \ldots, \nu_{\bf n} \ , \quad {\rm with}, 
%\\
%&&
\quad 
{\bf 1} = \{i_1\}\,,\quad
{\bf 2} = \{i_1,i_2\}\,,\quad
\ldots\,, \quad
{\bf n} = \{i_1,i_2,\ldots,i_n\}\,,
\qquad
\end{eqnarray}
 corresponding to the 
number of master integrals within each step of the iterative integration (respectively in  
$\{z_{i_1}\}$, in $\{z_{i_1},z_{i_2} \}$, $\ldots$ , in $\{z_{i_1},\ldots,z_{i_n}\}$).

It is interesting to observe that $\nu_{\bf n}$ is trivially independent
of the ordering of the integration variables. On the other hand, the sequence of the dimensions of all the subspaces
$\nu_{\bf k}$ may indeed change according to the chosen permutation of $\{ 1,2, \ldots, n\}$ that correspond to 
the ordering of nested integrations.
As a working principle, we choose the ordering that minimizes the sizes
of $\nu_{\bf k}$ for all $k$-forms ($k=1,\ldots,n$).

\subsection{Intersection numbers for $1$-forms}
\label{sec:InterX-1-forms}

Let us briefly recall the intersection number for $1$-forms \cite{cho1995}, discussed at length in \cite{Mastrolia:2018uzb,Frellesvig:2019kgj}.
Consider a generic integral with one integration variable, 
\begin{eqnarray}\label{I-integral-univariate}
I &= \int_{{\cal C}_R^{({\bf 1})}} \varphi_L^{(\bf{1})} \! \left( z_1\right) \, u(z_1)
= \langle \varphi_L^{(\bf{1})} | {\cal C}_R^{({\bf 1})} ] \, \, ,
\end{eqnarray} 
where ${\bf 1} = \{ 1\}$. 
Similarly, we consider dual integrals of the type,
\begin{eqnarray}
\tilde{I} &= \int_{{\cal C}_L^{({\bf 1})}} \varphi_R^{(\bf{1})} \! \left( z_1 \right)  \, u^{-1}(z_1) 
= [ {\cal C}_L^{({\bf 1})} |  \varphi_R^{(\bf{1})} \rangle \, \, .
\end{eqnarray}
We compute $\omega_1 = d {\rm log} u(z_1)  = {\hat \omega}_{1} dz_1$, and determine $\nu_{\bf 1}$ by counting the critical points of 
${\hat \omega}_{z_1}$. 
Simultaneously, we define ${\cal P}_{\omega_1}$ as the sets of its poles (including the pole at $\infty$). 
Then, the intersection number between $1$-forms can be computed as,
\begin{eqnarray}
\langle \varphi_L^{(\bf{1})} | \varphi_R^{(\bf{1})}\rangle 
= \sum_{p \in {\cal P}_{\omega_1} } {\rm Res}_{z_1=p} \Big[ 
\psi^{(p)} \, \varphi_R^{(\bf{1})} 
\Big] \ , 
\end{eqnarray}
where $\psi^{(p)}$ is the local solution of the differential equation
\begin{eqnarray}
\nabla_{\omega_{1}} \psi^{(p)} = \varphi_L^{(\bf{1})} \, ,
\end{eqnarray}
around the point $z_1 = p$.

\subsection{Intersection numbers for $2$-forms}

We consider instructive to show how the intersection numbers of 2-forms can be written recursively in terms of intersection numbers of 1-forms.

Consider an integral with two integration variables $\{z_1, z_2\}$, generically written as,
\begin{eqnarray}\label{I-integral}
I &= \int_{{\cal C}_R^{({\bf 2})}} \varphi_L^{(\bf{2})} \! \left( z_1, z_2\right) \, u(z_1, z_2)
= \langle \varphi_L^{(\bf{2})} | {\cal C}_R^{({\bf 2})} ] \, \, ,
\end{eqnarray}
where ${\bf 2} = \{ 1, 2 \}$, 
$\varphi_{L}^{(\mathbf{2})}$ is a differential $2$-form in the variables $z_1$ and $z_2$, 
{\it i.e.} $\langle \varphi_L| = {\hat \varphi}_L(z_1, z_2) \, dz_1 \wedge dz_2$,
while ${\cal C}^{(\mathbf{2})}_R$ is a two-dimensional integration domain embedded in some ambient space $X$ with complex dimension $2$.
We assume%
\footnote{This does not necessarily mean that $X = X_2 \times X_1$, since $X_1 = X_1(z_2)$ 
can depend on $z_2$ (but $X_2$ does not depend on $z_1$).}
that $X$ admits a fibration into one-dimensional spaces, say $X_2 \ni z_2$ and $X_1 \ni z_1$, 
yielding the corresponding decompositions of $\varphi_{L}^{(\mathbf{2})}$, ${\cal C}^{(\mathbf{2})}_R$. Similarly, we can consider a dual integral, given by
\begin{eqnarray}
\tilde{I} &= \int_{{\cal C}_L^{({\bf 2})}} \varphi_R^{(\bf{2})} \! \left( z_1, z_2\right)  \, u^{-1}(z_1, z_2) 
= [ {\cal C}_L^{({\bf 2})} |  \varphi_R^{(\bf{2})} \rangle \, \, .
\end{eqnarray}
Our goal is the evaluation of $\langle \varphi_L^{(\bf{2})} | \varphi_R^{(\bf{2})} \rangle$, in terms of intersection numbers for $1$-forms.

Given $u(z_1,z_2) \equiv u({\bf z})$, we define:
\begin{equation}
\omega=d \, \log u(\mathbf{z})= \sum_{i=1}^{2} \hat{\omega}_i \, dz_i \ .
\label{eq:omgdef2var}
\end{equation}
From $\omega$ we determine:  
the dimension $\nu_{\bf 1}$ with ${\bf 1} = \{1\}$, counting the solutions of ${\hat \omega_1} = 0$; 
and the dimension $\nu_{\bf 2}$ with ${\bf 2} = \{1,2\}$, counting the solutions of the system ${\hat \omega_1} = 0, {\hat \omega_2} = 0$.
The former number $\nu_{\bf 1}$ corresponds to the number of master 1-forms in $z_1$, which correspond to the MIs emerging from the integration in $z_1$, while the latter, $\nu_{\bf 2}$, to the number of master 2-forms in $z_1$ and $z_2$, therefore to the number of MIs of the integrals $I$  
in eq.~\eqref{I-integral}. 

We can therefore {\it choose} the bases of forms $\langle e_i^{({\bf 1})}|$ and $|h_i^{({\bf 1})}\rangle$ for $i=1,\ldots,\nu_{\bf 1}$, 
and compute the metric matrix ${\mathbf C}_{{\bf (1)}}$, {\it i.e.} the matrix of intersection numbers,
\begin{equation}
\big({\mathbf C}_{{\bf (1)}}\big)_{ij} \equiv\langle e_i^{({\bf 1})} | h_j^{({\bf 1})} \rangle \ \, .
\end{equation}

We make use of eqs. (\ref{eq:masterdeco},\ref{eq:masterdecodual}) to decompose the $2$-forms in terms of $1$-forms, by projecting the former on the chosen bases of $1$-forms,
\begin{eqnarray}
\label{varphi-L-2-projection}
\langle \varphi_{L}^{({\bf 2})} |
&=& \sum_{i=1}^{\nu_{\bf{1}}} \langle e^{{\bf (1)}}_i
| \wedge \langle \varphi_{L,i}^{(2)} | \ , \qquad 
| \varphi_{R}^{({\bf 2})} \rangle =
\sum_{i=1}^{\nu_{\bf{1}}} |h^{{\bf (1)}}_i \rangle  \wedge | \varphi_{R,i}^{(2)} \rangle  
\ ,
\end{eqnarray}
with 
\begin{align}
\label{eq:varphi-LR-decomp}
\langle \varphi_{L,i}^{(2)}| & = 
\langle \varphi_{L}^{(\mathbf{2})}|h_{j}^{(\mathbf{1})} \rangle \; 
\big( \mathbf{C}_{(\mathbf{1})}^{-1} \big)_{ji} \ , \\
&& \nonumber \\
| \varphi_{R,i}^{(2)} \rangle  & = \big({\mathbf C}_{({\bf 1})}^{-1}\big)_{ij}\;
\langle e_j^{({\bf 1})} | \varphi_R^{(\mathbf{2})} \rangle \ \,. 
\label{eq:varphi-LR-decomp-2}
\end{align}

To compute the intersection numbers of 2-forms, we also need
the $(\nu_{\bf 1} \times \nu_{\bf 1})$-matrix $\mathbf{\Omega}^{(2)}$ associated to the system of differential equations in $z_2$ obeyed by the bases $\langle e_i^{({\bf 1})} |$,
\begin{eqnarray}
\partial_{z_2} \langle e^{{\bf (1)}}_i | \equiv \langle (d_{z_2} + \omega_2 \wedge) e_i^{{\bf (1)}}|
= \mathbf{\Omega}^{(2)}_{ij} \langle e^{{\bf (1)}}_j| \ .
\end{eqnarray}
${\bf {\Omega}}^{(2)}$ is obtained by projecting the form $\langle (d_{z_2} + \omega_2 \wedge) e_i^{{\bf (1)}}|$ on the basis 
$\langle e^{{\bf (1)}}_i|$, using eq. (\ref{eq:masterdeco}),
\begin{eqnarray}
\mathbf{\Omega}^{(2)}_{ij} &=& 
\langle (d_{z_2} + \omega_2 \wedge) e_i^{(\bf{1})} | h_k^{(\bf{1})} \rangle ({\mathbf C}_{{\bf (1)}}^{-1})_{kj} \, .
\end{eqnarray}
The intersection number for 2-forms can be finally computed as \cite{Mizera:2019gea},
\begin{eqnarray}
    \langle \varphi_{L}^{({\bf 2})} | \varphi_{R}^{({\bf 2})} \rangle 
    &=&\sum_{i,j =1}^{\nu_{\bf 1}} \sum_{q \in \mathcal{P}_{\mathbf{\Omega}^{(2)}}} \text{Res}_{z_2=q} \left[ \psi_{i}^{(q)} \big({\mathbf C}_{{\bf (1)}}\big)_{ij} \varphi_{R,j}^{(2)}
      \right] \, ,
\label{eq:bivariatedef}
\end{eqnarray}
where $\psi_{i}^{(q)}$ is the local solution of the differential equation
\begin{eqnarray}
\nabla_{\mathbf{\Omega}^{(2)}} \psi_{i}^{(q)} = 
d_{z_2} \psi_{i}^{(q)} + \psi_{j}^{(q)} \wedge \mathbf{\Omega}^{(2)}_{ji} = \varphi_{L,i}^{(2)} \ , 
\end{eqnarray}
around each point $q \in {\cal P}_{{\bf \Omega}^{(2)}} \equiv \{ {\rm poles \ of\ } {\bf \Omega}^{(2)} \ ({\rm including}\ \infty)\}$ \, .

As shown in eq. \eqref{eq:bivariatedef}, the intersection number of $2$-forms  
$\langle \varphi_{L}^{({\bf 2})} | \varphi_{R}^{({\bf 2})} \rangle$ has been expressed  
in terms of quantities that are either intersection numbers of $1$-forms or can be derived through them.

\subsubsection*{Example}

Let us consider intersection numbers for integrals
of the type in eq. (\ref{I-integral}), where:
\begin{equation}
u(\mathbf{z}) = \big(z_1 z_2 (1{-}z_1{-}z_2) \big)^{\gamma} \, ,
\end{equation}
which gives
\begin{equation}
\hat{\omega}_1 = \gamma \left(\frac{1}{z_1} - \frac{1}{1-z_1-z_2} \right),\qquad
\hat{\omega}_2 = \gamma \left(\frac{1}{z_2} - \frac{1}{1-z_1-z_2} \right).
\end{equation}
We will focus on the steps required for the  computation of the self-intersection number 
of the 2-form $\langle 1| \equiv d z_1 \wedge d z_2$ (simply given as 1 times the wedge product of the two elementary differentials),
which, using the notation introduced above, can be written as
\begin{equation}
\langle \varphi_{L}^{(\mathbf{2})}| \varphi_{R}^{(\mathbf{2})} \rangle
\qquad \text{with} \qquad \hat{\varphi}_{L}^{(\mathbf{2})}=\hat{\varphi}_{R}^{(\mathbf{2})}=1 \, .  
\end{equation}
%These forms have poles only at infinities.

Within the iterative approach, we consider first the integration in $z_1$
and define ${\bf 1} = \{1\}$. Since $\hat{\omega}_1=0$ has one solution, 
\begin{equation}
\nu_{\bf 1}=1 \, ,
\end{equation}
implying that the number of master 1-forms in $z_1$ is just 1. Therefore,
we choose the inner basis for the left and right forms, denoted by $\langle e^{(\mathbf{1})} |$ and $| h^{(\mathbf{1})} \rangle$ respectively, as
\begin{equation}
\hat{e}^{(\mathbf{1})}=\hat{h}^{(\mathbf{1})}=z_1 \, .
\end{equation}
Given two arbitrary forms $\langle \varphi_{L}^{(\mathbf{2})}|$ and $|\varphi_{R}^{(\mathbf{2})} \rangle$, 
we decompose them as,
\begin{equation}
%\begin{split}
\langle \varphi_{L}^{(\mathbf{2})}|  = \langle e^{(\mathbf{1})}| \wedge \langle \varphi_{L}^{(2)}|\;, \qquad 
| \varphi_{R}^{(\mathbf{2})} \rangle  = | h^{(\mathbf{1})} \rangle \wedge | \varphi_{R}^{(2)} \rangle\;,
%\end{split}
\end{equation}
where $\langle \varphi_{L}^{(2)}|$ and $| \varphi_{R}^{(2)} \rangle$, are 1-forms in the variable $z_2$, and can be determined by projecting
the 2-forms on the bases of 1-forms, using eqs. \eqref{eq:varphi-LR-decomp} and \eqref{eq:varphi-LR-decomp-2}:
\begin{equation}
\langle \varphi_{L}^{(2)}|  = \langle \varphi_{L}^{(\mathbf{2})}|h^{(\mathbf{1})} \rangle \, \, \,  \mathbf{C}_{(\mathbf{1})}^{-1} \ , \qquad
|\varphi_{R}^{(2)} \rangle  =  \mathbf{C}_{(\mathbf{1})}^{-1} \, \,  \,\langle e^{(\mathbf{1})}|\varphi_{R}^{(\mathbf{2})} \rangle \ ,
\end{equation}
with
\begin{equation}
\mathbf{C}_{(\mathbf{1})}=\langle e^{(\mathbf{1})}| h^{(\mathbf{1})} \rangle.
\end{equation}
Within the recursive approach, the evaluation of the required intersection numbers of 1-forms w.r.t. $z_1$ constitutes the first step, and  
they are given by,
\begin{align}
\mathbf{C}_{(\mathbf{1})} &= \langle z_1 | z_1 \rangle = \frac{\gamma (z_2-1)^4}{8 (2 \gamma -1) (2 \gamma +1)}\,, \\
\hat{\varphi}_{L}^{(2)} &= \langle 1 | z_1 \rangle \,\, \mathbf{C}^{-1}_{(\mathbf{1})} = \frac{-2}{z_2-1}\,,\\
\hat{\varphi}_{R}^{(2)} &= \mathbf{C}^{-1}_{(\mathbf{1})} \,\, \langle z_1 | 1 \rangle = \frac{-2}{z_2-1}\, .
\end{align}
Univariate intersection numbers in $z_1$, are also needed to compute the 
$(1\times 1)$ connection matrix $\hat{\mathbf{\Omega}}^{(2)}$, 
\begin{equation}
\hat{\mathbf{\Omega}}^{(2)}=\langle (\partial_{z_2}+\hat{\omega}_2) \, z_1|z_1 \rangle \, \mathbf{C}^{-1}_{(\mathbf{1})}=\frac{(3 \gamma +2) z_2-\gamma }{\left(z_2-1\right) z_2},
\end{equation}
which is needed for the next step. 
We observe that the set of the poles of $\hat{\mathbf{\Omega}}^{(2)}$ is,
\begin{equation}
\mathcal{P}_2=\{ 0 , 1, \infty \}.\\
\end{equation}
Next, we consider the differential equation:
\begin{equation}
\left( \partial_{z_2}+\hat{\mathbf{\Omega}}^{(2)} \right) \psi^{(2)}= \hat{\varphi}_{L}^{(2)}.
\label{eq:DEQ_2vars_intersection_example}
\end{equation}
The full analytic solution of ($\ref{eq:DEQ_2vars_intersection_example}$) is not required, but rather a \emph{power series} around each $p \in \mathcal{P}_2$ is sufficient. 
Denoting by $y$ the local coordinate around the pole, the solutions of ($\ref{eq:DEQ_2vars_intersection_example}$) to leading orders in $y$ read:
\begin{itemize}
\item \textbf{Solution around $p=0$ ($y=z_2$)}: 
\begin{equation}
\psi^{(2)}_{0}(y)=\frac{2 y}{\gamma +1}+{\cal O}\left(y^2\right);
\end{equation}
\item \textbf{Solution around $p=1$ ($y=z_2-1$)}:
\begin{equation} 
\psi^{(2)}_{1}(y)=-\frac{1}{\gamma +1}+{\cal O}\left(y^1\right);
\end{equation}
\item \textbf{Solution around $p=\infty$ ($y=1/z_2$)}: 
\begin{equation}
\begin{split}
\psi^{(2)}_{\infty}(y)=c_{0,\infty }+ c_{1,\infty } \, y + c_{2,\infty } \,  y^2+c_{3,\infty } \, y^3+c_{4,\infty } \,  y^4+ {\cal O}\left(y^5\right)
\end{split}
\end{equation}   
with
\begin{align}
& c_{0,\infty } = \frac{-2}{3 \gamma +2}\,,
& c_{1,\infty } & = \frac{-2 \gamma }{(3 \gamma +1) (3 \gamma +2)}\,, \nonumber \\
& c_{2,\infty } = \frac{-2 (\gamma -1)}{3 (3 \gamma +1) (3 \gamma +2)}\,,
& c_{3,\infty } & = \frac{-2 (\gamma -2) (\gamma -1)}{3 (3 \gamma -1) (3 \gamma +1) (3 \gamma +2)}\,,  \nonumber \\
& c_{4,\infty } = \frac{-2 (\gamma -3) (\gamma -2) (\gamma -1)}{3 (3 \gamma -2) (3 \gamma -1) (3 \gamma +1) (3 \gamma +2)}\,. \!\!\!\! &&
\end{align}
\end{itemize}
Finally, we may evaluate the bi-variate intersection number as a sum of univariate residues, as given by eq. \eqref{eq:bivariatedef}:
\begin{equation}
\langle \varphi_{L}^{(\mathbf{2})}| \varphi_{R}^{(\mathbf{2})} \rangle= \sum_{p \in \mathcal{P}_2} \Res_{z_2=p} \left( \psi^{(2)} \,  \mathbf{C}_{(\mathbf{1})} \,  \varphi_{R}^{(2)} \right) \ , 
\end{equation}
yielding the final expression,
\begin{equation}
 \langle 1 | 1 \rangle=\frac{\gamma ^2}{3 (3 \gamma -2) (3 \gamma -1) (3 \gamma +1) (3 \gamma +2)} \ .
\end{equation}

\subsection{Intersection numbers for $n$-forms}

Following the above discussion, we can generalize the intersection number of 2-forms to the case of $n$-forms.  
In this case, we start by considering an integral with $n$ integration variables $(z_1, z_2, \ldots, z_n)$, written as
\begin{eqnarray}
I \left(z_1, z_2, \ldots, z_n \right) = \int_{{\cal C}_R^{({\bf n})}} \varphi_L^{(\bf{n})} \! \left( z_1, z_2, \ldots, z_n \right)  \, u(z_1, z_2, \ldots, z_n) 
= \langle \varphi_L^{(\bf{n})} | {\cal C}_R^{({\bf n})} ] 
\end{eqnarray}
with the notation ${\bf n} = \{1,\ldots,n \}$. The $\varphi_{L}^{(\mathbf{n})}$ is an $n$-variable differential form on some space $X$. Similarly, one can define a dual form $\varphi_{R}^{(\mathbf{n})}$. 
We assume that the $n$-complex-dimensional space with coordinates $(z_1,\ldots,z_n)$ admits a 
fibration into a $(n{-}1)$-dimensional subspace parametrized by 
$(z_1,\ldots,z_{n-1})$, denoted by ${\bf n{-}1}$,
which we call the {\it inner} space, and a one-dimensional subspace with $z_n$, 
which we refer to as the {\it outer} space. We have
\begin{equation}
\omega=d \, \log u(\mathbf{z})= \sum_{i=1}^{n} \hat{\omega}_i \, dz_i
\label{eq:omgdefnvar}
\end{equation}
and employing eq.~(\ref{eq:leepom}), we can count the number of MIs on 
the {\it inner space}, which we
define as
$\nu_{\bf{n-1}}$.
The aim is to express intersection number for $n-$forms 
$ \langle \varphi_L^{({\bf n})} | \varphi_R^{(\bf n)} \rangle$ in terms of 
intersection numbers for $(n{-}1)$-forms on the inner space, which are assumed to be known at this stage, following the recursive nature of the algorithm. 
The choice of the variables (and their ordering) parametrizing the inner and outer spaces is arbitrary: as before, we use the generic notation $\mathbf{k} \equiv \{i_1, i_2, \dots, i_k\}$ to denote the variables taking part in a specific computation. 

Thus, 
the original ${\bf n}$-forms can be decomposed according to 
 \begin{eqnarray}
 \label{varphi-L-projection}
\langle \varphi_{L}^{({\bf n})} |
 &=& \sum_{i=1}^{\nu_{\bf{n-1}}} \langle e^{{\bf (n-1)}}_i | \wedge 
 \langle \varphi_{L,i}^{(n)} | \ , \\ 
 \label{varphi-R-projection}
 | \varphi_{R}^{({\bf n})} \rangle &=& 
 \sum_{i=1}^{\nu_{\bf{n-1}}} |h^{{\bf (n-1)}}_i \rangle \wedge 
 | \varphi_{R,i}^{(n)} \rangle 
 \ ,
 \end{eqnarray}
where $\nu_{\bf{n-1}}$ is the number of master integrals on the inner space with arbitrary bases $\langle e^{{\bf (n-1)}}_i |$, $|h^{{\bf (n-1)}}_i \rangle$.
In the above expressions $\langle \varphi_{L,i}^{(n)} |$ and $| \varphi_{R,i}^{(n)} \rangle$ are one-forms in the variable $z_n$, and they treated as coefficients of the basis expansion.
They can be obtained by a projection similar to eq.~\eqref{eq:masterdeco}, giving
\begin{eqnarray}
\label{eq:varphi-LR}
\langle \varphi_{L,i}^{(n)}| 
&=& 
\langle \varphi_{L}^{(\mathbf{n})}|h_{j}^{(\mathbf{n-1})} \rangle \, 
\big( \mathbf{C}_{(\mathbf{n-1})}^{-1} \big)_{ji} \ ,\\
& & \nonumber \\
| \varphi_{R,i}^{(n)} \rangle 
&=& 
\big({\mathbf C}_{({\bf n-1})}^{-1}\big)_{ij} \, 
\langle e_j^{({\bf n-1})} | \varphi_R^{(\mathbf{n})} \rangle \ ,\label{eq:varphi-LR-2}
\end{eqnarray}
with
\begin{equation}
\big({\mathbf C}_{{\bf (n-1)}}\big)_{ij} =\langle e_i^{({\bf n-1})} | h_j^{({\bf n-1})} \rangle \ .
\end{equation}
It is important to remark that, within the recursive approach, the intersection numbers of ${\bf (n\!-\!1)}$-forms (depending on ${\bf (n\!-\!1)}$ variables) are assumed to be known.
The recursive formula for the intersection number reads \cite{Mizera:2019gea}:
\begin{eqnarray}
\hspace*{-0.5cm}
\langle \varphi_L^{({\bf n})} | 
\varphi_R^{({\bf n})} \rangle \;
{=} \sum_{p \in {\cal P}_{n} } \!
\Res_{z_n = p} \!
\Big( \psi_{i}^{(n)}  \, \left(\mathbf{C}_{(\mathbf{n-1})} \right)_{ij} \, \varphi_{R,j}^{(n)}
\Big) \ ,
\label{eq:multivarIntNumb}
\end{eqnarray}
where the functions $\psi^{(n)}_i$ are the solution of the system of differential equations
\begin{eqnarray}
\partial_{z_n} 
  \psi^{(n)}_i 
+\psi^{(n)}_j \; {\hat{\mathbf\Omega}}^{(n)}_{ji}= 
{\hat \varphi}^{(n)}_{L,i} \ ,
\label{eq:sysofdeq}
\end{eqnarray}
and $\hat{\varphi}_{L,i}$ are obtained through eq.~\eqref{eq:varphi-LR}.
Here, $\hat{\mathbf{\Omega}}^{(n)}$ is a $\nu_{\mathbf{n-1}} \times \nu_{\mathbf{n-1}}$ matrix, whose entries are given by
\begin{eqnarray}
\label{Omega-n-2}
\hat{\mathbf{\Omega}}^{(n)}_{ji}=\langle (\partial_{z_n}+\hat{\omega}_{n}) e^{(\mathbf{n-1})}_{j}|h^{(\mathbf{n-1})}_{k} \rangle \; \big( \mathbf{C}_{(\mathbf{n-1})}^{-1} \big)_{ki} \;
\end{eqnarray}
and finally ${\cal P}_{n}$ is the set of poles of ${\hat{\mathbf\Omega}}^{(n)}$ defined as the union
of the poles of its entries (including a possible pole at infinity).

We observe that the solution of eq.~\eqref{eq:sysofdeq} around $z_n{=}p$ can be formally written 
in terms of a path-ordered matrix exponential\\
\begin{equation}
\vec{\psi}^{(n)}(z_n)= \left( \mathcal{P}e^{-\int_{p}^{z_n} \mathbf{\Omega}^{(n)^{\text{T}}}(w)}\right) \left( \int_{p}^{z_n} \mathcal{P}e^{\int_{p}^{y} \mathbf{\Omega}^{(n)^{\text{T}}}(w)} \; \vec{\varphi}_{L}^{(n)}(y) \right)
\end{equation}
for a vector $\vec{\psi}^{(n)}$ with entries $\psi^{(n)}_i$. Nevertheless for its use in eq.~\eqref{eq:multivarIntNumb}, it is sufficient to know only a few leading orders of ${\vec{\psi}^{(n)}}$ around each $p \in {\cal P}_{n}$. 
Therefore, it is easier to find the solution of the system eq.~\eqref{eq:sysofdeq} by a {\it holomorphic} Laurent series expansion, using an ansatz for each component $\psi^{(n)}_i$, see \cite{Mastrolia:2018uzb,Frellesvig:2019kgj}. Such a solution exists if the matrix $\Res_{z_n = p} {\mathbf\Omega}^{(n)}$ does not have any non-negative integer eigenvalues, which we assume from now on (when this is not the case one can employ a regularization discussed in Sec.~\ref{subsec:straight_decomposition_method}).
Moreover, the number of critical points of the determinant of the ${\bf \Omega}^{(n)}$
provides
the dimension of that cohomology group, {\it i.e.} the number of the corresponding master forms~\cite{Weinzierl:2020xyy}.

The recursion terminates when $n{=}1$, in which case the inner space is trivial: $\nu_{\mathbf{0}} = \langle e_1^{(\mathbf{0})}| = | h_1^{(\mathbf{0})}\rangle = 1$, and we impose the initial conditions
\begin{eqnarray}
{\hat{\mathbf\Omega}}^{(1)}_{11} = \hat{\omega}_1 \, , \quad \mathbf{C}_{\mathbf{0}}=1 \, , \quad \varphi_{L,1}^{(1)} = \varphi_L^{(\bf{1})} ,\quad
\varphi_{R,1}^{(1)} = \varphi_R^{(\bf{1})} .
\end{eqnarray}
In this case eq.~(\ref{eq:multivarIntNumb}) reduces to a computation of an univariate intersection number~\cite{cho1995,matsumoto1998,Mastrolia:2018uzb,Frellesvig:2019kgj}, discussed in Sec. \ref{sec:InterX-1-forms}.

Let us observe that the matrix ${\bf \Omega}^{(n)}$ is important to define the equivalence classes, 
\begin{eqnarray}
 \varphi_{L}^{(n)} \sim \varphi_{L}^{(n)} + \nabla_{\bf \Omega}^{(n)} \xi(z_n) \ ,
\end{eqnarray}
where the covariant derivative, defined as
$\nabla_{\bf \Omega}^{(n)} \equiv {\mathbb I} \, d_{z_n} + {\bf \Omega}^{(n)} \ ,$ (${\mathbb I}$ is the $(\nu_{\bf n-1} \times \nu_{\bf n-1})$ identity matrix in the $({\bf n-1})$ subspace)
acts on any arbitrary function $\xi(z_n)$ -- see Appendix \ref{sec:App:masterdecoproofs} for a formal derivation.

\subsubsection{Explicit formula}
Let us notice also that combining eqs. \eqref{eq:multivarIntNumb} and \eqref{eq:varphi-LR-2} gives
\begin{align}
\hspace*{-0.5cm}
\langle \varphi_L^{({\bf n})} | \varphi_R^{({\bf n})} \rangle &=
\sum_{p \in {\cal P}_{n} } \! \Res_{z_n = p} \!
\Big( \psi_{i}^{(n)}  \, \langle e_i^{({\bf n-1})} | \varphi_R^{(\mathbf{n})} \rangle \Big) \,,
\end{align}
which is suitable for practical calculation purposes.
Using the above identity recursively, the intersection number can be expressed as,
\begin{equation}
\langle \varphi_L^{({\bf n})} 
|\varphi_R^{({\bf n})} \rangle 
= 
\!\!\!\sum_{p_{n} \in {\cal P}_{n}} \!\!\cdots\!\!\!
\sum_{ p_{1} \in {\cal P}_{1}} 
\Res_{z_n = p_n} \cdots
\Res_{z_1 = p_1}
\Big( 
\psi_{i_{\bf n-1}}^{(n)}
\,
\psi_{i_{\bf n-1}i_{\bf n-2}}^{(n-1)}
\cdots
\,
\psi_{ i_{\bf 2}i_{\bf 1}}^{(2)} 
\,
\psi_{i_{\bf 1} 1}^{(1)}
\,
\varphi_R^{({\bf n})}
\Big) ,
\label{eq:multivarIntNumb:IterativePsi}
\end{equation}
where the ranges of the summations are $i_{\bf m} = 1,\dots,\nu_{\bf m}$ and where the $\psi^{(m)}_{i_{\bf m}i_{\bf m-1}}$ are the solutions of
\begin{align}
\partial_{z_m} \psi^{(m)}_{i_{\mathbf{m}} i_{\mathbf{m-1}}} + \psi^{(m)}_{i_{\mathbf{m}} j_{\mathbf{m-1}}} \hat{\mathbf \Omega}^{(m)}_{j_{\mathbf{m-1}} i_{\mathbf{m-1}}} &= \hat{e}^{(m)}_{i_{\mathbf{m}} i_{\mathbf{m-1}}}
\label{eq:psi-m-DE}
\end{align}
for all $i_{\bf m}$ with $\langle e^{(m)}_{i_{\mathbf{m}}i_{\mathbf{m-1}} } | = \hat{e}^{(m)}_{i_{\mathbf{m}}i_{\mathbf{m-1}} } d z_m$ coming from the projection
\begin{align}
\langle e^{({\bf m})}_{i_{\bf m}} | &= 
\langle e_{i_{\bf m-1}}^{({\bf m-1})} | \wedge
\langle e_{i_{\bf m} i_{\bf m-1}}^{(m)} | \ ,
\end{align}
which may be computed initially, since the bases of all inner spaces are arbitrarily chosen. 
The matrices $\hat{\mathbf\Omega}^{(m)}$ needed in eq.~\eqref{eq:psi-m-DE} are computed analogously to eq.~\eqref{Omega-n-2}. 
Notice that all $\psi^{(m)}$ entering eq.~\eqref{eq:multivarIntNumb:IterativePsi} need to be computed only {\it once} for a given family of integrals.

\subsubsection{Dual formula}

Let us discuss an alternative recursive formula for intersection numbers, which uses the dual connection matrix $\widetilde{\bf \Omega}^{(n)}$ instead of ${\bf \Omega}^{(n)}$. 
This amounts to repeating the same steps presented in the former section, but using the decomposition of the differential dual-forms given in eq.~\eqref{varphi-R-projection} (instead of eq.~\eqref{varphi-L-projection}),
\begin{eqnarray}
    \langle \varphi_{L}^{({\bf n})} | \varphi_{R}^{({\bf n})} \rangle 
    &=& - \sum_{i,j =1}^{\nu_{\bf 1}} \, \sum_{q \in \mathcal{P}_{n} } 
          \text{Res}_{{z_n}=q} 
    \left[  \varphi_{L,i}^{(n)} 
            \big({\mathbf C}_{{\bf (n-1)}}\big)_{ij} \psi_{j}^{(q)}
      \right] \, ,
\end{eqnarray}
\\
where,
\begin{align}
     \widetilde{\mathbf{\Omega}}^{(n)}_{ij} =& -
     ({\mathbf C}_{{\bf (n-1)}}^{-1})_{ik} \, 
     \langle e_k^{(\bf{n-1})} |  (d_{z_n} - \omega_n \wedge) h_j^{(\bf{n-1})}\rangle  \, ,
\end{align}
${\cal P}_n$ is the set of its poles (including the pole at $\infty$),
and $\psi_{j}^{(q)}$ is the solution of 
\begin{eqnarray}
\nabla_{\widetilde{\mathbf{\Omega}}^{(n)}} \psi_{j}^{(q)} = 
d_{z_n} \psi_{j}^{(q)} - \widetilde{\mathbf{\Omega}}^{(n)}_{ji} \wedge \psi_{i}^{(q)}  = \varphi_{R,j}^{(n)} \, .
\end{eqnarray}

The dual formula provides a useful consistency check for the computation of intersection numbers, and it can be combined with the original formula
to devise efficient evaluation algorithms which can better exploit the pole structure of $\varphi_L$, $\varphi_R$, and of the chosen bases 
$e_i$ and $h_i$, in order to minimize the computational load.

\subsection{Simplifying the computation of intersection numbers}
\label{sec:equivalentsimplepoles}

The recursive algorithm for the computation of the multivariate intersection numbers presented in Sec.~$\ref{sec:multivariate}$
is applicable for any rational form. However, at each step of the recursive algorithm, the coefficients $\hat{\varphi}_{L,R}^{(n)}$ in eqs.~($\ref{varphi-L-projection}$), ($\ref{varphi-R-projection}$) are defined modulo the equivalence relations
\begin{eqnarray}
\hat{\varphi}^{(n)}_{L,i} &\;\sim\;& \hat{\varphi}^{\prime \, (n)}_{L,i}=\hat{\varphi}^{(n)}_{L,i}+ \Big( \partial_{z_n} \xi_{L,i} + \xi_{L,j} \,  \hat{\mathbf{\Omega}}^{(n)}_{ji} \Big) \, , \\[1mm]
\hat{\varphi}^{(n)}_{R,i} &\;\sim\;& \hat{\varphi}^{\prime \, (n)}_{R,i}=\hat{\varphi}^{(n)}_{R,i}+ \Big( \partial_{z_n} \xi_{R,i}- \hat{\widetilde{\mathbf{\Omega}}}^{(n)}_{ij} \, \xi_{R,j}  \Big) \, .
\end{eqnarray}
Thus, under the assumption that the connection matrices $\mathbf{\Omega}^{(n)}$ and $ \widetilde{\mathbf{\Omega}}^{(n)}$ contain only simple poles, its possible to replace the coefficients $\hat{\varphi}_{L,R}^{(n)}$ containing higher-degree poles, with a suitably chosen $\hat{\varphi}_{L,R}^{\prime (n)}$ belonging to the same equivalence class, but containing simple poles only. One may exploit this fact to compute intersection numbers in one variable as a univariate global residue, without introducing any algebraic extensions as observed in~\cite{Weinzierl:2020xyy}.

\section{Feynman integral decomposition}
\label{sec:feynintdeco}
As proposed in refs.~\cite{Mastrolia:2018uzb, Frellesvig:2019kgj, Frellesvig:2019uqt, Mizera:2019vvs, Weinzierl:2020xyy}, the use of multivariate intersection numbers yields a direct
decomposition of a given Feynman integral $I$ in terms of an a priori chosen set of
MIs $J_i$, with $i=1,\ldots,\nu$.\\
The decomposition given by eq.~\eqref{eq:decomposition_MIs} is on the form
\begin{align}
    I &= \sum_{i=1}^{\nu} c_i J_i,
\end{align}
where the determination of the coefficients $c_i$ is the goal of this section.
We identify {\it three} possible strategies which can be adopted in order to
achieve this task. They all employ the master projection formula from
eq.~(\ref{eq:masterdeco}), which is applied to differential forms constructed differently in the three cases. We name them the {\it straight decomposition}, the {\it bottom-up decomposition}, and the {\it top-down decomposition}.\\
All the approaches have the first step in common: \emph{finding the number} of MIs which appear in the decomposition and \emph{choosing} them accordingly.\\
We introduce the following definitions:
\begin{itemize}
\item $\Sigma$ denotes the set of integers used to label the full set of denominators;
\item $\sigma$ denotes a set of integers that label a subset of denominators, $\sigma \subseteq \Sigma$;
\item \emph{sector} is the set of integrals for which only the subset of propagators specified by $\sigma$ appear in the denominator (thus, a sector is unambiguously identified by $\sigma$).
\end{itemize}
There is a \emph{one-to-one} correspondence between sectors and (generalized unitarity) cuts. On the level of the function $u$, this correspondence is manifested by setting all $z_j$'s belonging to $\sigma$ to zero in the original $u({\bf z})$,
\begin{equation}
u_{\sigma} = u(\mathbf{z})|_{z_{j \in \sigma} \to 0},
\end{equation}
where we work in Baikov representation.
Given $u_{\sigma}$, the number of MIs in the corresponding sector, $\nu_{\sigma}$, can be determined through the criteria given in Sec.~\ref{subsec:Number_of_Master_Integrals}. The total number of MIs (without taking into account any symmetry relations) is then given by
\begin{equation}
\nu= \sum_{\sigma} \,  \nu_{\sigma},
\end{equation}
where the sum is over all sectors.
Finally we can choose the forms $\langle e_i|$ associated to the (arbitrarily chosen) MIs $J_i$, through the identification
\begin{equation}
J_i = \langle e_i| \mathcal{C}].
\end{equation}

\subsection{Straight decomposition}
\label{subsec:straight_decomposition_method}
We consider the following decomposition
\bea
\label{eq:decomposition_straight}
I = \int_{\cal C} u \, \varphi \,=\, 
\langle \varphi | {\cal C} ] \,=\, 
\sum_{i=1}^{\nu} c_{i} \, \langle e_i | {\cal C} ] \,=\,
\sum_{i=1}^{\nu} c_{i} \int_{\cal C} u \, e_i \,=\, 
\sum_{i=1}^{\nu} c_{i} \,  J_{i}
\eea
with
\begin{eqnarray}
c_{i} = \sum_{j=1}^\nu \langle \varphi | h_j \rangle \big({\mathbf C}^{-1}\big)_{ji} \ , 
\qquad
{\mathbf C}_{ij} = \langle e_i | h_j \rangle\, \ .
\end{eqnarray}
Here ${\hat \varphi}$ and ${\hat e_i}$ correspond simply to the integrands of the integral $I$ to decompose and of the chosen master integrals, $J_i$, respectively. In order to evaluate the intersection numbers, all the poles present in the differential forms must be regulated in $u$. If this assumption is violated, we can introduce a \emph{regulated} $u$, denoted by $u_{\rho}$, which contains a monomial $z_k^{\rho_k}$ for each (non-regulated) pole present in the differential forms,
that is
\begin{equation}\label{u-reg}
u_{\rho}(\mathbf{z}) = \left( \prod_{k \in \Sigma} z_{k}^{\rho_k} \right)  u(\mathbf{z})
\end{equation}
and correspondingly
\begin{equation}
\omega_{\rho}(\mathbf{z})= d \log u_{\rho}(\mathbf{z})=d \log u(\mathbf{z}) + \sum_{k \in \Sigma} \rho_k \, \frac{dz_k}{z_k} =\omega(\mathbf{z}) + \sum_{k \in \Sigma} \rho_k \, \frac{d z_k}{z_k} \, ,
\end{equation}
where we emphasized the action of regulators. By analogy, we also introduce a regularized version of $\hat{\mathbf{\Omega}}^{(n)}$, whenever $\text{Res}_{z_n=p} \,  \hat{\mathbf{\Omega}}^{(n)}$ has any non-negative integer eigenvalue. The regularized $\hat{\mathbf{\Omega}}^{(n)}$ reads:
\begin{equation}
    \hat{\mathbf{\Omega}}^{(n)}_{\Lambda}=\hat{\mathbf{\Omega}}^{(n)}+ \frac{\Lambda}{z_n-p} \, \mathbb{I} \, .
\end{equation}
Thus, we obtain a new system of differential equations, analogous to eq.~($\ref{eq:sysofdeq}$), which is, in this case, controlled by $\hat{\mathbf{\Omega}}^{(n)}_{\Lambda}$. We assume that the solution of the latter around a pole $p$, denoted by $\psi^{(n)}_{\Lambda,p}$, reproduces in the limit $\Lambda \to 0$, a solution for the original system (around the pole $p$).\\ 
The intersection numbers are computed through $\omega_{\rho}$, and lead to a set of coefficients, denoted by $c_{\rho,i}$, which depend on the set of regulators, collectively indicated by $\rho$. The coefficients $c_i$, which appear in the original decomposition eq.~(\ref{eq:decomposition_straight}), are recovered in the limit $\rho \to 0$,\footnote{Strictly speaking, we take it as an assumption that the limit $\rho \to 0$ is smooth, which turns out to be true in all practical examples we studied. This might seem reasonable given that the regularization used in \eqref{u-reg} is a version of analytic regularization for Feynman integrals \cite{speer1969theory} which cannot have poles in $\rho_k$ as long as dimensional regularization is also employed. However, there might exist situations where a MI $J_i$ has a \emph{zero} in $\rho$ compensated by a pole in $\rho$ of $c_i$, leading to a finite result: in this case, the product $c_i J_i$ has a smooth limit, but not each term individually.}
\begin{equation}
c_i= \lim_{\rho \to 0} c_{\rho,i} =
\lim_{\rho \to 0} \, 
\sum_{j=1}^\nu \langle \varphi | h_j \rangle_{\rho} \, 
\big({\mathbf C_{\rho}}^{-1}\big)_{ji} \ , 
\qquad
({\mathbf C}_\rho)_{ij} = \langle e_i | h_j \rangle_{\rho}\, \ .
\label{eq:masterdeco_regulated_straight}
\end{equation}
This approach requires the evaluation of intersection numbers, for which all the integration variables are present simultaneously.\\
For ease of notation, whenever the regulated $u$ is introduced, in the following we will omit the subscript $\rho$ from the individual intersection numbers $\langle \varphi| h_j \rangle_{\rho}$ and $\langle e_i|h_j \rangle_{\rho}$.

\subsection{Bottom-up decomposition}
\label{subsec:Bottom-up decomposition}
In this approach, proposed in~\cite{Frellesvig:2019uqt}, the decomposition is applied to the \emph{spanning set of cuts}, defined as the minimal set of cuts such that each MIs appears at least once~\cite{Larsen:2015ped, Frellesvig:2019uqt} (a cut
behave like a \emph{high-pass} filter, therefore MIs whose denominators do not contain {\it all} the cut-denominators will not contribute to the decomposition on that cut). We denote a given \emph{spanning cut} (i.e. an element in the spanning set of cuts) by $\tau$; moreover $\mathcal{S}_{\tau}$ is the set of sectors which survive on that spanning cut
\begin{equation}
\mathcal{S}_{\tau}= \{ \sigma \, | \sigma \supseteq \tau \} \, .
\end{equation}
Finally, the number of MIs which survive on the spanning cut $\tau$, denoted by $\nu_{\mathcal{S}_{\tau}}$ is
\begin{equation}
\nu_{\mathcal{S_{\tau}}}= \sum_{\sigma \in \mathcal{S}_{\tau}} \nu_{\sigma} \, .
\end{equation}
On the spanning cut $\tau$, we define
\begin{equation}
u_{\tau}=u(\mathbf{z})|_{z_{j \in \tau} \to 0}
\end{equation}
and we consider the following decomposition
\begin{equation}
\label{eq:decomposition_bottom_up}
\begin{split}
I_{\tau} \,=\, \int_{\mathcal{C}_{\tau}} u_{\tau} \, \varphi_{\tau} \,=\, \langle \varphi_{\tau}| \mathcal{C}_{\tau}] & =\sum_{i=1}^{\nu_{\mathcal{S}_{\tau}}} c_{i} \, \langle e_{i,\tau}| \mathcal{C}_{\tau}]\\
& = \sum_{i=1}^{\nu_{\mathcal{S}_{\tau}}} c_i \int_{\mathcal{C}_{\tau}} \! u_{\tau} \, e_{i,\tau} \, = \, \sum_{i=1}^{\nu_{\mathcal{S}_{\tau}}} c_i \, J_{i,\tau}
\end{split}
\end{equation}
with
\begin{equation}
c_i= \sum_{j=1}^{\nu_{\mathcal{S}_{\tau}}} \langle \varphi_{\tau}| h_{j,\tau} \rangle \left( \mathbf{C}^{-1} \right)_{ji} \, , \qquad \mathbf{C}_{ij}= \langle e_{i,\tau}| h_{j,\tau} \rangle \, .
\label{eq:ccoefficients_bottom_up}
\end{equation}
As expected, $\hat{\varphi}_{\tau}$ and $\hat{e}_{i,\tau}$ are inferred from the cut-integrals. As in any unitarity-based approach \cite{Primo:2016ebd, Frellesvig:2017aai, Harley:2017qut}, the coefficients $c_{i}$
determined from a cut decomposition are identical to those appearing in the original decomposition -- the coefficients are invariant under cuts. Therefore, the complete decomposition for the (uncut) integral $I$ can be obtained by combining the coefficients determined from the individual spanning cuts.\\
As described in Subsec.~\ref{subsec:straight_decomposition_method}, all the poles present in the differential forms must be regulated in $u_{\tau}$. If this is not the case, we can introduce the \emph{regularized} $u_{\tau}$, denoted by $u_{\rho, \tau}$
\begin{equation}
u_{\rho, \tau}= \left( \prod_{k \in \Sigma \setminus \tau} z_k^{ \rho_k} \right) u_{\tau},
\end{equation}
which leads to 
\begin{equation}
\omega_{\rho, \tau}= d \log u_{\rho,\tau} = d \log u(\mathbf{z}) + \sum_{k \in \Sigma \setminus \tau} \rho_k \, \frac{dz_k}{z_k}= \omega(\mathbf{z}) + \sum_{k \in \Sigma \setminus \tau} \rho_k \, \frac{d z_k}{z_k},
\end{equation}
used in the evaluation of the intersection number. 
We also use a regularized version of $\hat{\mathbf{\Omega}}^{(n)}$, whenever $\text{Res}_{z_n=p} \,  \hat{\mathbf{\Omega}}^{(n)}$ has any non-negative integer eigenvalue, as explained above.
Now, the coefficients of the decomposition, $c_{\rho,i}$ depend on the set of regulators $\rho$. The coefficients of the original decomposition~(\ref{eq:decomposition_bottom_up}) are recovered in the $\rho \to 0$ limit:
\begin{equation}
c_{i}= \lim_{\rho \to 0} c_{\rho,i}
=
\lim_{\rho \to 0}\sum_{j=1}^{\nu_{\mathcal{S}_{\tau}}} \langle \varphi_{\tau}| h_{j,\tau} \rangle_{\rho} \left( \mathbf{C}_{\rho}^{-1} \right)_{ji} \, , \qquad (\mathbf{C}_{\rho})_{ij}= \langle e_{i,\tau}| h_{j,\tau} \rangle_{\rho} \, .
\label{eq:coefficients_bottom_up:regulated}
\end{equation}
This procedure requires the evaluation of the intersection numbers only for the uncut variables, therefore it can be significantly less demanding than the previous case.\\
As before, whenever the regulated $u$ is introduced, we will omit the subscript $\rho$ from the individual intersection numbers.

\subsection{Top-down decomposition}
\label{sec:topdowndeco}
This approach is new and combines the advantages of the decomposition by intersection numbers with the top-down subtraction algorithm traditionally used in methods of {\it integrand decomposition} \cite{Ossola:2006us,Ellis:2007br,Ellis:2008ir,Mastrolia:2012bu,Zhang:2012ce,Mastrolia:2012an,Mastrolia:2011pr,Badger:2013gxa}
In particular, as for the integrand decomposition, one can determine the coefficients of the MIs systematically, beginning from the ones with the highest number of internal lines (the top sector) and moving downward, ending with the sector with a minimal number of lines equal to the number of the loops (built from product of tadpoles). At any step, the determination of the coefficients of a given MI, say $J_i$, is obtained on the corresponding cut, after {\it subtracting off} the known contributions coming from higher sectors, as the latter are written as a linear combination of the MIs with a higher number of internal lines (whose graph contain the one corresponding to $J_i$ as subdiagram), coming from the earlier steps of the decomposition.
 In particular, let us reconsider the complete decomposition,
\bea
I \,=\, \int_{\cal C} u \, \varphi \,=\, 
\langle \varphi | {\cal C} ] \,=\, 
\sum_{i=1}^{\nu} c_i \langle e_i | {\cal C} ] \,=\,
\sum_{i=1}^{\nu} c_i \! \int_{\cal C} \! u \, e_i \,=\, 
\sum_{i=1}^{\nu} c_i J_i \ ,
\eea
and assume that, within the top-down approach, after \emph{at most} $n$-steps,
the coefficients $c_i$, with $i=1,\ldots,n$ have been determined,
and can be considered as known. 
We can write,
\bea
I - \! \sum_{i=1}^{n} \! c_i J_i \,=\, 
\sum_{i=n+1}^{\nu} c_i J_i  
\ ,
\label{eq:integralsubtraction}
\eea
which, in terms of pairings, reads,
\bea
\langle \phi_{n} | {\cal C} ] = 
\sum_{i=1}^{n} c_i \langle e_i | {\cal C} ] \ , 
\eea
where $\langle \phi_n|$, defined as,
\bea
\langle \phi_n| \equiv \langle \varphi | - \! \sum_{i=1}^{n} \! c_i \langle e_i | 
\eea
is a {\it known} differential form.
By applying a cut $\tau$, namely $z_j = 0, \forall j \in {\tau}$,
we can then determine the coefficients $c_i$ for $i = n+1, \ldots n+\nu_\tau$,
where $\nu_\tau$ is the number of those MIs that have as denominators exclusively (all and only) the cut ones, namely $z_j$, with $j \in {\tau}$.
In fact,
on the cut $\tau$, we can define
\begin{equation}
u_{\tau}=u(\mathbf{z})|_{z_{j \in \tau} \to 0}
\end{equation}
and
\begin{equation}
\omega_{\tau}= d \log u_{\tau}    
\end{equation}
and the decomposition simplifies and becomes, 
\bea
\begin{split}
I_{\tau} \,=\, \int_{\mathcal{C}_{\tau}} \! u_{\tau} \, \phi_{n,\tau} \,=\,
\langle \phi_{n, \tau} |\mathcal{C}_{\tau}] & = \! \sum_{i=n+1}^{n+\nu_{\tau}} \! c_i
\langle e_{i,\tau}|\mathcal{C}_{\tau}] \\ 
& = \! \sum_{i=n+1}^{n+\nu_{\tau}} \! c_i \int_{\mathcal{C}_{\tau}} \! u_{\tau} \, e_{i,\tau} \,=\, \! \sum_{i=n+1}^{n+\nu_{\tau}} \! c_i \, J_{i, \tau}
\end{split}
\eea
with
\begin{eqnarray}
c_{n+i} = \sum_{j=1}^{\nu_{\tau} } 
\langle  \phi_{n, \tau} | h_{n+j, \tau}  \rangle \big({\mathbf C}^{-1}\big)_{ji} \ , 
\qquad
{\mathbf C}_{ij} = \langle e_{n+i,\tau}  | h_{n+j, \tau}  \rangle \ .
\end{eqnarray}
Two important observations are in order.
First, we notice that the subtraction in eq.~(\ref{eq:integralsubtraction}),
is similar in spirit to the subtraction performed in an integrand decomposition,
although the known coefficients depend also on $d$, and not only
on the kinematical variables. 
Second, after the subtraction of the known terms, the differential
form $\phi_{n,\tau}$ may contain {\it spurious poles}, which are
not regulated by $u_{\tau}$. 
By exploiting the equivalence class properties, we build an equivalent form, $\phi_{n,\tau}^\prime \sim \phi_{n, \tau}$, which is free of them, 
\bea
 \phi_{n,\tau}^\prime
\equiv
\phi_{n, \tau} + \nabla_{\omega_{\tau}} \xi_{L,\tau} \ ,
\eea
for a suitable choice of $\xi_{L,\tau}$. 
Thus, in this approach, the regulators are not introduced.
At this point the determination of the coefficients via intersection
numbers can proceed iteratively, {\it top-down}, until all sectors have had their $c_i$ coefficients determined.

We would like to observe that the top-down decomposition algorithm offers the advantage of a unitarity-based integrand-decomposition in terms of a minimal bases of MIs.

Let us finally remark that the exploitation of relations within equivalence class of differential forms for eliminating the contributions of poles that do not appear as being regulated is a novel idea which we plan to elaborate on in the future: this approach might be interestingly combined with the more recent mathematical idea of {\it relative twisted homology and cohomology groups} \cite{matsumoto2018relative}, to be used for computing intersection numbers without regulators, as well as, more generally, to investigate the finiteness of scattering amplitudes around explicit dimensions.

\section{Examples}
\label{sec:examples}
In this section we illustrate the previously-discussed decomposition algorithms on a few examples.

\subsection{One-loop massless box}
\begin{figure}[H]
    \centering
   \includegraphics[width=0.25\textwidth,clip=true, trim = 0 225 0 100]{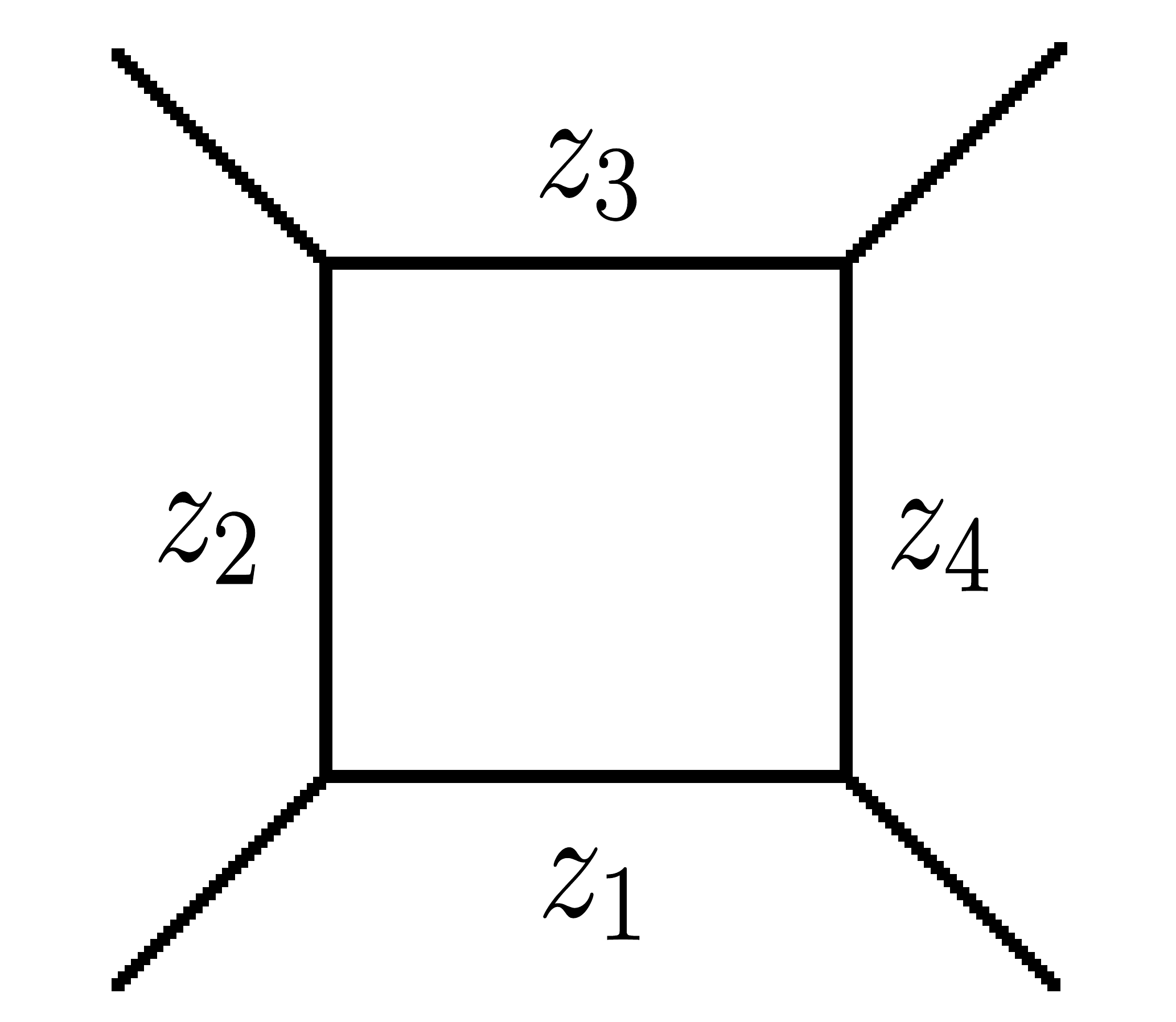}
    \caption{Massless Box}
    \label{fig:MasslessBox}
\end{figure}

As the first example we will discuss the one-loop massless box as shown in Fig.~\ref{fig:MasslessBox}. This diagram was discussed in the context of intersection theory already in ref.~\cite{Frellesvig:2019uqt}, but we will here add further details, and go through the reduction with each of the three methods presented in Sec. \ref{sec:feynintdeco}.

The kinematics is such that
\begin{align}
D_1 &= k^2, & D_2 &= (k+p_1)^2, \nonumber \\
D_3 &= (k+p_1+p_2)^2, & D_4 &= (k+p_1+p_2+p_3)^2,
\end{align}
with $p_i^2=0$, $(p_1+p_2)^2 = s$, $(p_2+p_3)^2 = t$, $(p_1+p_3)^2 = -s-t$.

Performing the Baikov parametrization yields
\begin{equation}
u = {\mathcal{B}}^{(d-5)/2}
\end{equation}
with
\begin{align}
\mathcal{B} &= 2 s t \big( s (z_2 + z_4) + t (z_1 + z_3) - z_1 z_2 - z_2 z_3 - z_3 z_4 - z_4 z_1 + 2 z_1 z_3 + 2 z_2 z_4 \big) \nonumber \\[1mm]
& \;\;\; - s^2 t^2 - t^2 (z_1 - z_3)^2 - s^2 (z_2 - z_4)^2
\end{align}
and performing the sector-by-sector analysis described in the beginning of Sec. \ref{sec:feynintdeco} yields $\nu_{\sigma} = 1$ for the sectors
\begin{align}
\sigma \in \big\{\{1,2,3,4\}\,,\;\{1,3\}\,,\;\{2,4\} \big\}
\end{align}
and $\nu_{\sigma}=0$ for the remaining sectors, corresponding to the well-known set of master integrals: the box and the $s$- and the $t$-channel bubble:
\begin{equation}
J_1=
\begin{gathered}
\includegraphics[width=0.12\textwidth,valign=c]{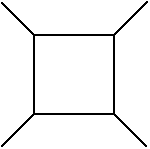}
\end{gathered}
, \qquad
J_2=
\begin{gathered}
\includegraphics[width=0.12\textwidth,valign=c]{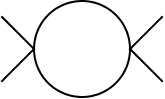}
\end{gathered},
 \qquad
J_3=
\begin{gathered}
\includegraphics[width=0.12\textwidth, angle=90,valign=c]{figure/bubble_s_channel.png}
\end{gathered}
\ .
\end{equation}
The corresponding differential forms read
\begin{align}
\hat{e}_1 = \frac{1}{z_1 z_2 z_3 z_4}\,,\;\; \hat{e}_2 = \frac{1}{z_1 z_3}\,,\;\; \hat{e}_3 = \frac{1}{z_2 z_4}\,.
\end{align}
In the following we will decompose the example
\begin{align}
\begin{gathered}
\includegraphics[width=0.12\textwidth,valign=c]{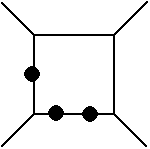}
\end{gathered}
=
\int u \,  \frac{d^{4} \mathbf{z}}{z_1^3 z_2^2 z_3 z_4} \, ,
\end{align}
which can be expressed in terms of the chosen master integrals as
\begin{equation}\label{box-reduction}
\begin{gathered}
\includegraphics[width=0.12\textwidth,valign=c]{figure/Box_3dots.png}
\end{gathered} 
=
c_1 \, 
\begin{gathered}
\includegraphics[width=0.12\textwidth,valign=c]{figure/Box_Nodots.png}
\end{gathered}
+ 
c_2 \, 
\begin{gathered}
\includegraphics[width=0.12\textwidth,valign=c]{figure/bubble_s_channel.png}
\end{gathered}
+
c_3 \,  \, 
\begin{gathered}
\includegraphics[width=0.12\textwidth, angle=90,valign=c]{figure/bubble_s_channel.png}
\end{gathered} \ .
\end{equation}
We will determine these coefficients with the three methods presented in Sec.~$\ref{sec:feynintdeco}$.

\subsubsection{Straight decomposition}
\label{subsubsec:masslesbox_straight}
As prescribed in Sec.~\ref{subsec:straight_decomposition_method} we may construct the regulated $u$ as
\begin{align}
u_{\rho} = u \times z_1^{\rho} z_2^{\rho} z_3^{\rho} z_4^{\rho},
\end{align}
where in this case we pick the regulators to be all equal.
From this definition we may construct the corresponding $\omega$ as
\begin{align}
\omega_{\rho} &= \sum_{i=1}^4 \hat{\omega}_i \, \id z_i \qquad \text{with} \qquad \hat{\omega}_i = \partial_{z_i} \text{log}\, u_{\rho}.
\end{align}
Choosing the variable ordering to be, from the innermost to the outermost, $z_4, z_3, z_2, z_1$, we can compute the dimensions of the twisted cohomology groups corresponding to the individual layers of the fibration. The result is
\begin{align}
\nu_{\{4321\}}=3 \,,\;\; \nu_{\{432\}}=4 \,,\;\; \nu_{\{43\}}=3 \,,\;\; \nu_{\{4\}}=2 \,.
\end{align}
Corresponding to the order of variables given above, we pick the basis for each level to be
\begin{align}
\hat{e}^{(4321)} = \hat{e} &= \left\{ \frac{1}{z_1 z_2 z_3 z_4}  , \frac{1}{z_1 z_3} , \frac{1}{z_2 z_4}\right\} , &  \hat{e}^{(432)} &= \left\{\frac{1}{z_2},\frac{1}{z_3},\frac{1}{z_2 z_3},\frac{1}{z_2 z_3 z_4}\right\} , \nonumber \\
\hat{e}^{(43)} &= \left\{\frac{1}{z_4},\frac{1}{z_3},\frac{1}{z_3 z_4}\right\} , & 
 \hat{e}^{(4)} &= \left\{\frac{1}{z_4},1\right\} .
\end{align}
We choose the dual bases to be $\hat{h}_i=\hat{e}_i$.
In the following, we will decompose
\begin{align}
\hat{\varphi} &= \frac{1}{z_1^3 z_2^2 z_3 z_4}.
\end{align}
The required intersection numbers are
\begin{equation}
    \mathbf{C}_{ij}= \langle e_i | h_j \rangle, \quad 1 \leq i,j \leq 3,
\end{equation}
and
\begin{equation}
    \langle \varphi | h_k \rangle, \quad 1 \leq k \leq 3.
\end{equation}
The individual intersection numbers, up to the leading order in $\rho$, are presented in App.~\ref{appendix:Box}.\\
Combining the intersection numbers as dictated by eq.~($\ref{eq:masterdeco_regulated_straight}$), we obtain, after taking the limit $\rho \rightarrow 0$, the coefficients
\begin{align}
    c_1 &= \frac{-(d-7) (d-6) (d-5)}{2 s^2 t} \, ,  &  c_2 &= \frac{2 (d-7) (d-5) (d-3) }{s^4 t} \, , \nonumber \\[-8mm] \nonumber
\end{align}
\begin{align}
    c_3 &= \frac{2 (d-7) (d-5) (d-3) (2 s + (d-8) t)}{(d-8) s^2 t^4} \, .
    \label{eq:olboxcoefs}
\end{align}
These results are in agreement with the values obtained with \textsc{FIRE}~\cite{Smirnov:2014hma}.

\subsubsection{Bottom-up decomposition}
\label{subsubsec:masslessbox_bottomup}
The first step of a bottom-up decomposition is to identify a spanning set of cuts $\tau$. That set is easily seen to be the cuts corresponding the two bubbles
\begin{align}
\tau \in \big\{ \{1,3\}\,,\;\{2,4\} \big\} \, .
\end{align}
$\bullet$ \textbf{Cut} $\tau= \{ 1,3\}$. Let us first consider the $\tau = \{1,3\}$ cut.\\
On this cut, the decomposition reads:
\begin{equation}
\begin{gathered}
\includegraphics[width=0.14\textwidth,valign=c]{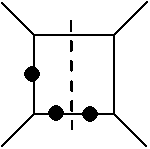}
\end{gathered} 
=\;\;
c_{1}
\begin{gathered}
\includegraphics[width=0.14\textwidth,valign=c]{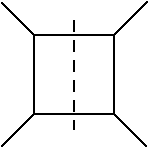}
\end{gathered}
+ 
c_{2} \, 
\begin{gathered}
\includegraphics[width=0.14\textwidth,valign=c]{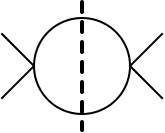}
\end{gathered}
\, .
\end{equation}
We have 
\begin{align}
u_{\rho,\tau} &= z_2^{\rho} \, z_4^{\rho} \, \mathcal{B}_{\tau}^{(d-5)/2} \, (-s)^{(d-5)/2}
\end{align}
where
\begin{align}
\mathcal{B}_{\tau} &=  \Big(s t^2 + s (z_2 - z_4)^2 - 2 t \big( s (z_2 + z_4) + 2 z_2 z_4 \big) \Big) \,,
\end{align}
and $\omega_{\rho,\tau} = \hat{\omega}_2 \, \id z_2 + \hat{\omega}_4 \, \id z_4$ with
\begin{align}
    \hat{\omega}_2=\partial_{z_2} \log u_{\rho,\tau} \, , \qquad \hat{\omega}_4=\partial_{z_4} \log u_{\rho,\tau} \, .
\end{align}
The variable ordering, from the innermost to the outermost, is chosen as $z_2, z_4$. The dimensions of the cohomology groups read:
\begin{equation}
    \nu_{\{24\}}=2 \, , \quad \nu_{\{2\}}=2 \, .
\end{equation}
The basis elements, on the cut, are:
\begin{equation}
    \hat{e}_{\tau}^{(24)}=\hat{e}_{\tau}= \Big\{ \frac{1}{z_2 z_4}, 1 \Big\} \, , \quad \hat{e}_{\tau}^{(2)}= \Big\{1, \frac{1}{z_2} \Big\} \, .
\end{equation}
The dual basis elements are chosen as $\hat{h}_{i,\tau}=\hat{e}_{i,\tau}$.\\
We will show the decomposition, on the cut, of:
\begin{align}
\hat{\varphi}_{\tau} &= \frac{\frac{1}{2} \, \partial_{z_1}^2 u}{u \, z_2^2 z_4} \big|_{z_1,z_3=0}
= \frac{(d-5) t^2 \big( (d-6) s (z_2 + z_4 - t)^2 -4 (s + t) z_2 z_4 \big)}{2 s z_2^2 z_4 \, \mathcal{B}_{\tau}^2} \,.
\end{align}
This requires the intersection numbers
\begin{equation}
    \mathbf{C}_{ij}= \langle e_{i,\tau} | h_{j,\tau} \rangle \,, \quad 1 \leq i,j \leq 2 \,,
\end{equation}
and
\begin{equation}
 \langle \varphi_{\tau} | h_{k,\tau} \rangle \,, \quad 1 \leq k \leq 2 \,.    
\end{equation}
Expressions for the individual intersection numbers are presented in Appendix $\ref{appendix:Box}$.
Combining them as prescribed by eq.~($\ref{eq:coefficients_bottom_up:regulated}$), and considering the limit $\rho \to 0$, we obtain the coefficients $c_1$ and $c_2$ in agreement with eq.~($\ref{eq:olboxcoefs}$).\\
\\
$\bullet$ \textbf{Cut} $\tau=\{2,4\}$. Performing instead the decomposition on the second of the spanning cuts, $\tau = \{2,4\}$ will allow us to reconstruct $c_1$ and $c_3$ in eq.~($\ref{eq:olboxcoefs}$), which means that in total all of the master integral coefficients $c_i$ have been extracted.
\subsubsection{Top-down decomposition}
\label{subsubsec:masslessbox_topdown}

The first step in the top-down decomposition is the extraction of the box-coefficient.
\begin{equation}
\begin{gathered}
\includegraphics[width=0.14\textwidth,valign=c]{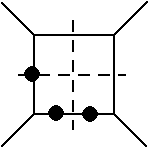}
\end{gathered}
=c_1 \, 
\begin{gathered}
\includegraphics[width=0.14\textwidth,valign=c]{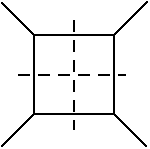}
\end{gathered}\ .
\end{equation}
The coefficient $c_1$ can be computed as $\varphi/e_1$ on the maximal cut:
\begin{align}
c_1 &= \frac{\frac{1}{2} \partial_{z_1}^2 \partial_{z_2} u}{u} \Big|_{z_i \rightarrow 0} = \frac{-(d-7) (d-6) (d-5)}{2 s^2 t}\,,
\end{align}
in agreement with eqs.~\eqref{eq:olboxcoefs}.

We then consider the $s$-channel bubble corresponding to the cut $\tau = \{1,3\}$. 
\begin{equation}
\begin{gathered}
\includegraphics[width=0.14\textwidth,valign=c]{figure/Box_3dots_s_channel_cut.png}
\end{gathered}
-c_1 \, 
\begin{gathered}
\includegraphics[width=0.14\textwidth,valign=c]{figure/box_s_channel_cut.png}
\end{gathered}
=
c_{2} \, 
\begin{gathered}
\includegraphics[width=0.13\textwidth,valign=c]{figure/bubble_s_channel_cut.png}
\end{gathered}\ .
\end{equation}
Here we have
\begin{align}
u_{\tau} &= \mathcal{B}_{\tau}^{(d-5)/2} \, (-s)^{(d-5)/2} \qquad \text{with} \qquad \mathcal{B}_{\tau} = \Big(s t^2 + s (z_2 {-} z_4)^2 - 2 t \big( s (z_2 {+} z_4) + 2 z_2 z_4 \big) \Big)\,,
\end{align}
and
\begin{align}
\hat{\varphi} &= \frac{\frac{1}{2} \, \partial_{z_1}^2 u}{u \, z_2^2 z_4} \big|_{z_1,z_3=0}
\,=\, \frac{(d-5) t^2 \big( (d-6) s (z_2 + z_4 - t)^2 -4 (s + t) z_2 z_4 \big)}{2 s z_2^2 z_4 \, \mathcal{B}_{\tau}^2} \, .
\end{align}
We also get
\begin{align}
\omega &= \frac{-(d-5) \Big( \big( t (z_4{-}z_2) {+} s (t {+} 2 z_4) \big) \, \id z_2 + \big(s (t {+} 2 z_2) {+} t (z_2 {-} z_4) \big) \, \id z_4 \Big)}{\mathcal{B}_{\tau}}
\end{align}
from which we can extract $\nu_{\tau} = 1$ corresponding to the $s$-channel bubble.

We know that
\begin{align}
\begin{gathered}
\includegraphics[width=0.12\textwidth,valign=c]{figure/Box_3dots_s_channel_cut.png}
\end{gathered}
-c_1 \, 
\begin{gathered}
\includegraphics[width=0.12\textwidth,valign=c]{figure/box_s_channel_cut.png}
\end{gathered}
= & 
\int u_{\tau}
\underbrace{
\left(
\hat{\varphi} - \frac{c_1}{z_2 z_4}
\right)
}_{\equiv \, \textstyle  \hat{\phi}}
\,
dz_2 \wedge dz_4 \label{eq:boxsub}
\end{align}
has to be reducible to the $s$-channel bubble. 
This property is not manifest because $\hat{\phi}$ contains a double pole in $z_2$ and a simple pole in $z_4$ which do not belong to the $s$-bubble sector. However, by exploiting the equivalence class properties, $\phi$ can be made equivalent to a form $\phi'$ free of these poles. Accordingly, we define $\phi' \;\sim\; \phi$, such as
\begin{align}
%\phi \;\sim\; 
\phi' \equiv \phi - \nabla_{\omega} \xi
\label{eq:boxsubxi}
\end{align}
with the following ansatz for $\xi$,
\begin{align}
\xi = \frac{\; \sum_{i=-1, j=-1}^{2,2} \! \kappa_{1,i,j} z_2^i z_4^j \, \id z_4 \; + \; \sum_{i=-2, j=0}^{2,2} \! \kappa_{2,i,j} z_2^i z_4^j \, \id z_2 \; }{\mathcal{B}_{\tau}}.
\label{eq:boxxiansatz}
\end{align}
Fitting the free coefficients $\kappa$ with the requirement that all poles of $\phi'$ in $z_2$ or $z_4$ vanish, gives a solution
\begin{align}
\kappa_{1,-1,-1} &= \tfrac{-(d-6) (d-5) t^2}{2 s}\,, & \kappa_{1, -1, 0} &= \tfrac{(d-6) (d-5) t}{2 s}\,, \nonumber \\
\kappa_{1, -1, 1} &= 0\,, & \kappa_{1, -1, 2} &= 0\,, \nonumber \\
\kappa_{1, 0, -1} &= \tfrac{(3 d^2 - 36 d + 107) t}{2 s}\,, & \kappa_{1, 1, -1} &= \tfrac{-(d-7) (3d-17)}{2 s}\,, \nonumber \\
\kappa_{1, 2, -1} &= \tfrac{(d-7) (d-6)}{2 s t}\,, & \kappa_{2, -2, 0} &= \tfrac{-(d-5) t^2}{2 s}\,, \nonumber \\
\kappa_{2, -2, 1} &= \tfrac{(d-5) t}{2 s}\,, & \kappa_{2,-2,2} &= 0\,, \\ 
\kappa_{2, -1, 0} &= \tfrac{t (71 s - 24 d s + 2 d^2 s + 35 t - 12 d t + d^2 t)}{s^2}\,, \!\!\! &
\kappa_{2, -1, 1} &= \tfrac{-(d-7) (3d-17)}{2 s}\,, \nonumber \\
\kappa_{2, -1, 2} &= \tfrac{(d-7) (d-6)}{2 s t}\,, & \kappa_{\text{remain.}} &= 0\,. \nonumber
\end{align}
The corresponding ${\hat \phi}'$ is of the form
\begin{align}
{{\hat \phi}'} &= \frac{\mathcal{P}(z_2,z_4)}{\mathcal{B}_{\tau}^2} \, ,
\end{align}
where $\mathcal{P}$ is a polynomial, so we see explicitly that the $z_2$ and $z_4$ poles are gone, and that no poles are present in $\phi$ that are not poles of $\omega$.
With this we may compute the intersection number for $2$-forms, and we get
\begin{align}
c_2 &= \frac{\langle \phi' | 1 \rangle}{\langle 1 | 1 \rangle} = \frac{2 (d-7) (d-5) (d-3) }{s^4 t}
\end{align}
in agreement with eqs. \eqref{eq:olboxcoefs}.
The expressions for the two intersection numbers are listed in App. \ref{appendix:Box}, and please note that they are much simpler than the one computed in the other two approaches, due to the absence of the regulator.

For the $t$-channel cut one may proceed likewise, and extract the coefficient of the $t$-channel bubble, again in agreement with eqs. \eqref{eq:olboxcoefs}.

Let us note that one could use the subtraction 
\begin{align}
\hat{\phi} &= \hat{\varphi} - \frac{\kappa_{1}}{z_2 z_4} \ ,
\end{align}
in eq. \eqref{eq:boxsub},
where $\kappa_{1}$ is a free coefficient. 
Then, the fitting of the unknown coefficients of eq. \eqref{eq:boxxiansatz} generates a system whose solution does require the value $\kappa_{1} = c_1$. In other words, $\kappa_{1}$, which in this case corresponds to the coefficient of a master integral in the higher sector (the box function) may be fixed together with the remaining  $\kappa$-parameters\footnote{In principle such a procedure generalises beyond this example, to cases where more masters are present in the higher sectors.}.
\subsubsection*{Discussion}
Considering the three methods for the intersection-based reduction of the one-loop massless box, we observe that the straight decomposition required the computation of 12 intersection numbers for 4-forms, the bottom-up decomposition required 12 intersection numbers for 2-forms, and the top-down decomposition required $4$ intersection numbers for $2$-forms.
Due to the recursive nature of the multivariate algorithm of Sec. \ref{sec:multivariate}, the computation of intersection numbers for $2$-forms is much less demanding than the one of $4$-forms, thereby showing the efficiency of the bottom-up algorithm compared to the straight decomposition. On the other hand, in the top-down decomposition, we compute fewer intersection numbers than in the other two approaches. 
Neverthless, within this approach, an extra effort is taken by the fit of the extra $\kappa$-coefficients appearing in the subtraction term, see eq. \eqref{eq:boxsubxi} for the one-loop massless box, which might become computationally expensive in a generic case.

\subsection{One-loop QED triangle}
\begin{figure}[H]
    \centering
   \includegraphics[width=0.40\textwidth,clip=true, trim = 0 350 0 250]{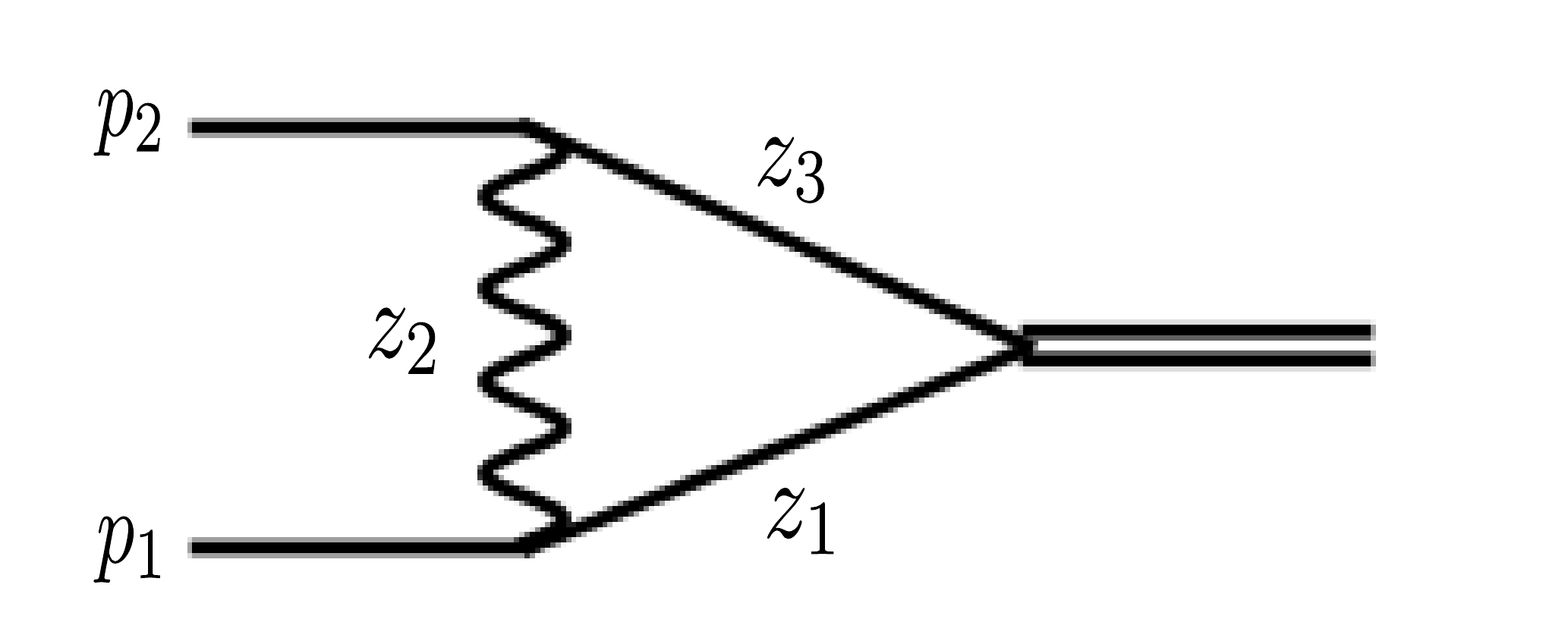}
    \caption{QED triangle.}
    \label{fig:QED_triangle}
\end{figure}

In this subsection we discuss the one-loop QED triangle as shown in Fig.~\ref{fig:QED_triangle}.
The denominators are
\begin{equation}
D_1=k^{2}-m^2, \quad D_2=(k+p_1)^2, \quad D_3=(k+p_1+p_2)^2-m^2
\end{equation}
and the kinematics is such that $p_1^2=p_2^2=m^2$, $(p_1+p_2)^2=s$. The Baikov parametrization yields:
\begin{equation}
u=\mathcal{B}^{(d-4) /2}
\end{equation}
with
\begin{align}
\mathcal{B} &= m^2 \big( 4 s z_2 - (z_1 - z_3)^2 \big) - s \big( s z_2 + (z_1 - z_2) (z_3 - z_2) \big).
\end{align}
Performing the sector-by-sector analysis described in the beginning of Sec. \ref{sec:feynintdeco} we obtain $\nu_{\sigma} = 1$ for the sectors
\begin{align}
\sigma \in \big\{\{1,3\}\,,\;\{1\}\,,\;\{3\} \big\}
\end{align}
and $\nu_{\sigma}=0$ for the remaining ones.\\
The master integrals are chosen as:
\begin{equation}
J_1=
\begin{gathered}
 \includegraphics[width=0.14\textwidth,valign=c]{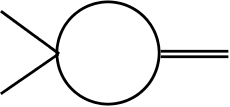}
\end{gathered}
\;,
\quad
J_2=
\begin{gathered}
 \includegraphics[width=0.1\textwidth,valign=c]{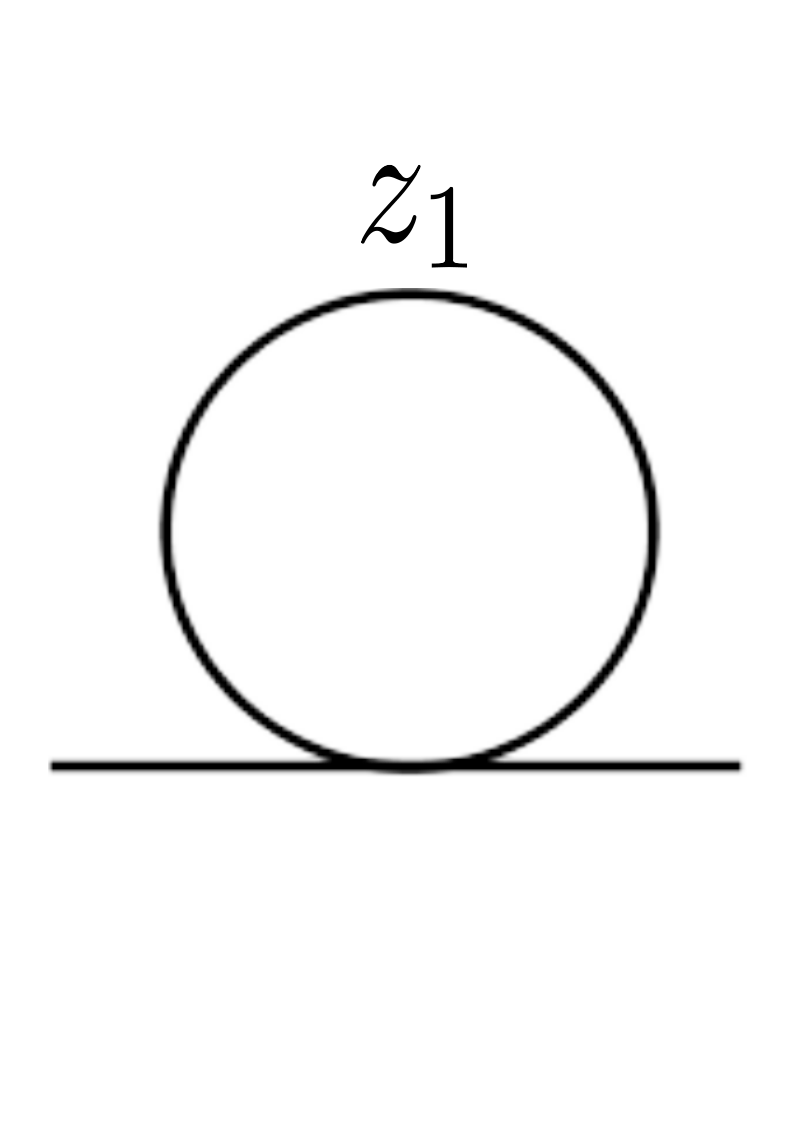}
\end{gathered}
\,
,
\quad
J_3=
\begin{gathered}
\includegraphics[ width=0.1\textwidth,valign=c]{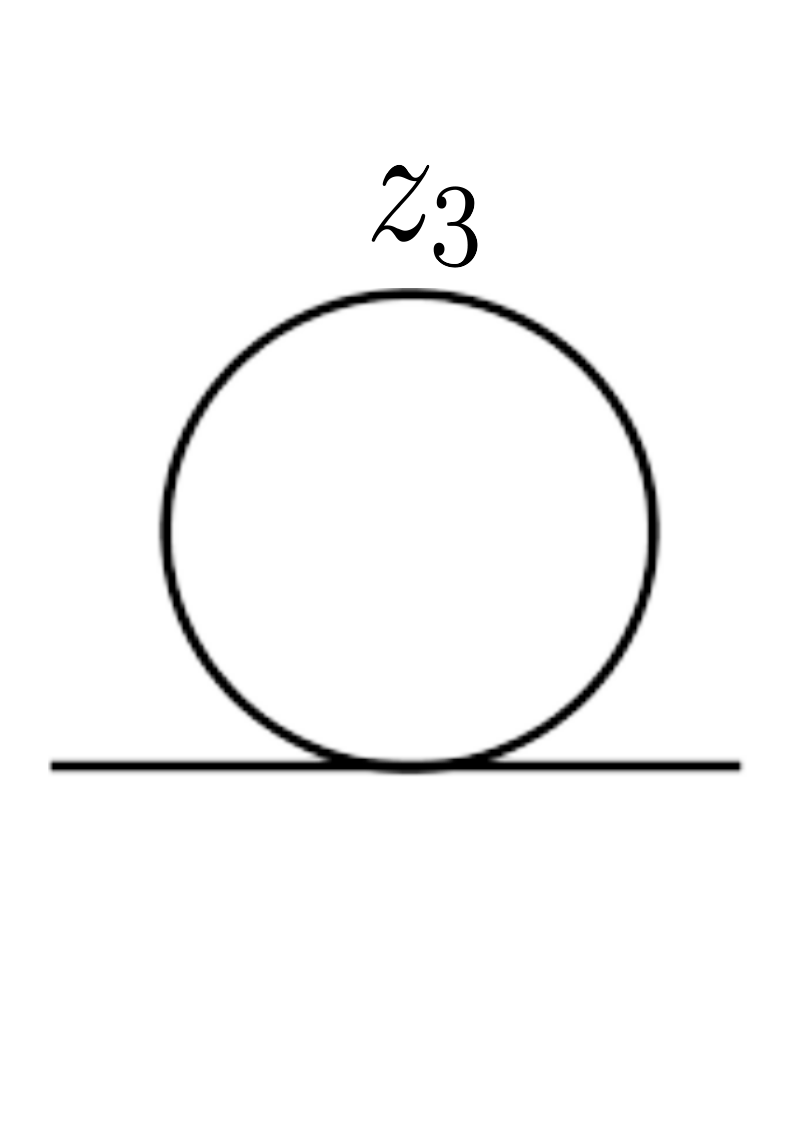}
\end{gathered}
,
\end{equation}
and the corresponding differential forms read
\begin{equation}
\hat{e}_1= \frac{1}{z_1 z_3} \, , \quad
\hat{e}_{2}=\frac{1}{z_1} \, , 
\quad \hat{e}_{3}=\frac{1}{z_3}
\, .
\end{equation}
In the following we will decompose:
\begin{equation}
\begin{gathered}
\includegraphics[ width=0.14\textwidth,valign=c]{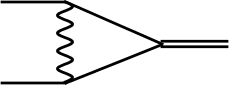}
\end{gathered}
= \int u \frac{d^3 \mathbf{z}}{z_1 z_2 z_3} \, ,
\end{equation}
which can be expressed in terms of the chosen master integrals as 
\begin{equation}
\begin{gathered}
\includegraphics[ width=0.14\textwidth,valign=c]{figure/Triangle_QED.png}
\end{gathered}
\;=\; c_1 \, 
\begin{gathered}
 \includegraphics[width=0.14\textwidth,valign=c]{figure/Bubble_QED.png}
\end{gathered}
+ c_2 \, 
\begin{gathered}
 \includegraphics[width=0.1\textwidth,valign=c]{figure/tadpole_QED_z1.png}
\end{gathered}
+ c_3 \, 
\begin{gathered}
 \includegraphics[width=0.1\textwidth,valign=c]{figure/tadpole_QED_z3.png}
\end{gathered}\ .
\end{equation}
\subsubsection*{Straight decomposition}
\label{subsubsec:QED_Triangle}

We introduce a regularized $u$ given by
\begin{equation}
u_{\rho}=u \times z_1^{\rho} z_2^{\rho} z_3^{\rho}
\end{equation}
and then 
\begin{equation}
    \omega_{\rho}=\sum_{i=1}^{3} \hat{\omega}_{i} \, dz_i \qquad \text{with}
    \qquad
    \hat{\omega}_i= \partial_{z_i} \log u_{\rho}\,.
\end{equation}
We consider the ordering of the variables, from the innermost to the outermost layer, as $z_3, z_1, z_2$ and the dimension of the twisted cohomology groups are
\begin{equation}
    \nu_{\{312\}}=3 \,,\;\; \nu_{\{31\}}=4\,,\;\; \nu_{\{3\}}=2\,.
\end{equation}
Given the order of variables considered above, we chose the basis elements to be
\begin{equation}
\hat{e}^{(312)}=\hat{e}= \left\{ \frac{1}{z_1 z_3}, \frac{1}{z_1}, \frac{1}{z_3} \right\}, \quad 
\hat{e}^{(31)}= \left\{ \frac{1}{z_1},\frac{1}{z_1 z_3}, \frac{1}{z_3}, 1\right\}, \quad
\hat{e}^{(3)}=\left\{\frac{1}{z_3}, 1 \right\},
\end{equation}
while the dual basis elements are chosen as $\hat{h}_i=\hat{e}_i$.\\
The required intersection numbers are
\begin{equation}
    \mathbf{C}_{ij}= \langle e_i | h_j \rangle, \quad 1 \leq i,j \leq3
\end{equation}
and
\begin{equation}
    \langle \varphi | h_{k} \rangle, \quad 1 \leq k \leq 3.
\end{equation}
Explicit expressions for the individual intersection numbers, up to the leading order in $\rho$, are presented in App.~\ref{appendix:QEDTriangle}.\\ 
Combining the intersection numbers, and taking the $\rho \to 0$ limit as in eq.~\eqref{eq:masterdeco_regulated_straight}, we obtain 
\begin{align}
c_1 &= \frac{2 (d-3)}{(d-4) (4 m^2 - s)} \,, \quad  c_2 = \frac{2-d}{2 (d-4) m^2 (4 m^2 - s)} \,, \nonumber \\[-7mm] \nonumber
\end{align}
\begin{align}
c_3 &= \frac{2-d}{2 (d-4) m^2 (4 m^2-s)} \, .
\label{eq:oldtrianglecoefficients}
\end{align}
These coefficients are in agreement with the result obtained from \textsc{FIRE}~\cite{Smirnov:2014hma} (before applying any symmetry relations).
\subsection{Two-loop QED sunrise}
\begin{figure}[H]
    \centering
   \includegraphics[width=0.25\textwidth,clip=true, trim = 0 275 0 200]{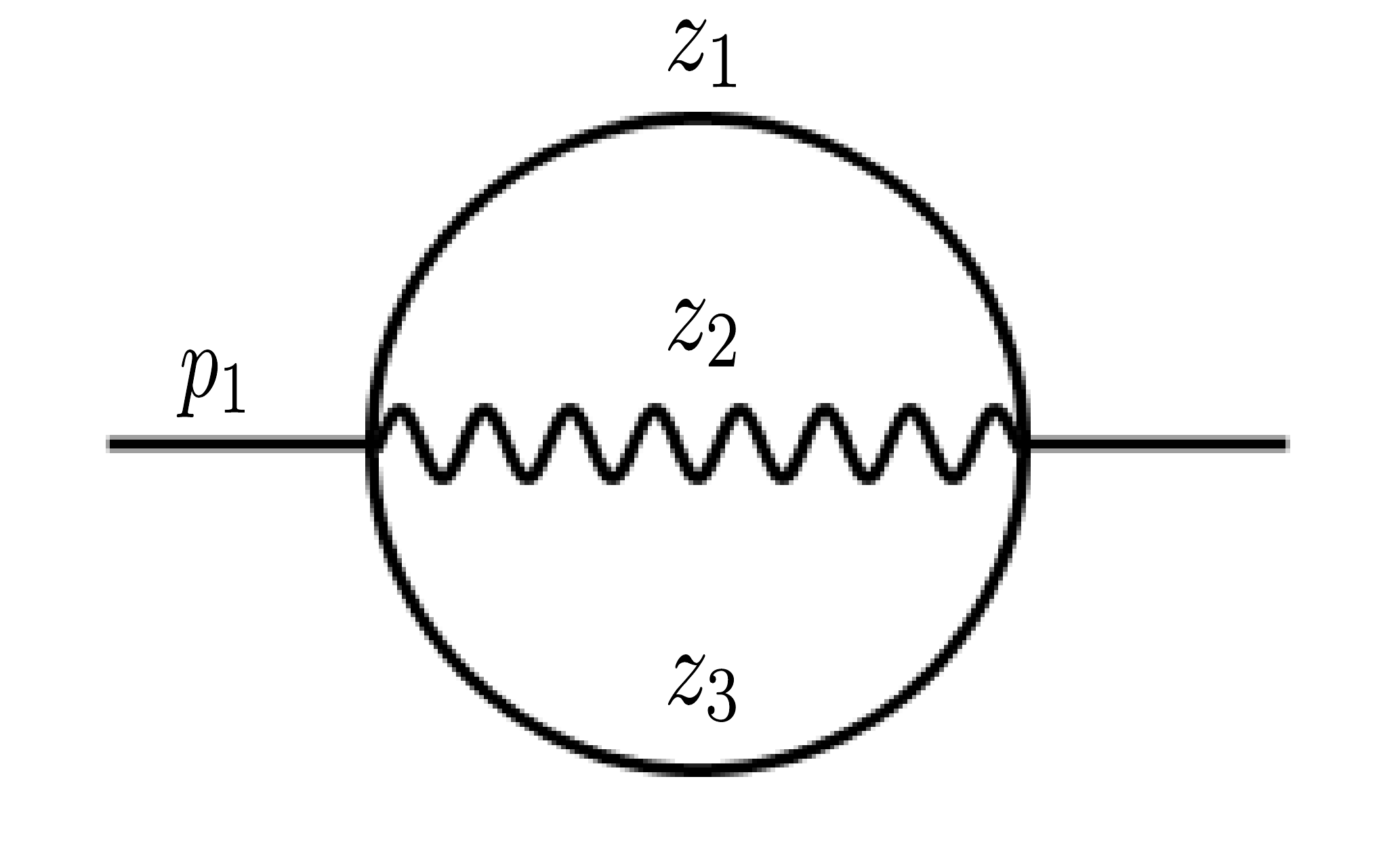}
    \caption{QED Sunrise.}
    \label{fig:qed_sunrise}
\end{figure}
Here, we consider 2-loop QED sunrise diagram as shown in Fig.~\ref{fig:qed_sunrise}.
The denominators are:
\begin{equation}
D_{1}=z_1=k_1^2-m^2, \quad D_{2}=z_2=\left(k_1-k_2\right){}^2, \quad D_{3}=z_{3}=\left(k_2-p_1\right){}^2-m^2,
\end{equation}
while the ISPs are chosen as:
\begin{equation}
z_{4}=k_2^2-m^2, \quad z_{5}=\left(k_1-p_1\right){}^2-m^2.
\end{equation}
The Baikov parametrization gives:
\begin{align}
u(\mathbf{z}) &= \frac{1}{s} \mathcal{B}^{\gamma},
\end{align}
where
\begin{align}
\mathcal{B} &= \frac{-1}{4 s} \Big( m^2 \big( (z_1 {+} z_3 {-} z_4 {-} z_5)^2 - 4 s z_2 \big) \nonumber \\
&\qquad - s \big( (z_1 {-} z_4) (z_3 {-} z_5) + z_2 (z_1 {+} z_3 {+} z_4 {+} z_5) - z_2^2 \big) + s^2 z_2  \\
&\qquad + (z_1 {+} z_3 {-} z_4 {-} z_5) (z_1 z_3 {-} z_4 z_5) - z_2 (z_3 {-} z_4) (z_1 {-} z_5) \Big)\,, \nonumber \\[1mm]
\gamma &= (d-4)/2 \,.
\end{align}
We choose the invariant $p_1^2 =s$ and normalise it by the squared internal mass effect $m^2$, effectively setting $m^2 = 1$, and the $m^2$ dependence can be recovered later by power counting.
We perform the sector-by-sector analysis for each of the $7$($=2^3-1$) sectors as described in Sec.~\ref{sec:feynintdeco}, and obtain zero MIs in all sectors except for
\begin{align}
    \sigma \in \big\{\{1,2,3\},\, \{1,3\} \big\}
\end{align}
where for the sector $\{1,2,3 \}$ we obtain 3 MIs and for the sector $\{1,3 \}$ 1 MI, 
amounting to a total of 4 MIs.
The MIs are chosen as the following: \\[-10mm]
\begin{align}
J_1=
\begin{gathered}
\includegraphics[scale=0.1,valign=c]{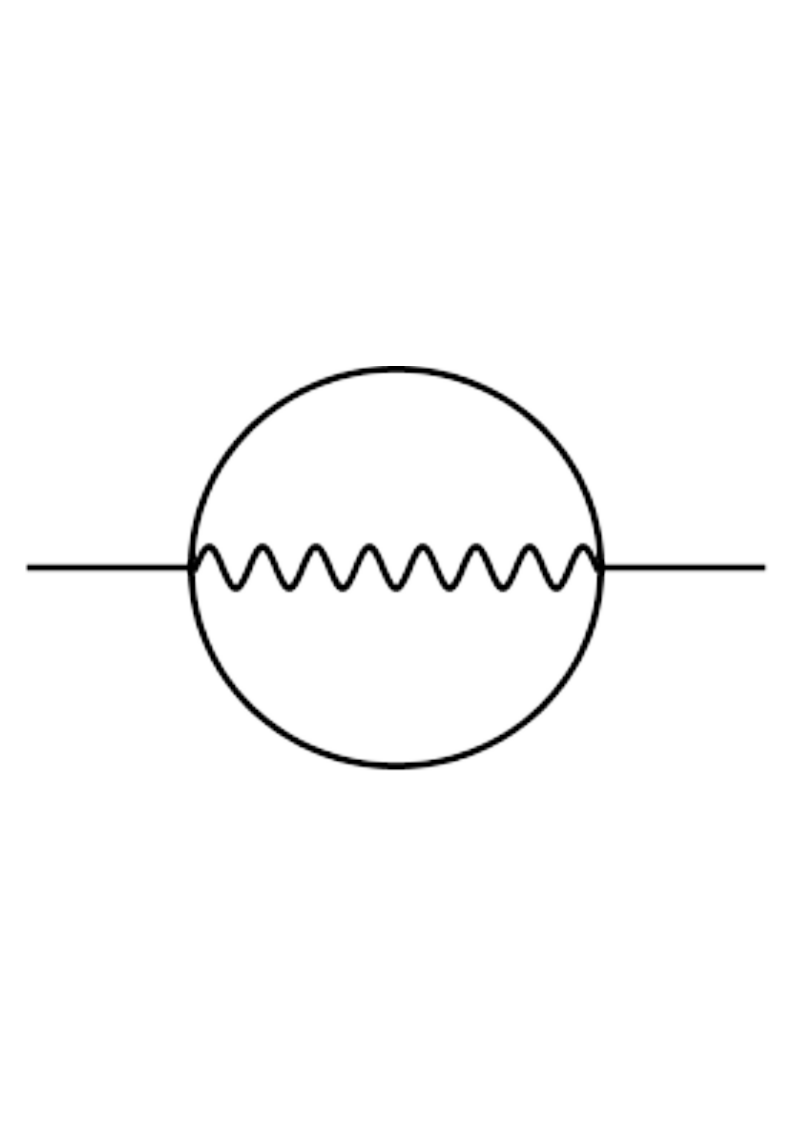}
\end{gathered}
,\quad J_2=
\begin{gathered}
\includegraphics[scale=0.1,valign=c]{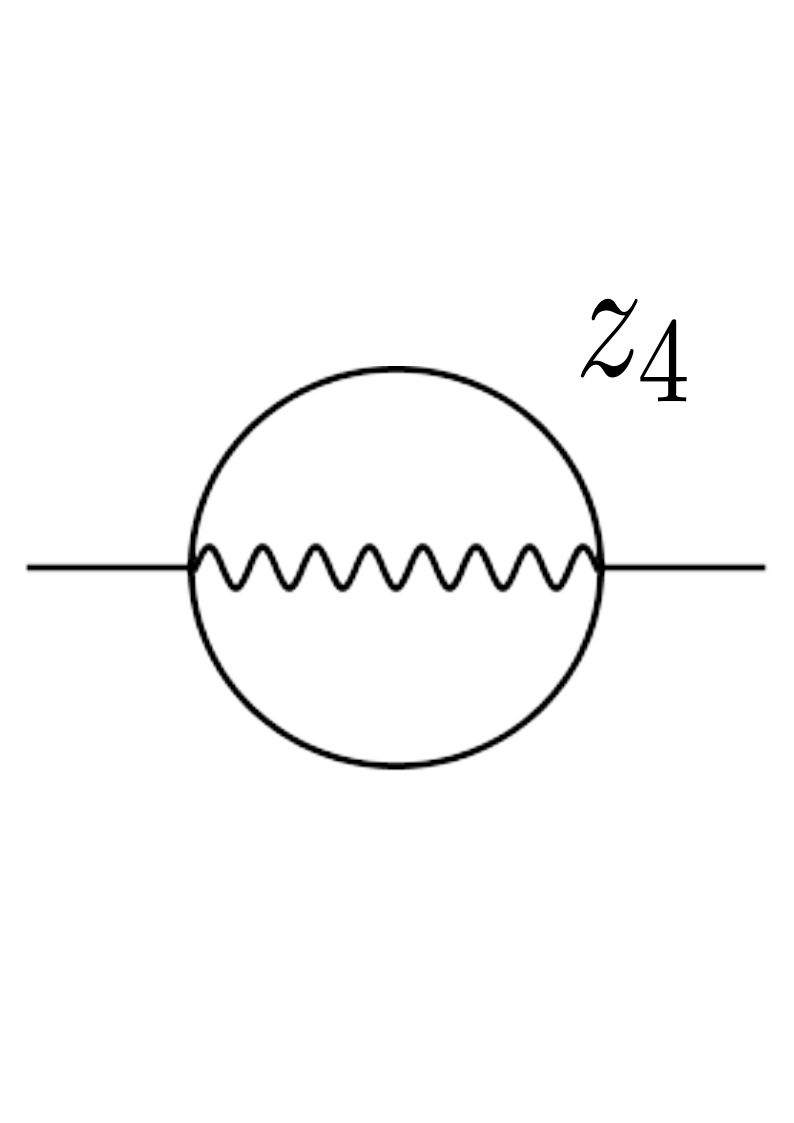}
\end{gathered}, \quad
J_{3}=
\begin{gathered}
\includegraphics[scale=0.1,valign=c]{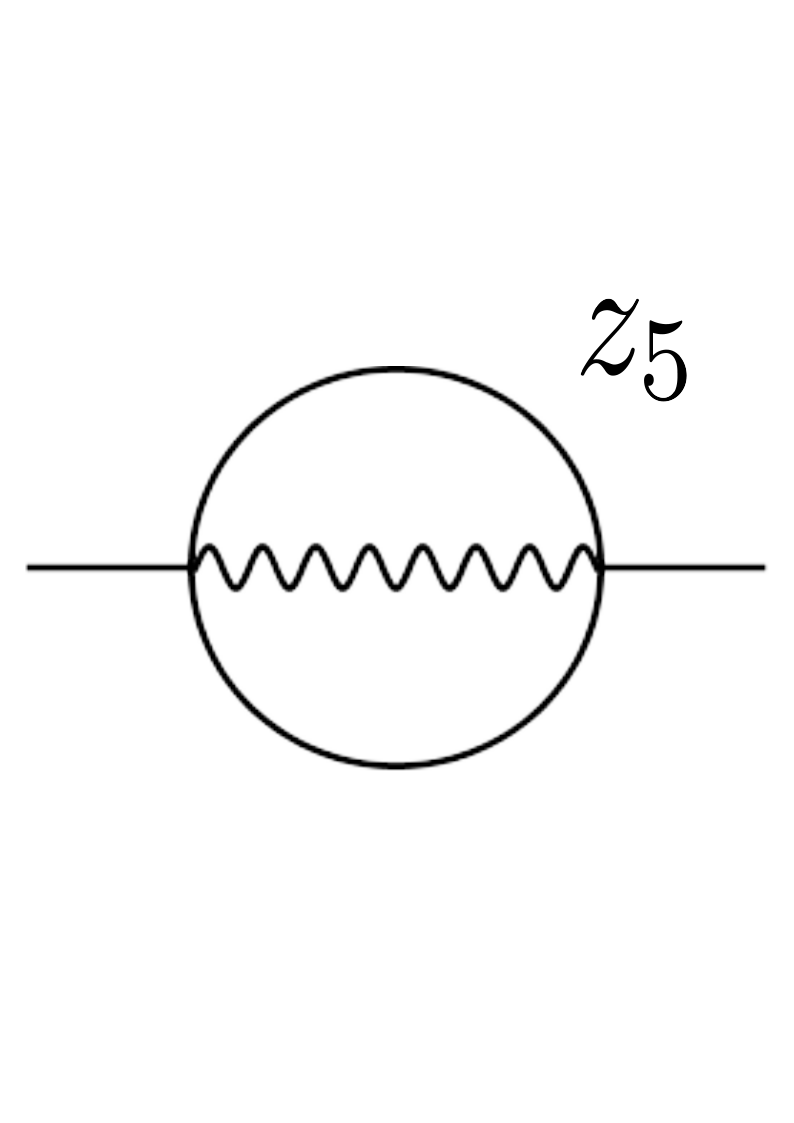}
\end{gathered}
, \quad J_{4}=
\begin{gathered}
\includegraphics[scale=0.2,valign=c]{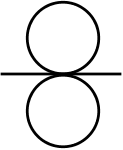}
\end{gathered},
\\[-12mm] \nonumber
\end{align}
and the corresponding differential forms read
\begin{align}
    \hat{e}_1 = \frac{1}{z_1 z_2 z_3} \,, \quad \hat{e}_2 = \frac{z_4}{z_1 z_2 z_3} \,, \quad \hat{e}_3 = \frac{z_5}{z_1 z_2 z_3}  \,, \quad \hat{e}_4 = \frac{1}{z_1 z_3} \,. 
\end{align}
Here, we will build the differential equation for the set of master integrals
\begin{equation}
\mathbf{J}=
\begin{pmatrix}
\begin{gathered}
\includegraphics[scale=0.09,clip=true, trim = 0 250 0 225]{figure/QEDSunriseNoisp.png}
\end{gathered}
\;,\quad
\begin{gathered}
\includegraphics[scale=0.09,clip=true, trim = 0 250 0 225]{figure/QEDSunrisez4.png}
\end{gathered}
\;,\quad
\begin{gathered}
\includegraphics[scale=0.09,clip=true, trim = 0 250 0 225]{figure/QEDSunrisez5.png}
\end{gathered}
\;,\quad
\begin{gathered}
\includegraphics[scale=0.20]{figure/DoubleTadpole.png}
\end{gathered}
\end{pmatrix}^{T}
\end{equation}
namely
\begin{equation}
    \partial_{s} \, \mathbf{J}= \mathbf{\Omega} \,  \, \mathbf{J} \, .
\end{equation}
We will now determine $\mathbf{\Omega}$ using the {\it bottom-up decomposition}
as described in Sec.~$\ref{sec:feynintdeco}$

\subsubsection*{Bottom-up decomposition}
\label{subsubsec:QEDSunrise}

First we identify a spanning set of cuts $\tau$. That set is easily seen to only contain the cut corresponding to the double tadpole:
\begin{align}
\tau \in \big\{ \{1,3\} \big\}.
\end{align}
On this specific cut, we use:
\begin{align}
u_{\rho,\tau}=\frac{1}{s} \, z_{2}^{\rho} \, \mathcal{B}_{\tau}^{\gamma}
\end{align}
with
\begin{align}
\mathcal{B}_{\tau} &= \frac{1}{4 s} \left( (z_5 + z_4 - z_2 - s) (s z_ 2 - z_ 4 z_ 5) + 4 s z_ 2 - (z_ 4 + z_ 5)^2 \right)
\end{align}
and $\omega_{\rho,\tau} = \hat{\omega}_2 \, \text{d}z_2 + \hat{\omega}_4 \,  \text{d} z_4 + \hat{\omega}_5 \, \text{d} z_5$ with
\begin{align}
\hat{\omega}_2 = \partial_{z_2}\,\, \text{log} \, u_{\rho,\tau}\,,\quad
\hat{\omega}_4 = \partial_{z_4}\,\, \text{log} \, u_{\rho,\tau}\,,\quad
\hat{\omega}_5 = \partial_{z_5}\,\, \text{log} \, u_{\rho,\tau}.
\end{align}
We consider the ordering of the variables, from the innermost to the outermost, as $z_4,z_2,z_5$ and the corresponding numbers of independent forms read:
\begin{align}
\nu_{\{425\}}=4, \quad\,
\nu_{\{42\}}=2,   \quad\,
\nu_{\{4\}}=1.   
\end{align}
On the cut we have
\begin{equation}
\mathbf{J}_{\rho, \tau}=
\begin{pmatrix}
\begin{gathered}
\includegraphics[scale=0.09,clip=true, trim = 0 250 0 225]{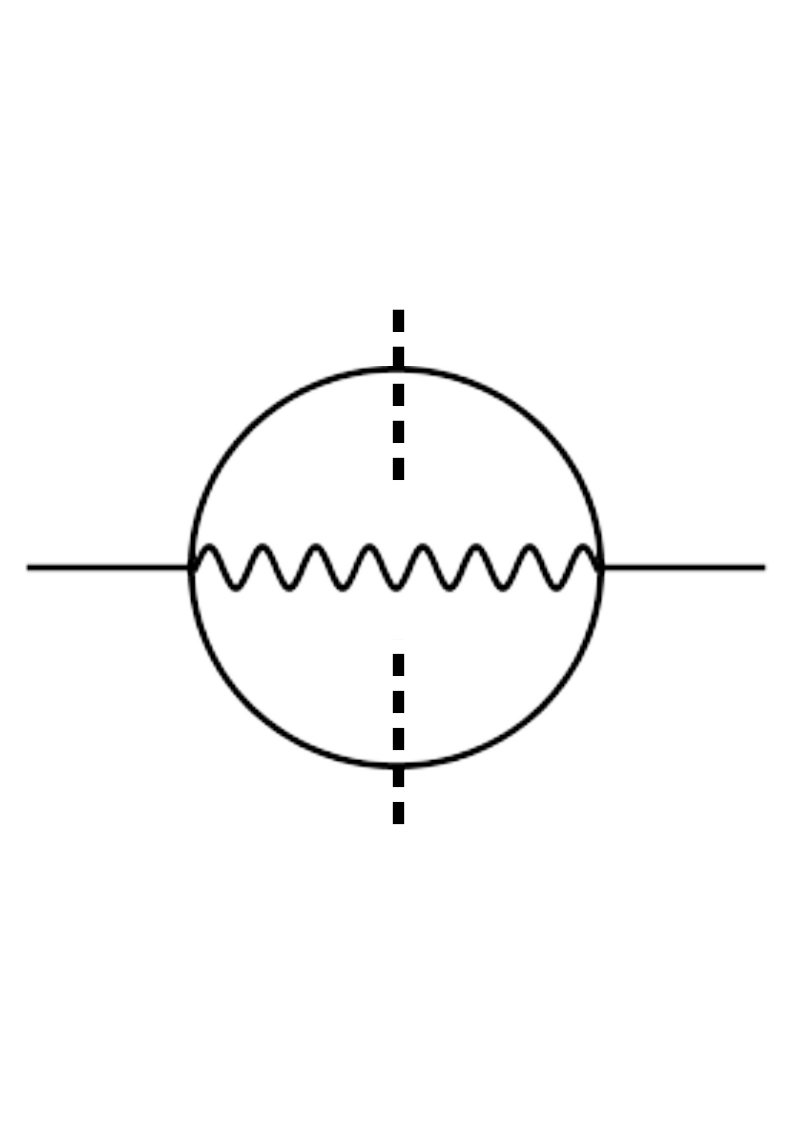}
\end{gathered}
\;,\quad
\begin{gathered}
\includegraphics[scale=0.09,clip=true, trim = 0 250 0 225]{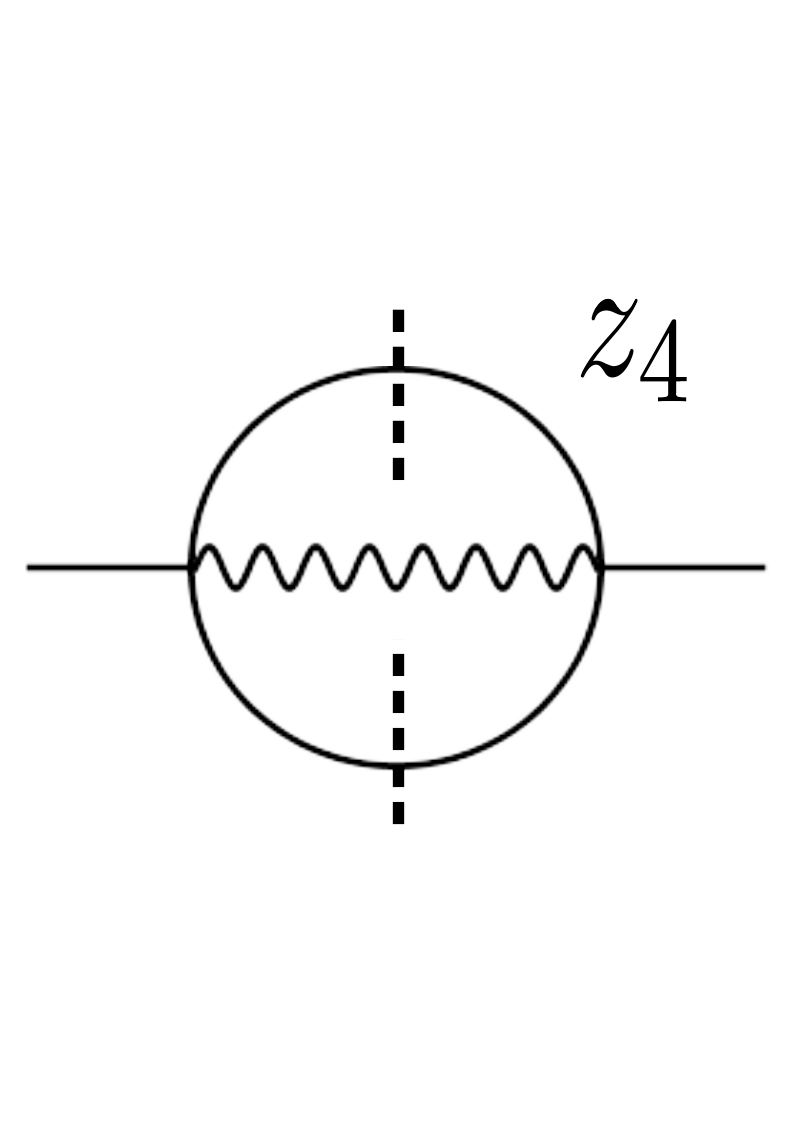}
\end{gathered}
\;,\quad
\begin{gathered}
\includegraphics[scale=0.09,clip=true, trim = 0 250 0 225]{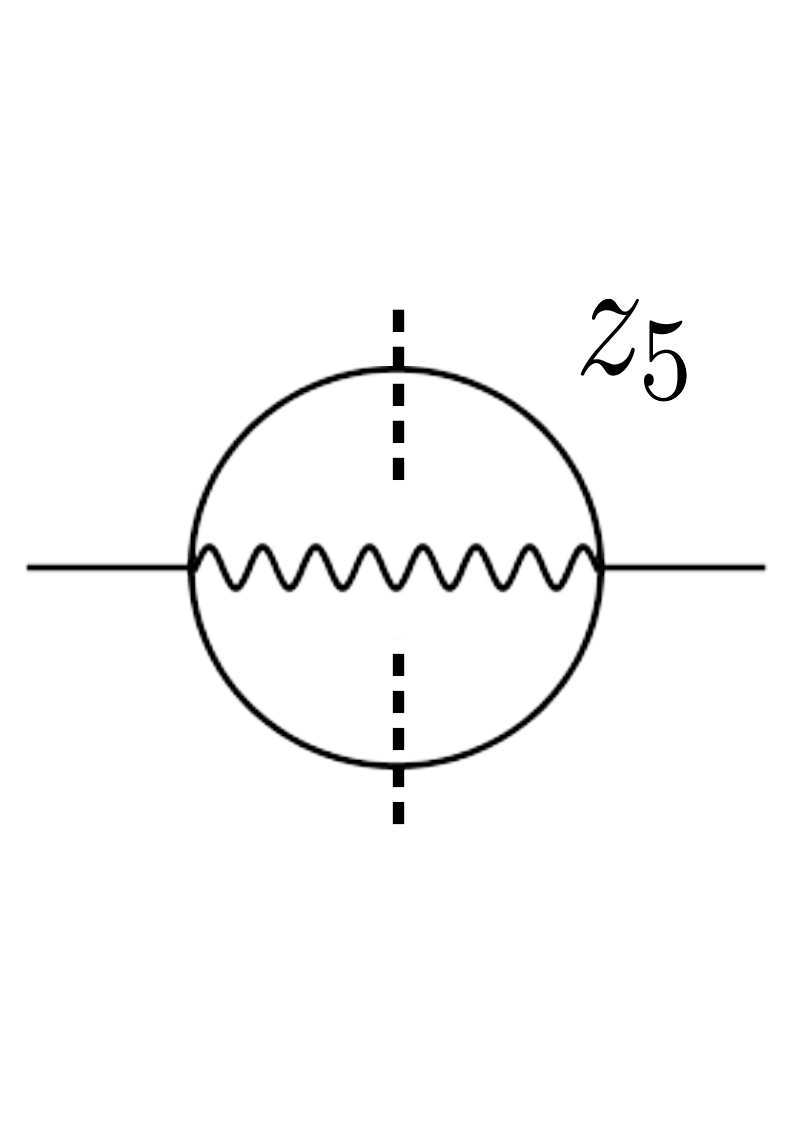}
\end{gathered}
\;,\quad
\begin{gathered}
\includegraphics[scale=0.20]{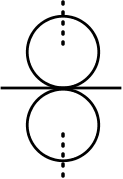}
\end{gathered}
\end{pmatrix}^{T},
\label{eq:DEQEDSunrise}
\end{equation}
where
$\mathbf{J}_{\rho, \tau} = \int_{\mathcal{C}} u_{\rho, \tau} \, \boldsymbol{e}_{\tau}$
and the differential equation reads
\begin{align}
    \partial_{s} \, \mathbf{J}_{\rho,\tau}&= \mathbf{\Omega}_{\rho} \,  \, \mathbf{J}_{\rho, \tau} \, , 
\end{align}
The twisted cocycles corresponding to the individual MIs on the cut are
\begin{equation}
    \hat{e}^{(425)}_{\tau}=\hat{e}_{\tau}= \Big\{ \frac{1}{z_2}, \frac{z_4}{z_2}, \frac{z_5}{z_2},1 \Big\}.
\end{equation}
Following eq.~(\ref{eq:DEQ:sigma}) we define $\sigma = \partial_s \, \text{log} \, u_{\rho, \tau}$
and the corresponding twisted cocycles for the decomposition of eq. (\ref{eq:DEQEDSunrise}) read:
\begin{align}
\hat{\varphi}_{\tau}=(\partial_s + \sigma) \hat{e}_{\tau}= \Big\{ \sigma \frac{1}{z_2}, \sigma \frac{z_4}{z_2}, \sigma \frac{z_5}{z_2}, \sigma \Big\}
\end{align}
\\
For the inner spaces, we choose the basis elements as:
\begin{align}
\hat{e}^{(42)}=\Big\{ 1, \frac{1}{z_2}  \Big\},  \quad 
\hat{e}^{(4)}= \Big\{1 \Big\},
\end{align}
and the dual basis elements are chosen as $\hat{h}_i=\hat{e}_i$.\\
Then, we compute the metric matrix defined as
\begin{align}
    \mathbf{C}_{ij} = \langle e_{i,\tau} | h_{j,\tau} \rangle, \quad 1 \leq i,j \leq 4
\end{align}
and the individual projections
\begin{equation}
    \langle \varphi_{k,\tau} | e_{l,\tau} \rangle, \quad 1 \leq k,l \leq 4.
\end{equation}
Using eq.~\eqref{eq:multivarIntNumb} we may then get the individual entries of the differential equation matrix
\begin{equation}
    (\mathbf{\Omega}_{\rho})_{ij} = \sum_{k=1}^{4} \langle \varphi_{i} | e_{k,\tau} \rangle( \mathbf{C}^{-1})_{kj},
    \quad 1 \leq i,j \leq 4.
\end{equation}
The individual multivariate intersection numbers are provided in App. \ref{appendix:QEDSunrise}.
Using these intersection numbers, 
we obtain after taking the limit $\rho \rightarrow 0$
\begin{align}
\mathbf{\Omega} = \left(
\begin{array}{cccc}
 \frac{2 d (s-1)-5 s+6}{(s-4) s} & -\frac{3 (d-2)}{2 (s-4) s} & \
-\frac{3 (d-2)}{2 (s-4) s} & \frac{d-2}{(s-4) s} \\
 \frac{d-2}{2} & 0 & -\frac{d-2}{2 s} & 0 \\
 \frac{d-2}{2} & -\frac{d-2}{2 s} & 0 & 0 \\
 0 & 0 & 0 & 0 \\
\end{array}
\right),
\label{eq:QEDSunriseDEcoeffs}
\end{align}
which is in agreement with the result obtained from \textsc{LiteRed}~\cite{Lee:2012cn}.
\\
\, \\

\subsection{Further examples}

In the following, we present the key information useful to perform the reduction by means of intersection theory, in a set of cases all corresponding to physically relevant Feynman integrals. 
In particular, for each case, we provide a table containing: the definition of the integral family; the spanning cuts ($\tau$); 
the dimensions of the vector spaces at each step of the recursive algorithm ($\nu$) and the corresponding bases ($e$), for the evaluation of multivariate intersection numbers; 
a pictorial decomposition of a generic integral, whose coefficients can be determined by means of our master decomposition formula eq.~\eqref{eq:masterdeco}.
In all these cases, the reduction and/or the differential equations were computed successfully, in agreement with the results of public IBP codes~\cite{Smirnov:2014hma, Lee:2012cn, vonManteuffel:2012np,Maierhoefer:2017hyi}.

\clearpage

\subsection*{Box with four different masses}

\begin{table}[!h]
\centering
\begin{tabular}{|c|c|}
\hline $\;\;$ Integral family &   Denominators \\ 
\hline 
\hline 
\hline
\shortstack{
\includegraphics[width=0.15\textwidth,clip=true, trim = 0 175 0 175]{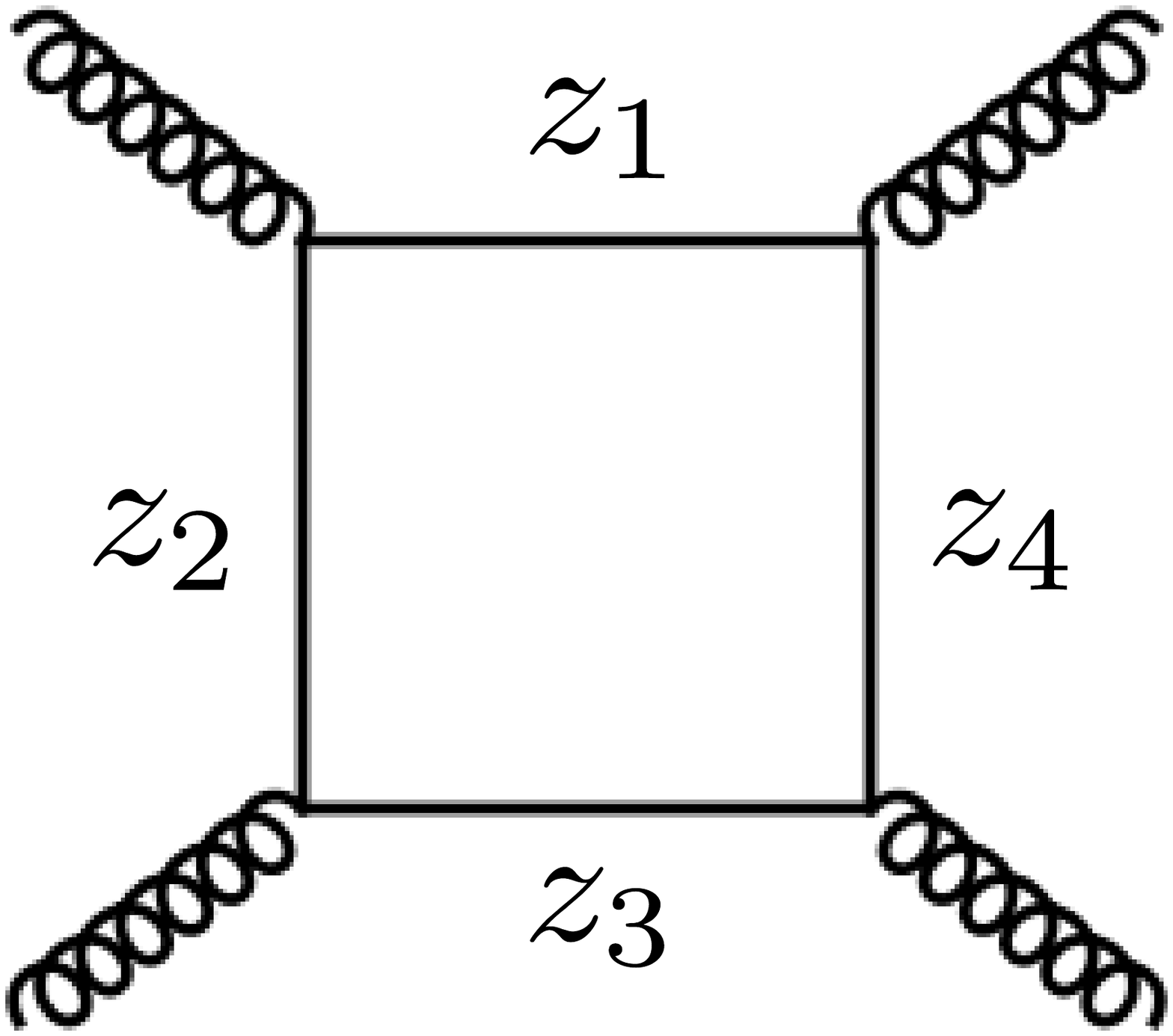}
\\ $s=(p_1+p_2)^2, \quad t=(p_2+p_3)^2$ }
& \shortstack{ $z_1=k^2 - m_1^2$ \\
  $z_2=(k+p_1)^2 - m_2^2$ \\
  $z_3=(k+p_1+p_2)^2 - m_3^2$ \\
  $z_4=(k+p_1+p_2+p_3)^2 - m_4^2$
  \, \\ \, \\ \,  } \\
  \hline
\end{tabular}
\\[2mm]
\begin{tabular}{|c|c|c|}
\hline $\tau$ & $\nu$ & $e$ \\ 
\hline 
\hline 
\hline
\multirow{3}{*}{
$
z_4=0 
$}
& $\nu_{\{3\}}=2$
&    $e^{(3)}= \left\{1, \frac{1}{z_3}\right\}$ \\
& $\nu_{\{32\}}=3$
&  $e^{(32)}= \left\{\frac{1}{z_2}, \frac{1}{z_3}, \frac{1}{z_2 z_3}\right\}$ \\
& $\nu_{\{321\}}=6$ 
&  $\; e^{(321)}=\left\{1, \frac{1}{z_2}, \frac{1}{z_1 z_2}, \frac{1}{z_1 z_3}, \frac{1}{z_2 z_3}, \frac{1}{z_1 z_2 z_3}\right\} \; $ \\ \hline

\multirow{3}{*}{$
z_3=0 
$}
& $\nu_{\{4\}}=2$ 
&    $e^{(4)}= \left\{1, \frac{1}{z_4}\right\}$\\
& $\nu_{\{41\}}=3$
&  $e^{(41)}= \left\{\frac{1}{z_1}, \frac{1}{z_4}, \frac{1}{z_1 z_4}\right\}$ \\
& $\nu_{\{412\}}=6$ 
&  $e^{(412)}=\left\{1, \frac{1}{z_1}, \frac{1}{z_1 z_2}, \frac{1}{z_1 z_4}, \frac{1}{z_2 z_4}, \frac{1}{z_1 z_2 z_4}\right\}$ \\ \hline

\multirow{3}{*}{$
z_2=0 
$}
& $\nu_{\{4\}}=2$ 
&    $e^{(4)}= \left\{1, \frac{1}{z_4}\right\}$ \\
& $\nu_{\{43\}}=3$
&  $e^{(43)}= \left\{\frac{1}{z_3}, \frac{1}{z_4}, \frac{1}{z_3 z_4}\right\}$ \\
& $\nu_{\{431\}}=6$
&  $e^{(431)}=\left\{1, \frac{1}{z_4}, \frac{1}{z_1 z_3}, \frac{1}{z_1 z_4}, \frac{1}{z_3 z_4}, \frac{1}{z_1 z_3 z_4}\right\}$ \\ \hline

\multirow{3}{*}{$
z_1=0 
$}
& $\nu_{\{4\}}=2$
&    $e^{(4)}= \left\{1, \frac{1}{z_4}\right\}$ \\
& $\nu_{\{43\}}=3$
&  $e^{(43)}= \left\{\frac{1}{z_3}, \frac{1}{z_4}, \frac{1}{z_3 z_4}\right\}$ \\
& $\nu_{\{432\}}=6$ 
&  $e^{(432)}=\left\{1, \frac{1}{z_3}, \frac{1}{z_2 z_3}, \frac{1}{z_2 z_4}, \frac{1}{z_3 z_4}, \frac{1}{z_2 z_3 z_4}\right\}$ \\ \hline
\end{tabular}
\end{table}

\begin{align}
\begin{gathered}
\includegraphics[scale=0.24, valign=c]{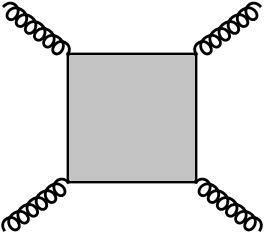}
\end{gathered}
=\;\;
&
c_1 \, 
\begin{gathered}
\includegraphics[scale=0.18, valign=c]{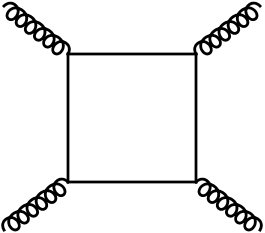}
\end{gathered}
+
c_2 \, 
\begin{gathered}
\includegraphics[scale=0.18, angle=90, valign=c]{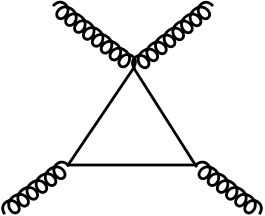}
\end{gathered}
+
c_3 \!\!
\begin{gathered}
\includegraphics[scale=0.18, valign=c]{figure/massivetriangle.png}
\end{gathered}
+
c_4 \!\!\!
\begin{gathered}
\vspace*{+0.2cm}
\includegraphics[scale=0.18, angle=180, valign=c]{figure/massivetriangle.png}
\end{gathered}
+
c_5 \, 
\begin{gathered}
\vspace*{+0.2cm}
\includegraphics[scale=0.18, angle=270, valign=c]{figure/massivetriangle.png}
\end{gathered}
\nonumber \\[-8\jot]
&
+ c_6 \, \, \, \, 
\begin{gathered}
\includegraphics[scale=0.2, valign=c]{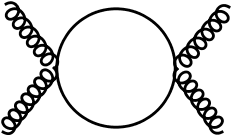}
\end{gathered}
+
c_7 \, \, \, \, 
\begin{gathered}
\includegraphics[scale=0.2, angle=90, valign=c]{figure/massivebubble.png}
\end{gathered}
+
c_8 \!\!\!\!
\begin{gathered}
\includegraphics[scale=0.1, valign=c]{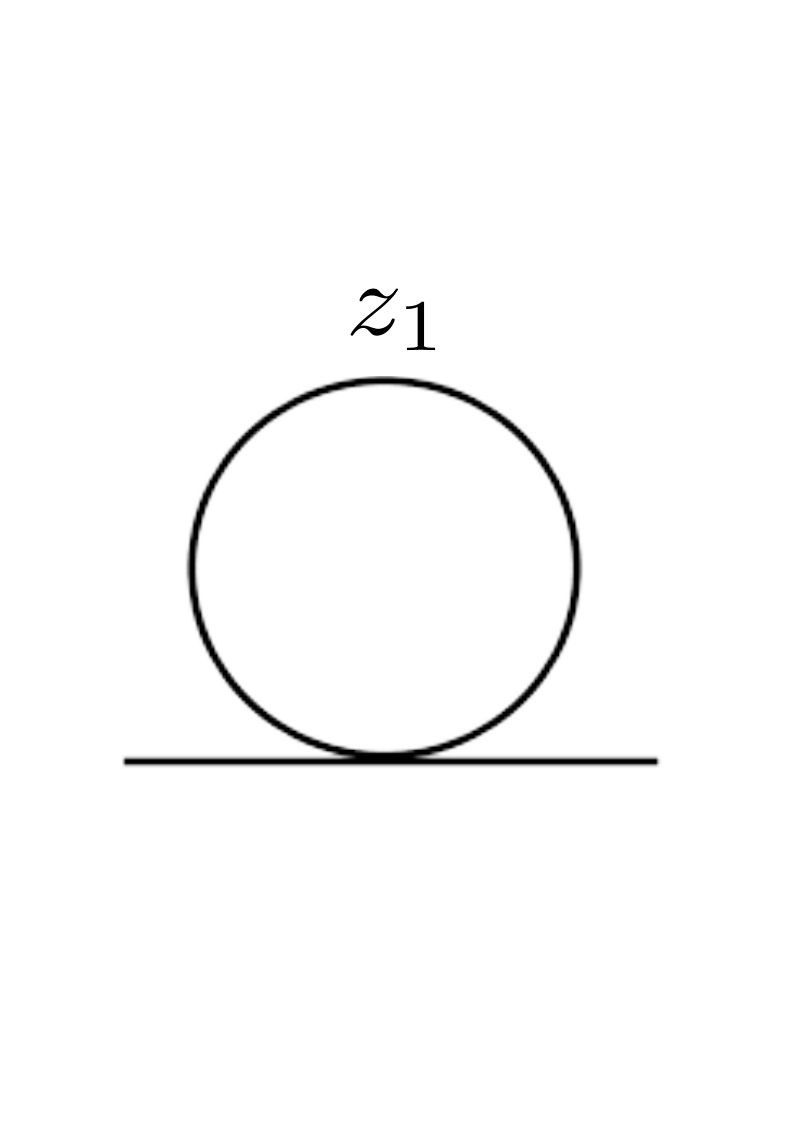}
\end{gathered}
+
c_9 \!\!\!\!
\begin{gathered}
\includegraphics[scale=0.1, valign=c]{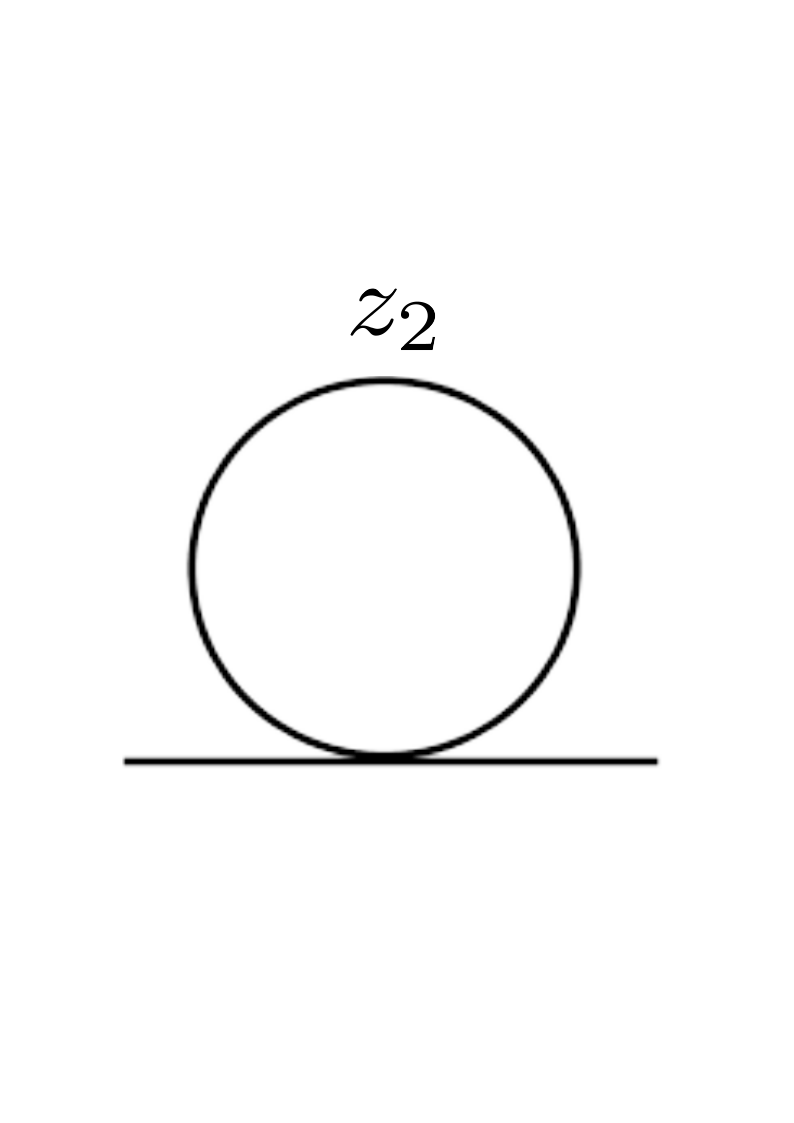}
\end{gathered}
\nonumber  \\[-8\jot]
& + c_{10} \!\!\!\!
\begin{gathered}
\includegraphics[scale=0.1,clip=true, trim = 0 225 0 150, valign=c]{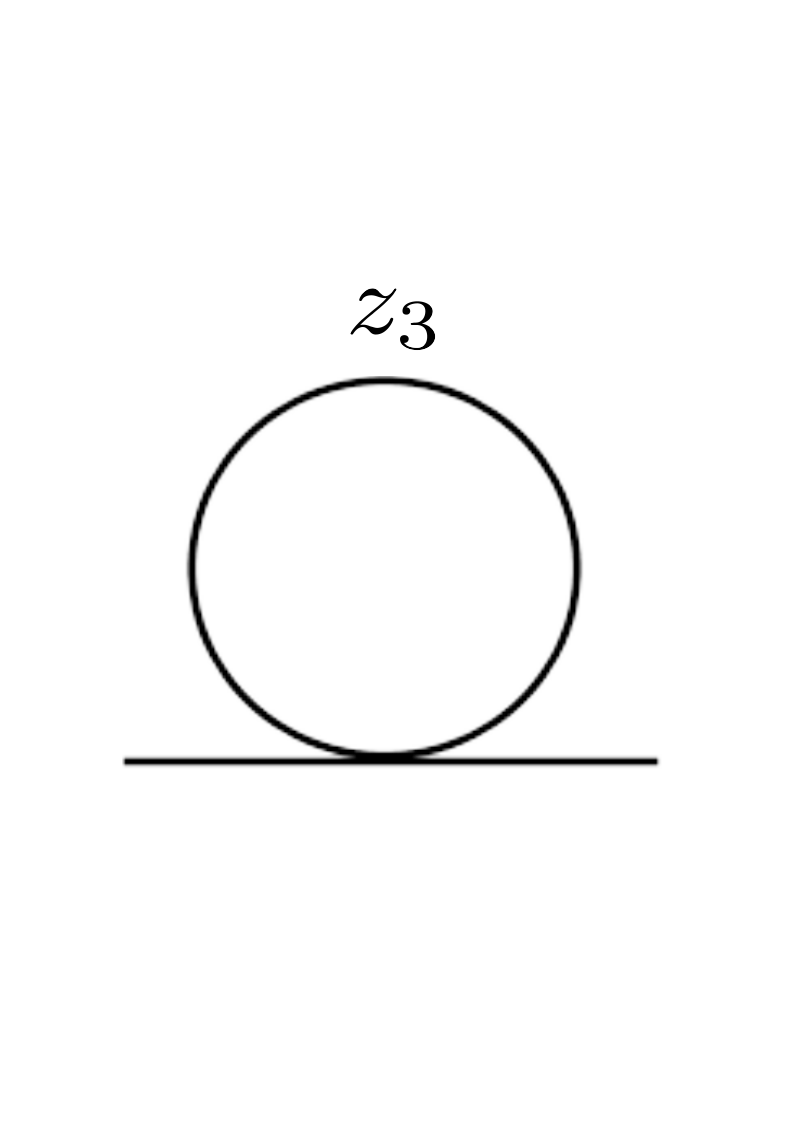}
\end{gathered}
+ c_{11} \!\!\!\!
\begin{gathered}
\includegraphics[scale=0.1,clip=true, trim = 0 225 0 150, valign=c]{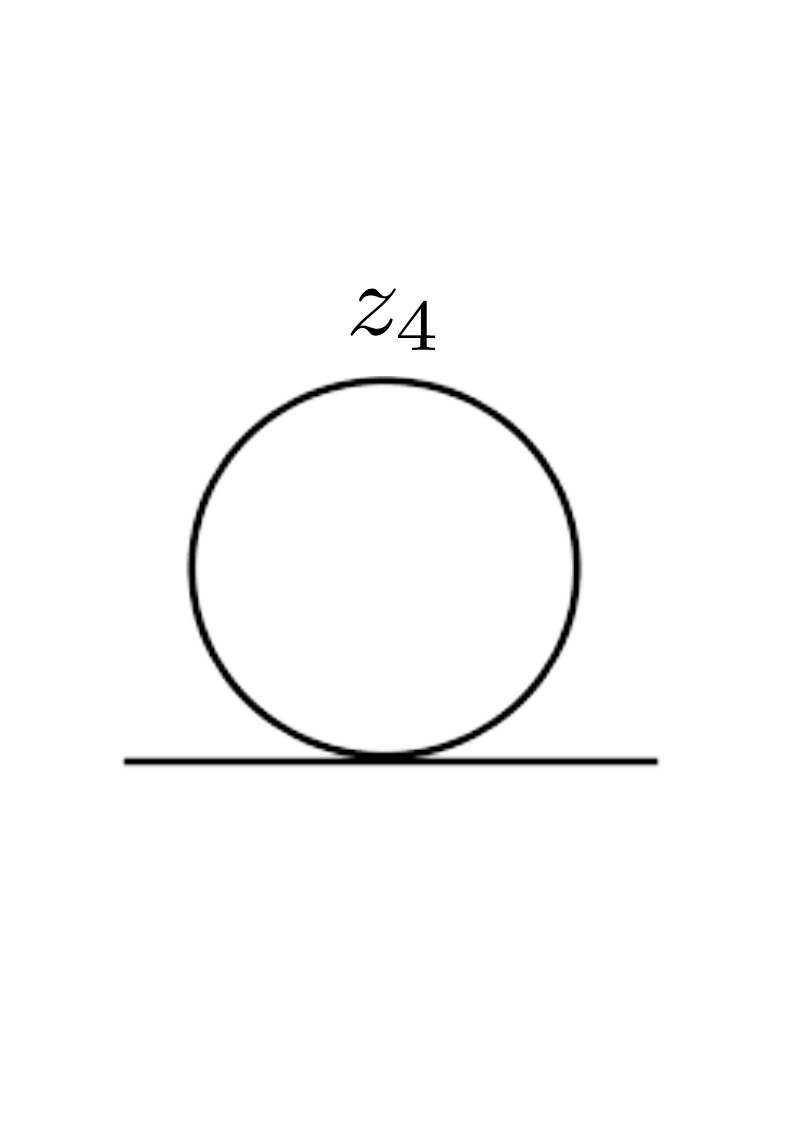} 
\end{gathered}
\, .
\end{align}

\clearpage

\subsection*{Sunrise with different masses}

\begin{table}[!h]
\centering
\begin{tabular}{|c|c|}
\hline $\;\;$ Integral family & Denominators \\ \hline 
\hline
\hline 
\shortstack{ \, \\ \, \\ \includegraphics[width=0.2\textwidth,clip=true, trim = 0 325 0 150]{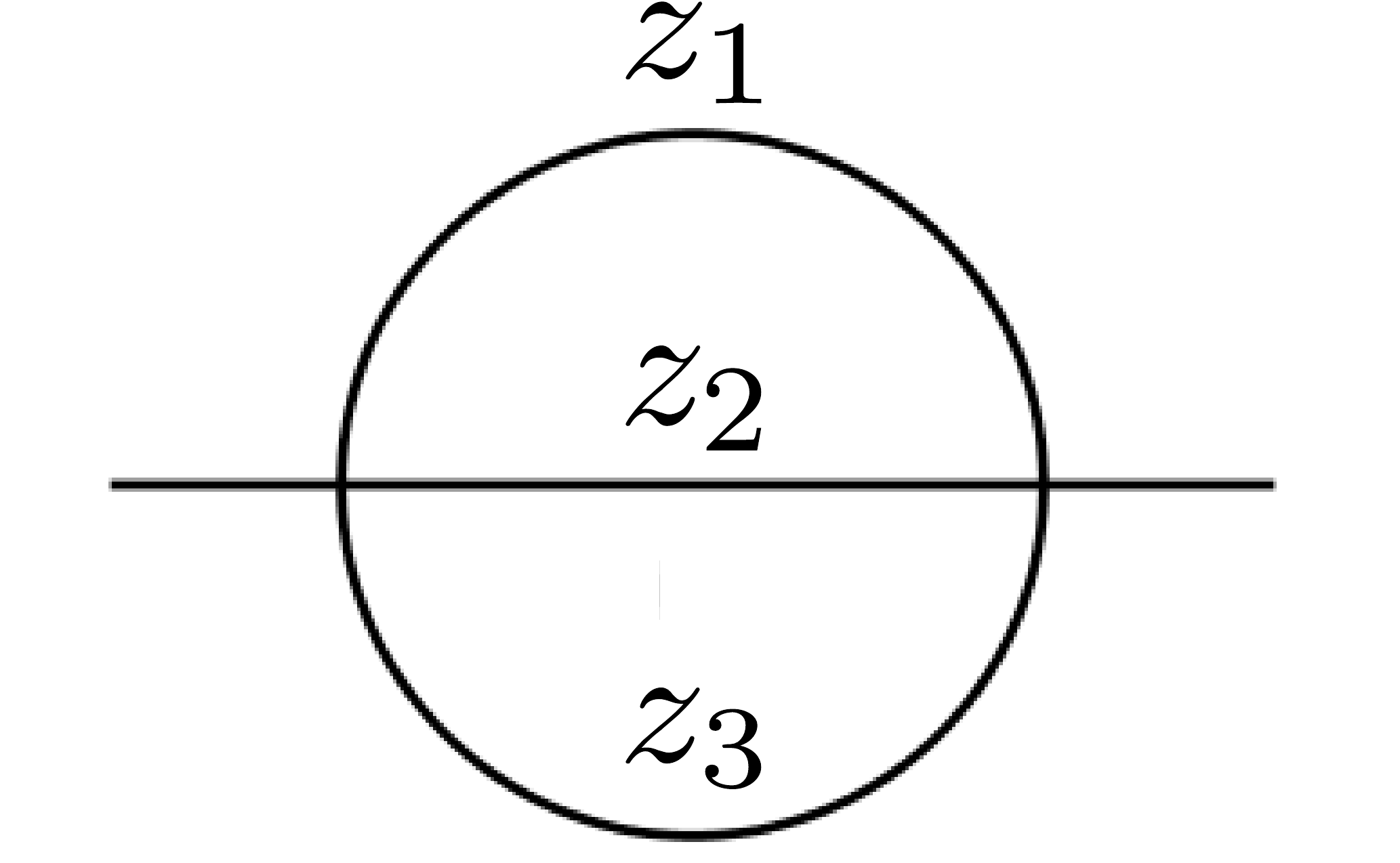}
\\   \\ $\; \, \, \quad \qquad s=p_1^2 \quad \qquad \, \, \; $  }
& \shortstack{ $z_1=k_1^2 - m_1^2$ \\
  $z_2=(k_1-k_2)^2 - m_2^2$ \\
  $z_3=(k_2 - p_1)^2 - m_3^2$ \\
  $z_4=k_2^2 - m_1^2$\\ 
  $\, \quad z_5=(k_1 - p_1)^2 - m_3^2 \quad \,$} \\
   \hline
\end{tabular}
\\[2mm]
\begin{tabular}{|c|c|c|}
\hline $\tau$ & $\nu$ & $e$ \\ \hline 
\hline
\hline 
 \multirow{3}{*}{
 \shortstack{%\{ 
 \vspace{2ex} \\
$z_1=0$%\} 
\\
%\{
$z_2=0 $
%\}
}
}
& $\nu_{\{5\}}=1$
&    $e^{(5)}= \left\{ 1 \right\}$ \\
& $\nu_{\{53\}}=2$
&  $e^{(53)}= \left\{1,\frac{1}{z_3} \right\}$ \\
& $\nu_{\{534\}}=5$
&  $e^{(534)}=\left\{1, \frac{1}{z_3}, \frac{z_4}{z_3}, \frac{z_5}{z_3}, \frac{z_4^2}{z_3} \right\}$ \\ \hline

\multirow{3}{*}{\shortstack{
\vspace{2ex} \\
$ z_1=0 $ \\
$z_3=0 $}}
& $\nu_{\{5\}}=1$
&    $e^{(5)}= \left\{1\right\}$ \\
& $\nu_{\{52\}}=2$
&  $e^{(52)}= \left\{1, \frac{1}{z_2} \right\}$\\
& $\nu_{\{524\}}=5$ 
&  $e^{(524)}=\left\{1, \frac{1}{z_2}, \frac{z_4}{z_2}, \frac{z_5}{z_2}, \frac{z_4^2}{z_2}\right\}$ \\ \hline

\multirow{3}{*}{\shortstack{
\vspace{2ex} \\
$ z_2=0 $ \\
$z_3=0 $}}
& $\nu_{\{5\}}=1$
&    $e^{(5)}= \left\{1\right\}$ \\
& $\nu_{\{51\}}=2$
&  $e^{(51)}= \left\{1,\frac{1}{z_1}\right\}$ \\
& $\nu_{\{514\}}=5$ 
&  $e^{(514)}=\left\{1, \frac{1}{z_1}, \frac{z_4}{z_1}, \frac{z_5}{z_1}, \frac{z_4^2}{z_1}\right\}$ \\ \hline
\end{tabular}
\end{table}

\begin{align}
\begin{gathered}
\includegraphics[scale=0.12,clip=true, trim = 0 130 0 100]{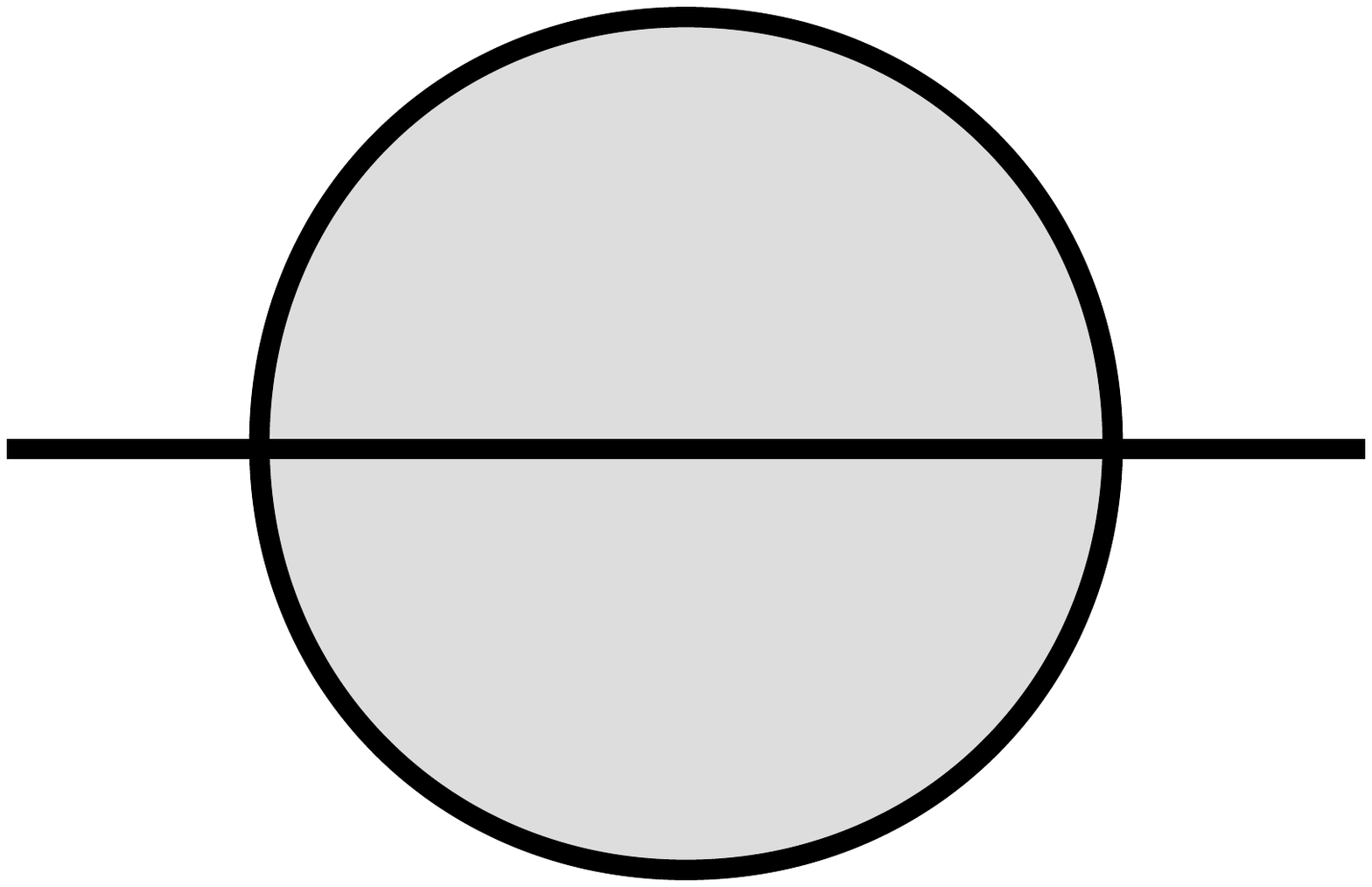}
\end{gathered}   
\; = \;\; & c_1 \!
\begin{gathered}
\includegraphics[scale=0.12,clip=true, trim = 0 180 0 100]{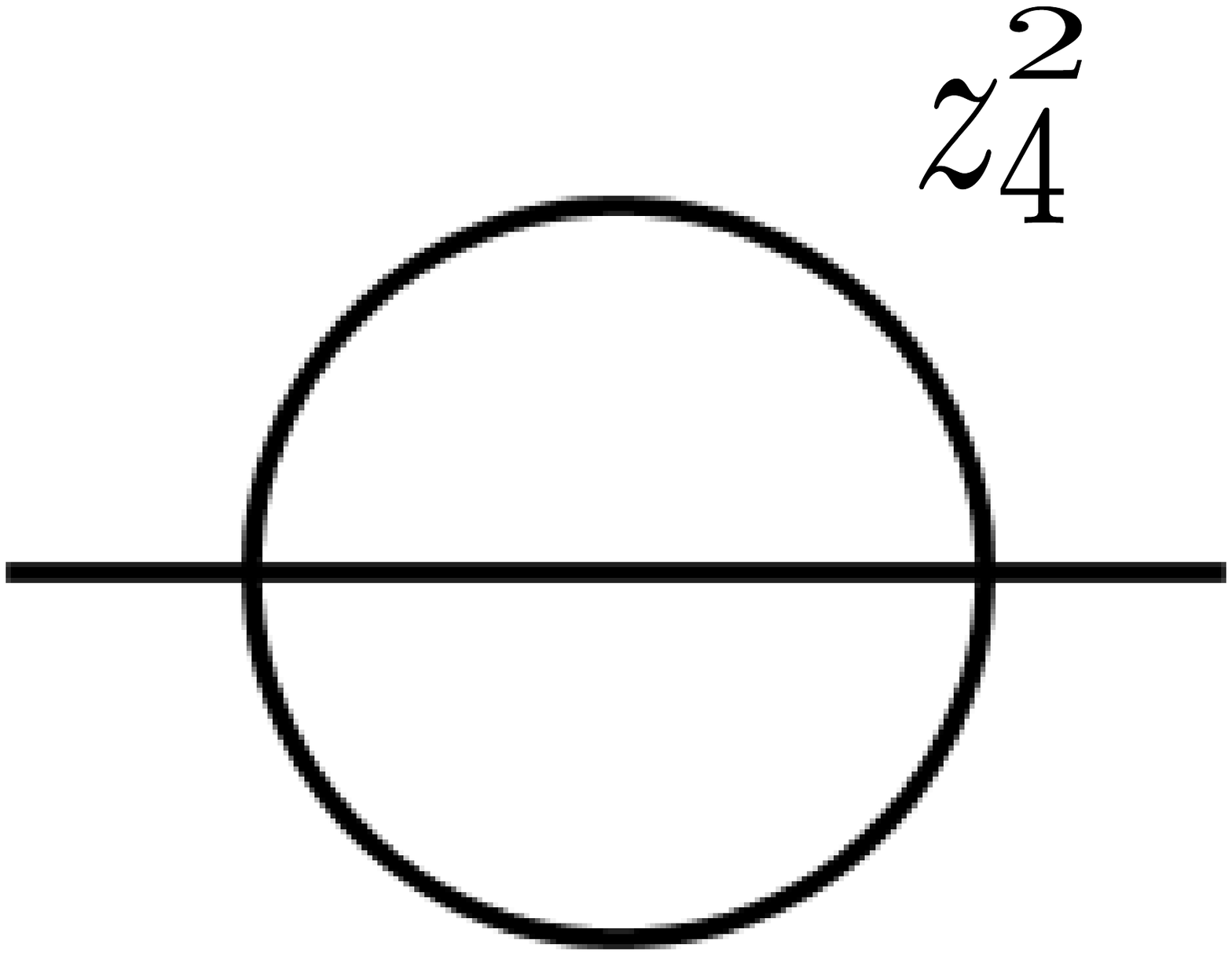}
\end{gathered}
+ c_2 \!
\begin{gathered}
\includegraphics[scale=0.12,clip=true, trim = 0 180 0 100]{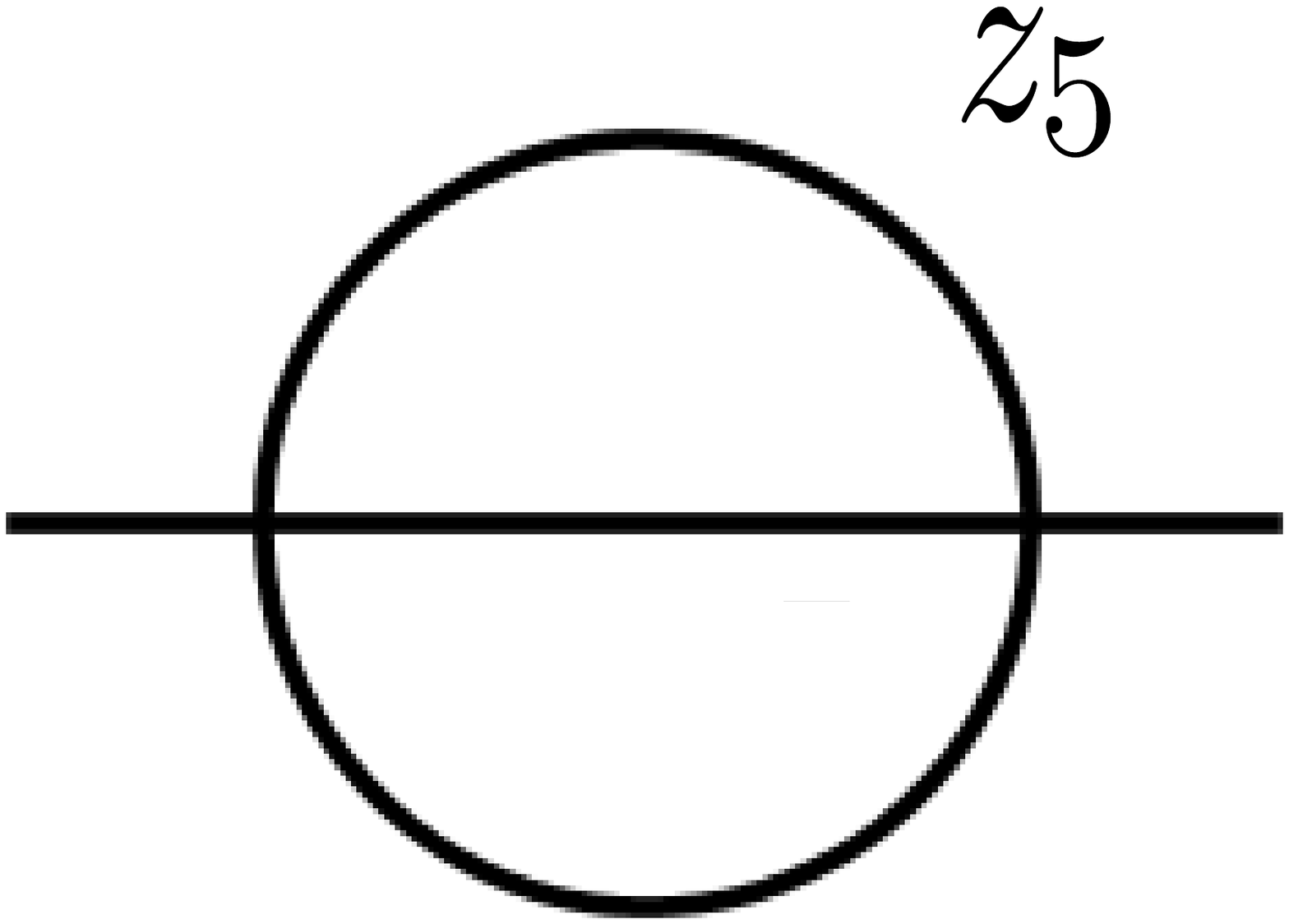}
\end{gathered}
\nonumber \\[-5\jot]
& + c_3 \!
\begin{gathered}
\includegraphics[scale=0.12,clip=true, trim = 0 180 0 100]{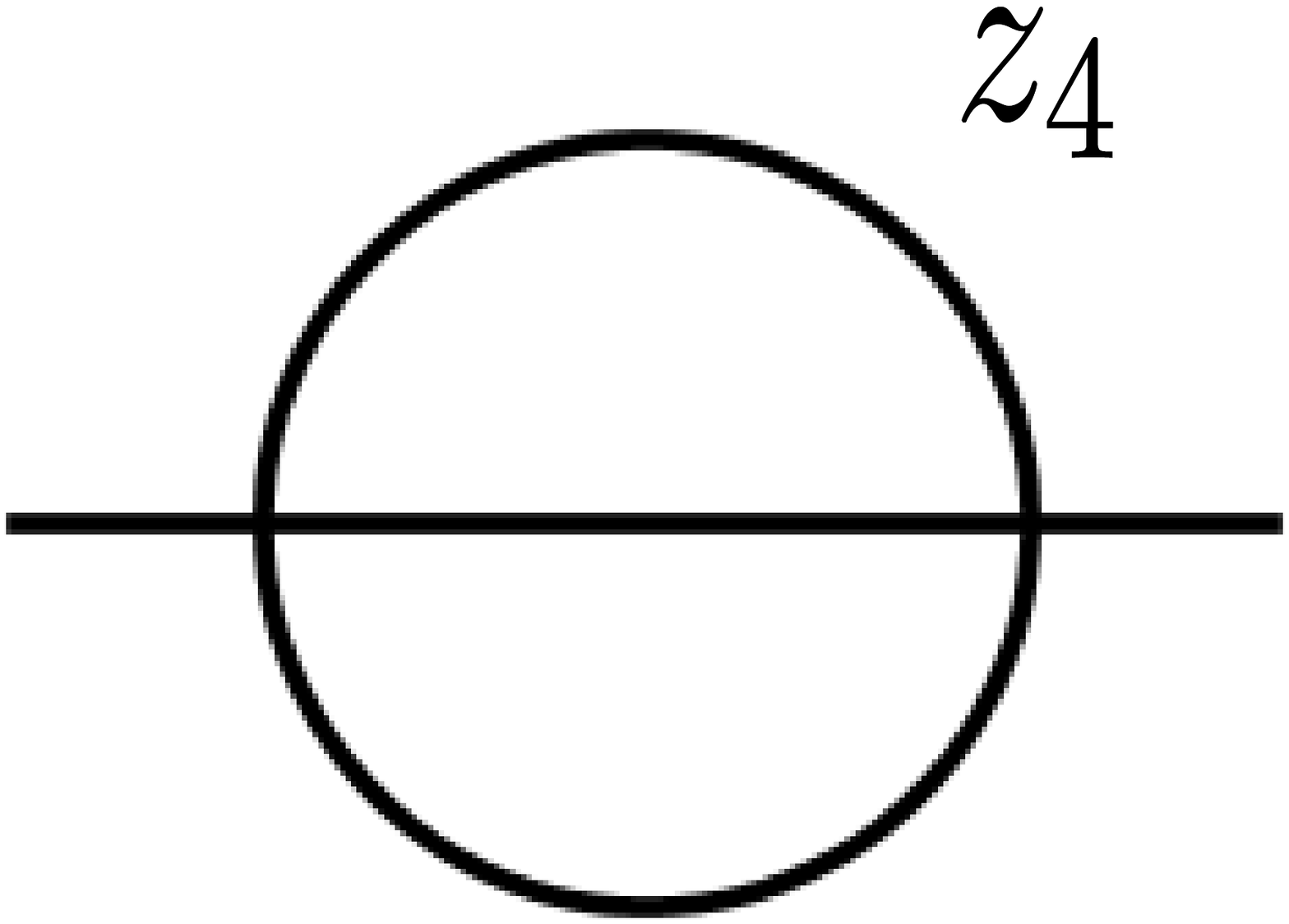}
\end{gathered}
+ c_4 \!
\begin{gathered}
\includegraphics[scale=0.12,clip=true, trim = 0 180 0 100]{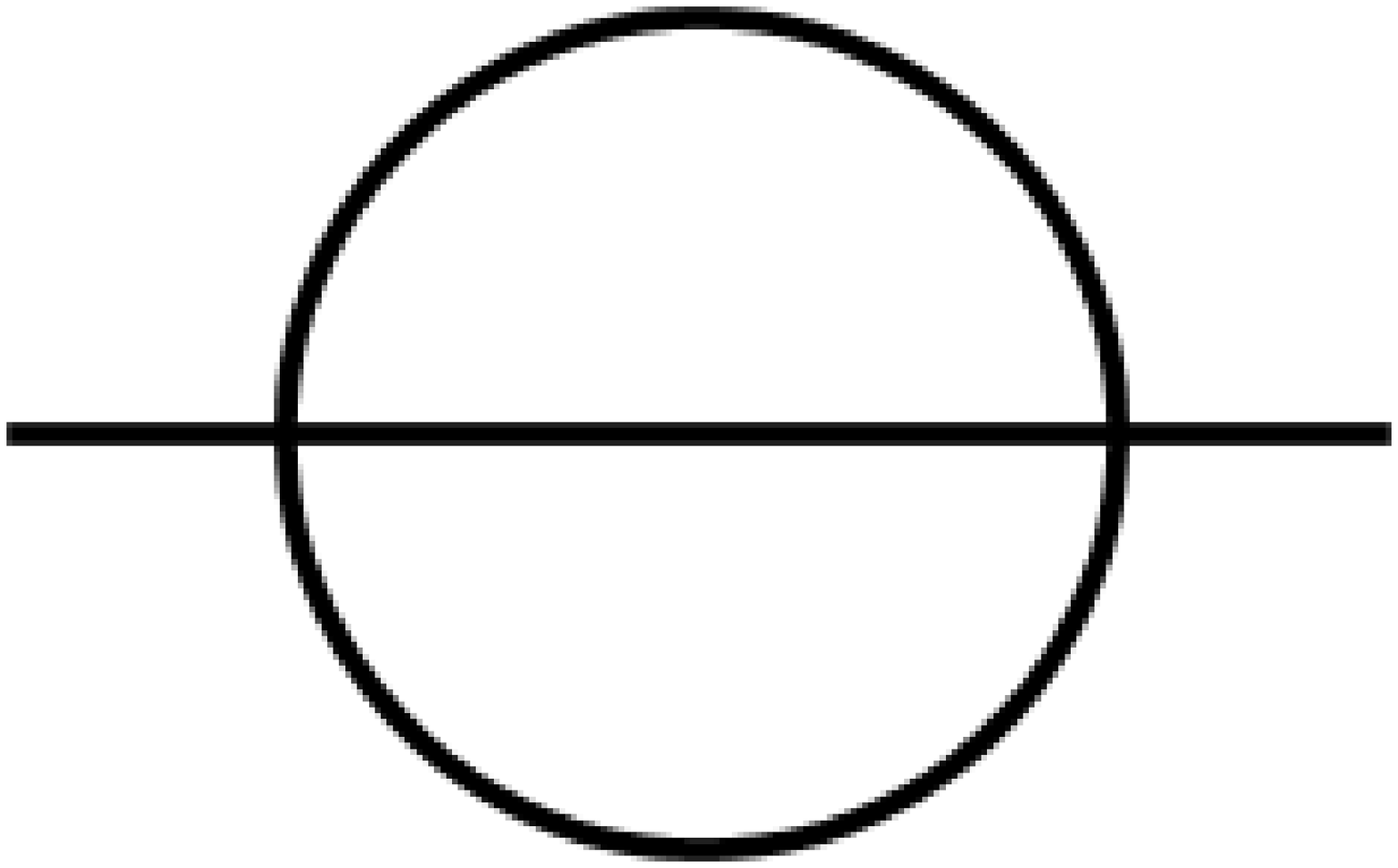}
\end{gathered}
\nonumber \\[-5\jot]
& + c_5 \!
\begin{gathered}
\includegraphics[scale=0.1]{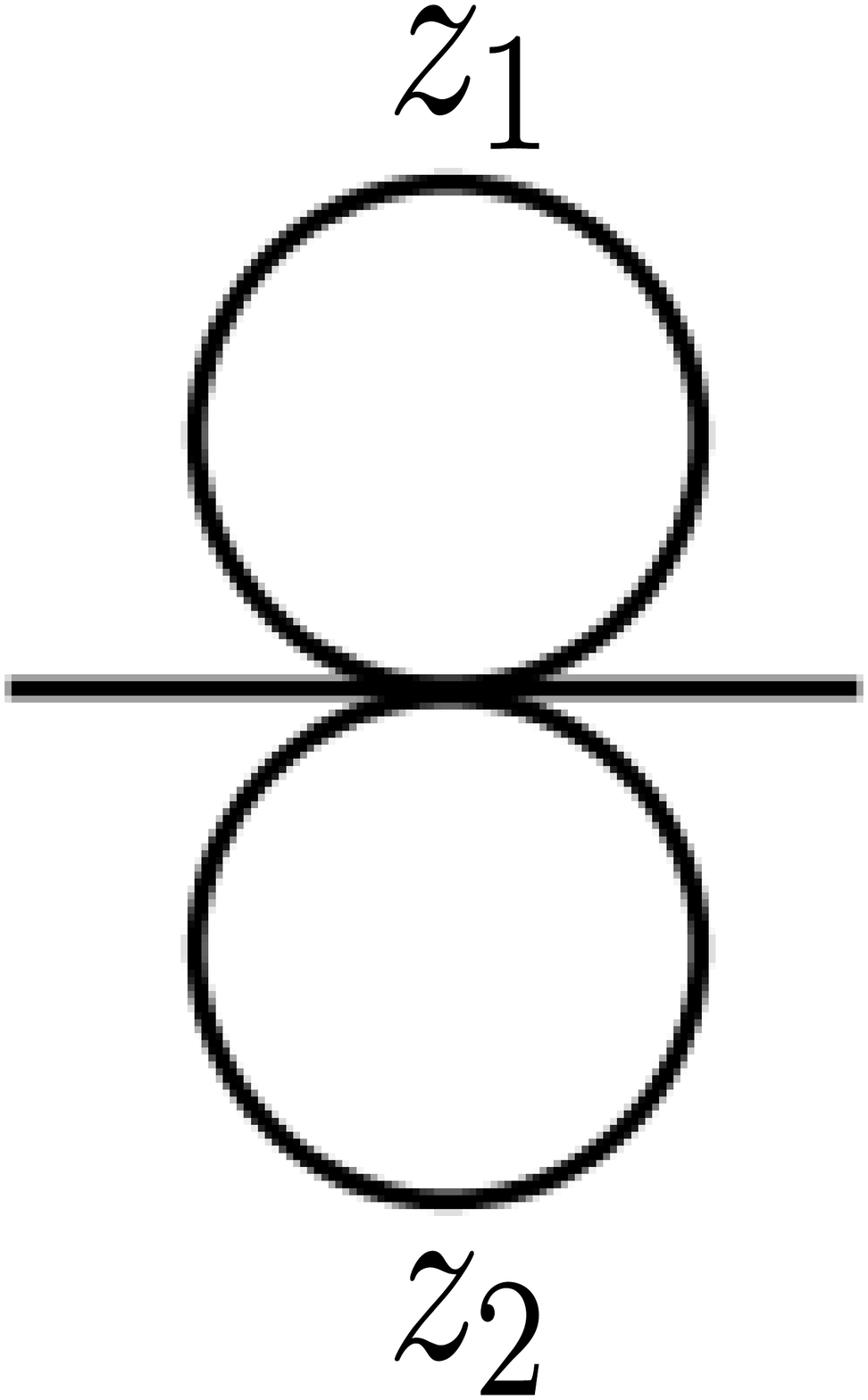}
\end{gathered}
+ c_6 \!
\begin{gathered}
\includegraphics[scale=0.1]{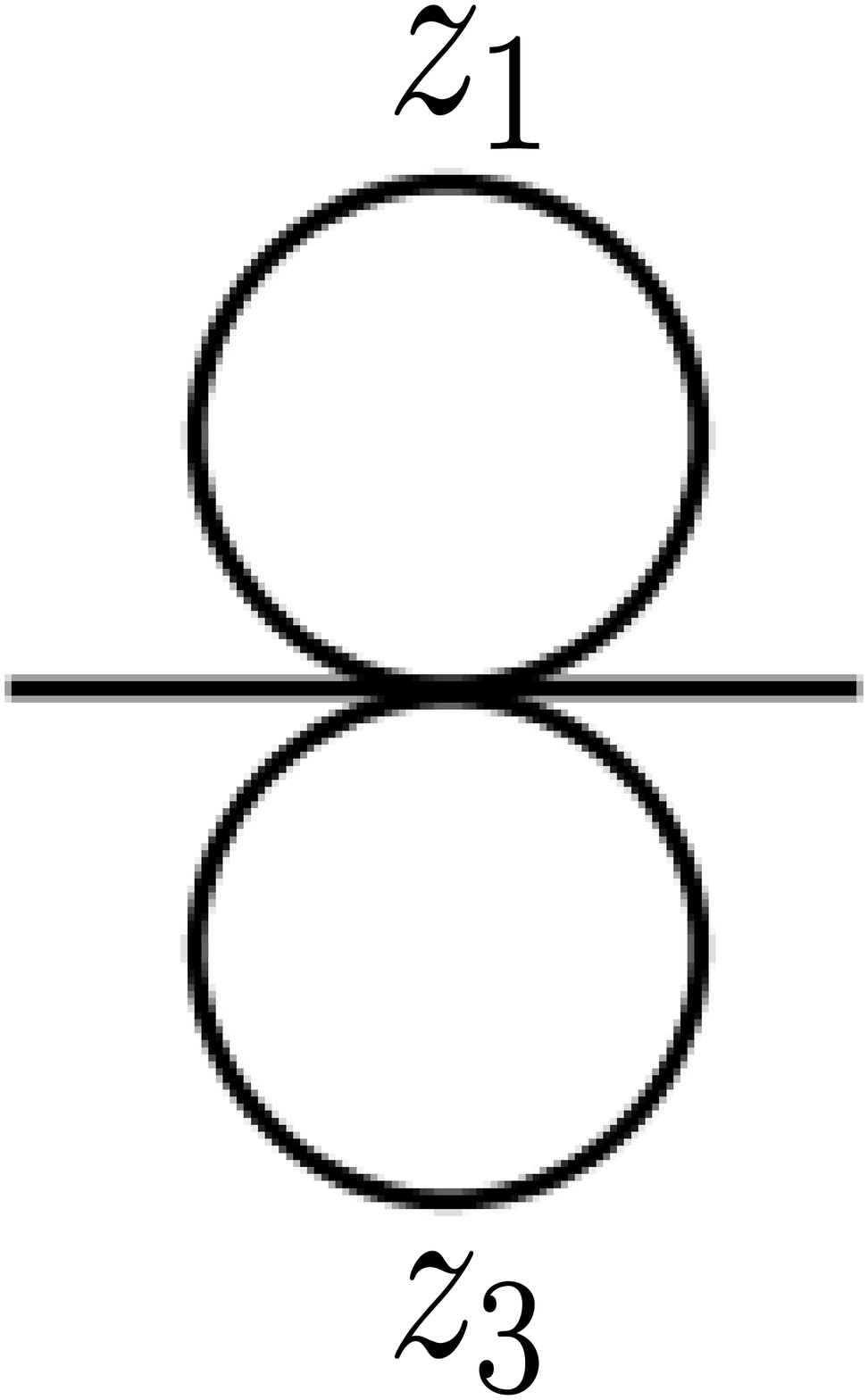}
\end{gathered}
+ c_7 \!
\begin{gathered}
\includegraphics[scale=0.1]{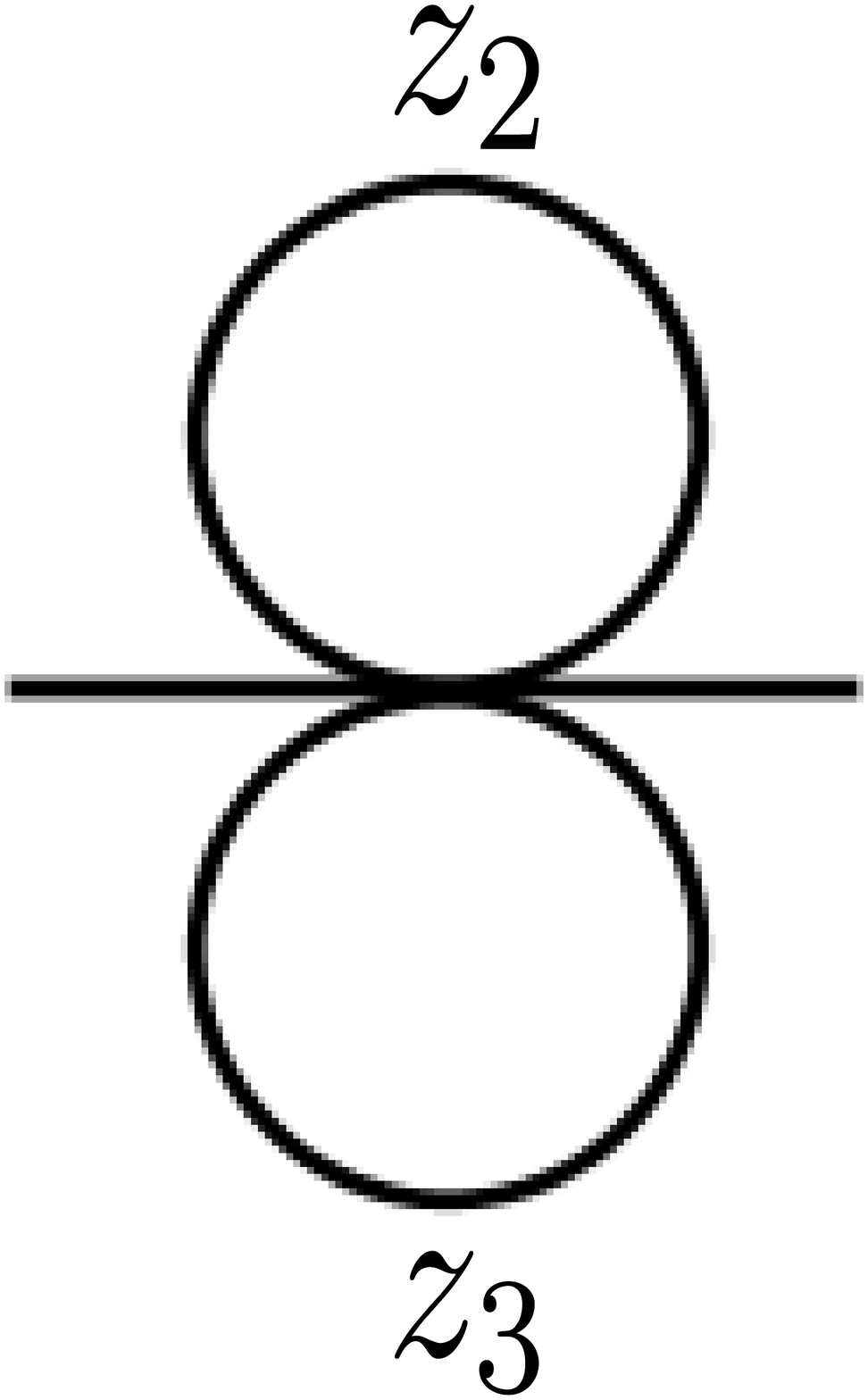}
\end{gathered}
\! .
\end{align}

\clearpage

\subsection*{Massless planar box-triangle}

\begin{table}[!h]
\centering
\begin{tabular}{|c|c|}
\hline $\;\;$ Integral family & Denominator \\ \hline 
\hline
\hline 
\shortstack{
\includegraphics[width=0.22\textwidth,clip=true, trim = 0 300 0 200]{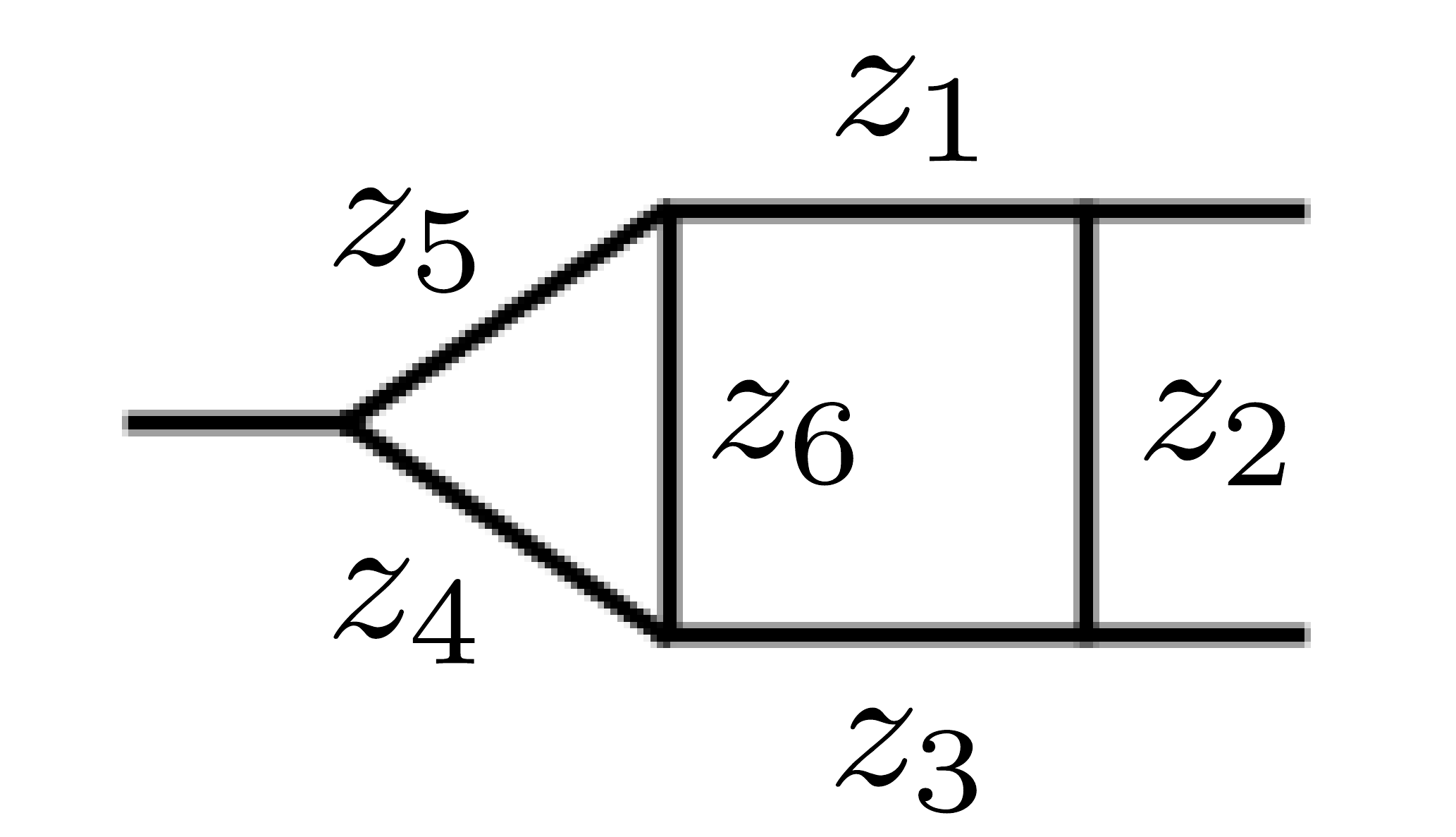}
\\ \, \\ $\,  s=(p_1+p_2)^2  \,$ }
&  \shortstack{\, \\ \, \\ $z_1=k_1^2 \qquad z_2=(k_1+p_1)^2 $ \\
  $z_3=(k_1 + p_1 + p_2)^2  $ \\
  $z_4=(k_2 + p_1 + p_2)^2$\\
  $ z_5=k_2^2 \qquad z_6=(k_1 - k_2)^2$ \\ 
  $z_7=(k_2+p_1)^2$  \\ \, } \\
   \hline
\end{tabular}
\\[2mm]
\begin{tabular}{|c|c|c|}
\hline $\tau$ & $\nu$ & $e$ \\ \hline 
\hline
\hline 
\multirow{3}{*}{\shortstack{
\, \\
\, \\
$z_2=0$\\
$z_4=0 $ \\
$z_5=0 $\\
$z_6=0 $  }} 
&  \multirow{3}{*}{\shortstack{
\, \\ \, \\ \, \\  $\nu_{\{7\}}=1$ \\
$\nu_{\{73\}}=2$ \\
$\nu_{\{731\}}=1$  }}
&   \multirow{3}{*}{\shortstack{  \, \\
\, \\
$e^{(7)}= \left\{ 1 \right\}$ \\
  $e^{(73)}= \left\{1, \frac{1}{z_3} \right\}$ \\
 $e^{(731)}=\left\{1 \right\}$ }}\\
 & & \\ 
 & & \\
 & & \\\hline

\multirow{3}{*}{\shortstack{
\, \\
\, \\
\; \; $z_1=0$  \; \; \\
$z_3=0 $ \\
$z_4=0 $\\
$z_5=0 $  }} 
&  \multirow{3}{*}{\shortstack{
\, \\ \, \\ \, \\  $\nu_{\{7\}}=1$ \\
$\nu_{\{76\}}=1$ \\
\, $\nu_{\{762\}}=1$ \,   }}
&   \multirow{3}{*}{\shortstack{   \, \\
  \, \\
$e^{(7)}= \left\{ 1 \right\}$ \\
  $e^{(76)}= \left\{\frac{1}{z_6} \right\}$ \\
 $e^{(762)}=\left\{1 \right\}$ }}\\
 & & \\ 
 & & \\
 & & \\\hline

\multirow{4}{*}{\shortstack{
\, \vspace{3ex}\\
$z_3=0$\\
$z_5=0 $ \\
$z_6=0 $ }}
& $\nu_{(7)}=1$
&    $e^{(7)}= \left\{ 1 \right\}$ \\
& $\nu_{\{74\}}=1$
&  $e^{(74)}= \left\{\frac{1}{z_4} \right\}$ \\
& $\nu_{\{742\}}=1$
&  $e^{(742)}= \left\{\frac{1}{z_2 z_4} \right\}$ \\
& $\nu_{\{7421\}}=1$ 
&  $e^{(7421)}=\left\{1  \right\}$ \\ \hline

\multirow{4}{*}{\shortstack{
\, \vspace{3ex}\\
$z_1=0$\\
$z_4=0 $ \\
$z_6=0 $ }}
& $\nu_{\{7\}}=1$
&    $e^{(7)}= \left\{ 1 \right\}$ \\
& $\nu_{\{75\}}=1$
&  $e^{(75)}= \left\{\frac{1}{z_5} \right\}$ \\
&  $\nu_{\{752\}}=1$ 
& \, $e^{(752)}= \left\{\frac{1}{z_2 z_5} \right\}$ \, \\
& \, $\nu_{\{7523\}}=1$ \,
&  $e^{(7523)}=\left\{1 \right\}$ \\ \hline
\end{tabular}
\end{table}

\begin{align}
\begin{gathered}
\includegraphics[scale=0.13,clip=true, trim = 0 180 0 100]{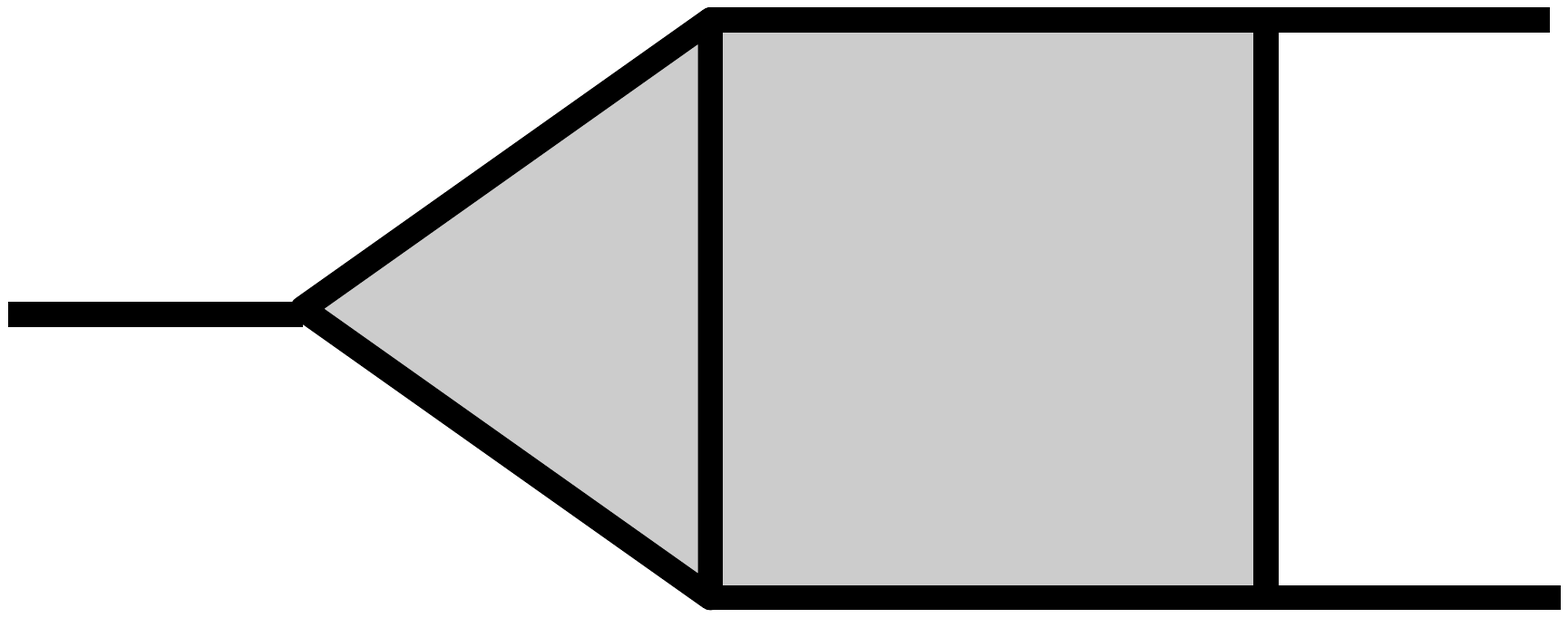}
\end{gathered}
= \;\; & c_1 \,
\begin{gathered}
\includegraphics[scale=0.4]{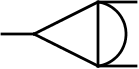}
\end{gathered} 
+ c_2 \, 
\begin{gathered}
\includegraphics[scale=0.4]{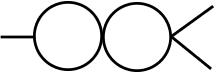}
\end{gathered} 
+ c_3 \!
%\begin{gathered}
%\includegraphics[scale=0.12,clip=true, trim = 0 200 0 100]{figure/Sunrisez3z5z6.pdf}
%\end{gathered}
\begin{gathered}
\includegraphics[scale=0.12,clip=true, trim = 0 200 0 100]{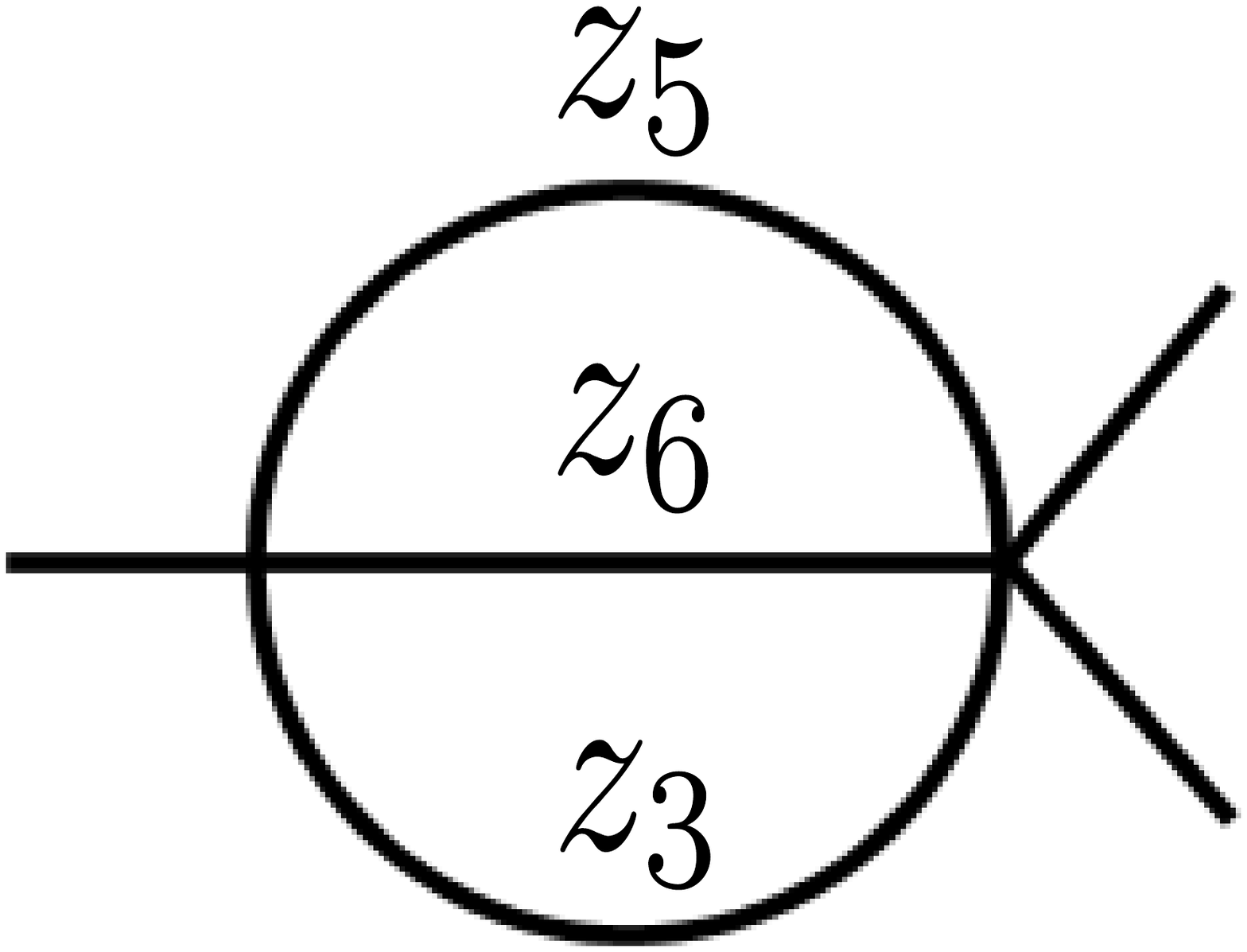}
\end{gathered}
\nonumber \\[-2mm]
& + c_4 \!
\begin{gathered}
\includegraphics[scale=0.12,clip=true, trim = 0 200 0 100]{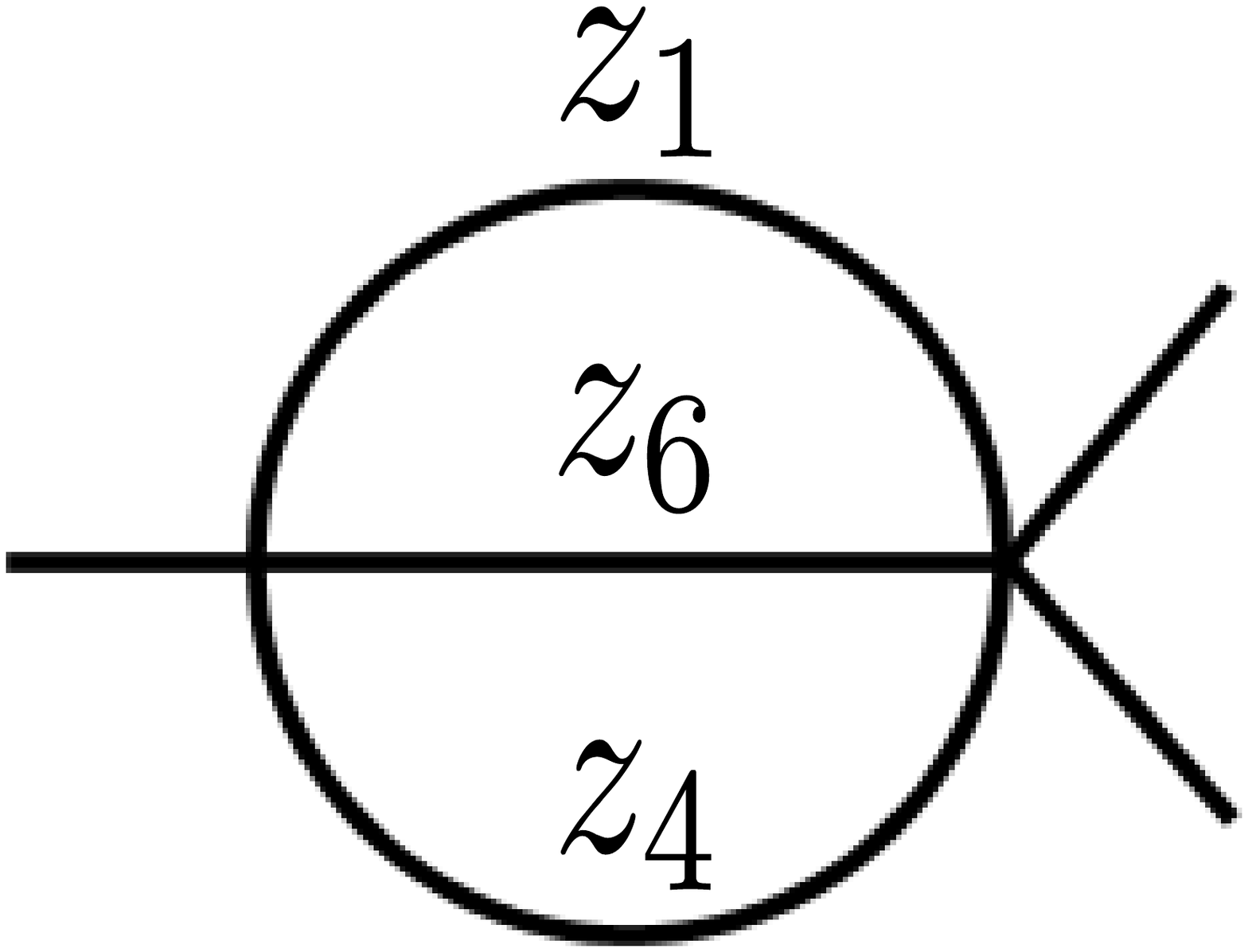}
\end{gathered}
\, .
\end{align}

\clearpage

\subsection*{Massless non-planar triangle-box}

\begin{table}[!h]
\centering
\begin{tabular}{|c|c|}
\hline $\;\;$ Integral family & Denominators \\ \hline 
\hline
\hline 
\shortstack{
\includegraphics[width=0.22\textwidth,clip=true, trim = 0 275 0 200]{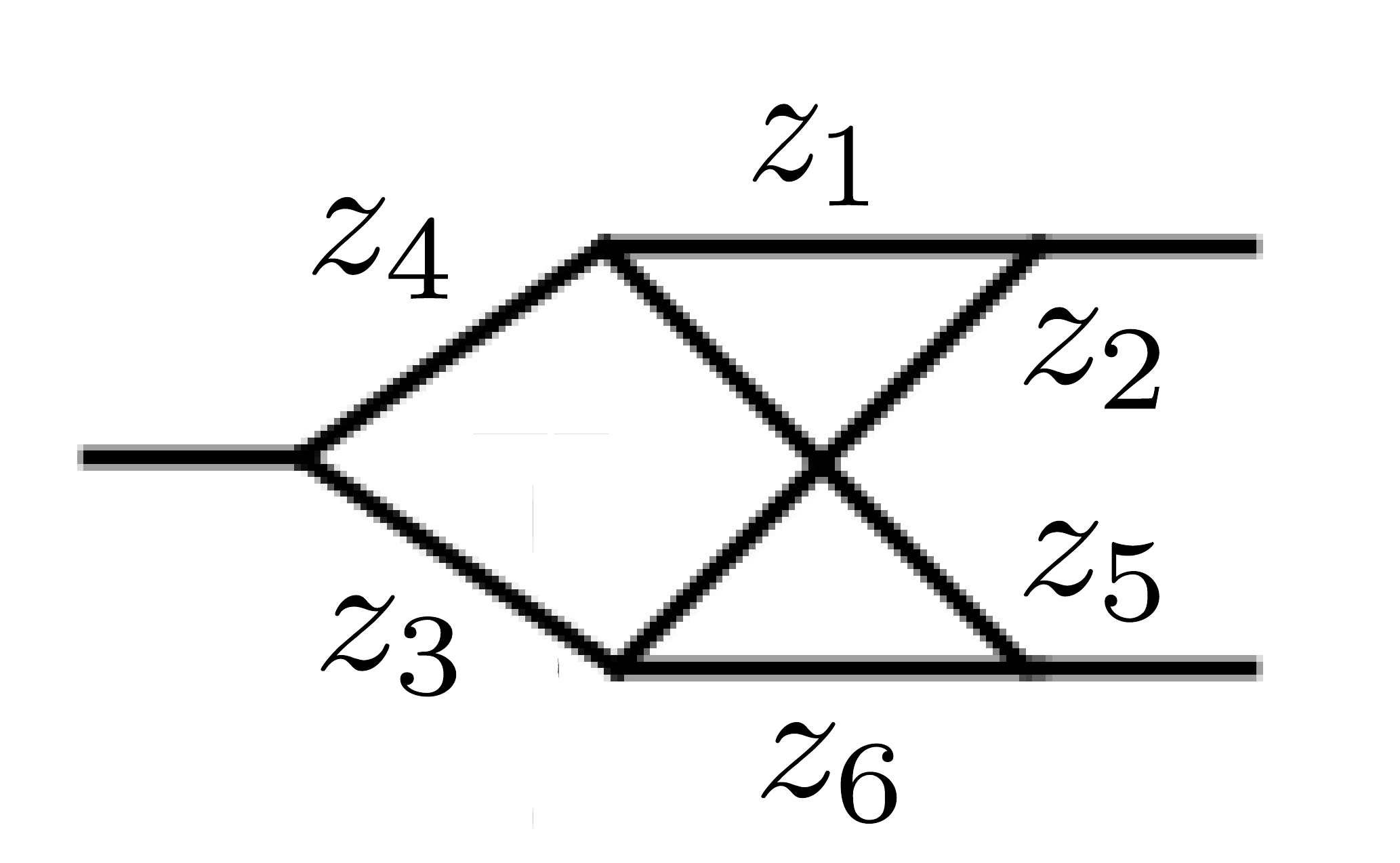}
\\ \, \vspace{2ex} \\ $\, \qquad s=(p_1+p_2)^2 \qquad \,$ }
& \shortstack{ $z_1=k_1^2 \qquad z_2=(k_1+p_1)^2 $ \\
  $z_3=(k_2 + p_1 + p_2)^2  $ \\
  $ z_4=k_2^2 \qquad z_5=(k_1 - k_2)^2$ \\
  $z_6=(k_1 - k_2 - p_2)^2$\\ 
  $z_7=(k_2+p_1)^2$  \\ \, } \\
   \hline
\end{tabular}
\\[2mm]
\begin{tabular}{|c|c|c|}
\hline $\tau$ & $\nu$ & $e$ \\ \hline 
\hline
\hline 
 \multirow{3}{*}{
 \shortstack{
 \, \\
$z_2=0$\\
$z_3=0 $ \\
$z_4=0 $\\
$z_5=0 $ 
}
}
& $\nu_{\{1\}}=2$ 
&    $e^{(1)}= \left\{ 1,\frac{1}{z_1} \right\}$ \\
& $\nu_{\{16\}}=2$
&  $e^{(16)}= \left\{\frac{1}{z_6}, \frac{1}{z_1 z_6} \right\}$ \\
& $\nu_{\{167\}}=2$
&  $e^{(167)}=\left\{1, \frac{1}{z_1 z_6} \right\}$ \\ \hline

\multirow{3}{*}{\shortstack{
\, \\
$z_1=0$\\
$z_3=0 $ \\
$z_4=0 $\\
$z_6=0 $ }}
& $\nu_{\{2\}}=2$
&    $e^{(2)}= \left\{ 1,\frac{1}{z_2} \right\}$ \\
& $\nu_{\{25\}}=2$
&  $e^{(25)}= \left\{\frac{1}{z_5}, \frac{1}{z_2 z_5} \right\}$ \\
& $\nu_{\{257\}}=2$
&  $e^{(257)}=\left\{1, \frac{1}{z_2 z_5} \right\}$ \\ \hline

\multirow{4}{*}{\shortstack{
\, \vspace{3ex}\\
$z_1=0$\\
$z_3=0 $ \\
$z_5=0 $ }}
& $\nu_{\{2\}}=2$
&    $e^{(2)}= \left\{ 1,\frac{1}{z_2} \right\}$ \\
& $\nu_{\{24\}}=2$
&  $e^{(24)}= \left\{\frac{1}{z_4}, \frac{1}{z_2 z_4} \right\}$ \\
& $\nu_{\{246\}}=3$
&  $e^{(246)}= \left\{\frac{1}{z_6}, \frac{1}{z_4 z_6} , \frac{1}{z_2 z_4 z_6} \right\}$ \\
& $\nu_{\{2467\}}=2$
&  $e^{(2467)}=\left\{1, \frac{1}{z_2 z_4 z_6} \right\}$ \\ \hline
\multirow{4}{*}{\shortstack{
\, \vspace{3ex}\\
$z_2=0$\\
$z_4=0 $ \\
$z_6=0 $ }}
& $\nu_{\{1\}}=2$
&    $e^{(1)}= \left\{ 1,\frac{1}{z_1} \right\}$ \\
& $\nu_{\{15\}}=2$
&  $e^{(15)}= \left\{\frac{1}{z_5}, \frac{1}{z_1 z_5} \right\}$ \\
& $\nu_{\{153\}}=3$
&  $e^{(153)}= \left\{\frac{1}{z_3}, \frac{1}{z_3 z_5} , \frac{1}{z_1 z_3 z_5} \right\}$ \\
& $\nu_{\{1537\}}=2$
&  $e^{(1537)}=\left\{1, \frac{1}{z_1 z_3 z_5} \right\}$ \\ \hline
\end{tabular}
\end{table}
\begin{align}
\begin{gathered}
\includegraphics[scale=0.54]{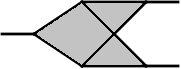}
\end{gathered} 
= \;\; & c_1\;
\begin{gathered}
\includegraphics[scale=0.4]{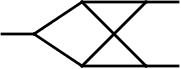}
\end{gathered} 
+ c_2 \!
\begin{gathered}
\includegraphics[scale=0.13,clip=true, trim = 0 180 0 100]{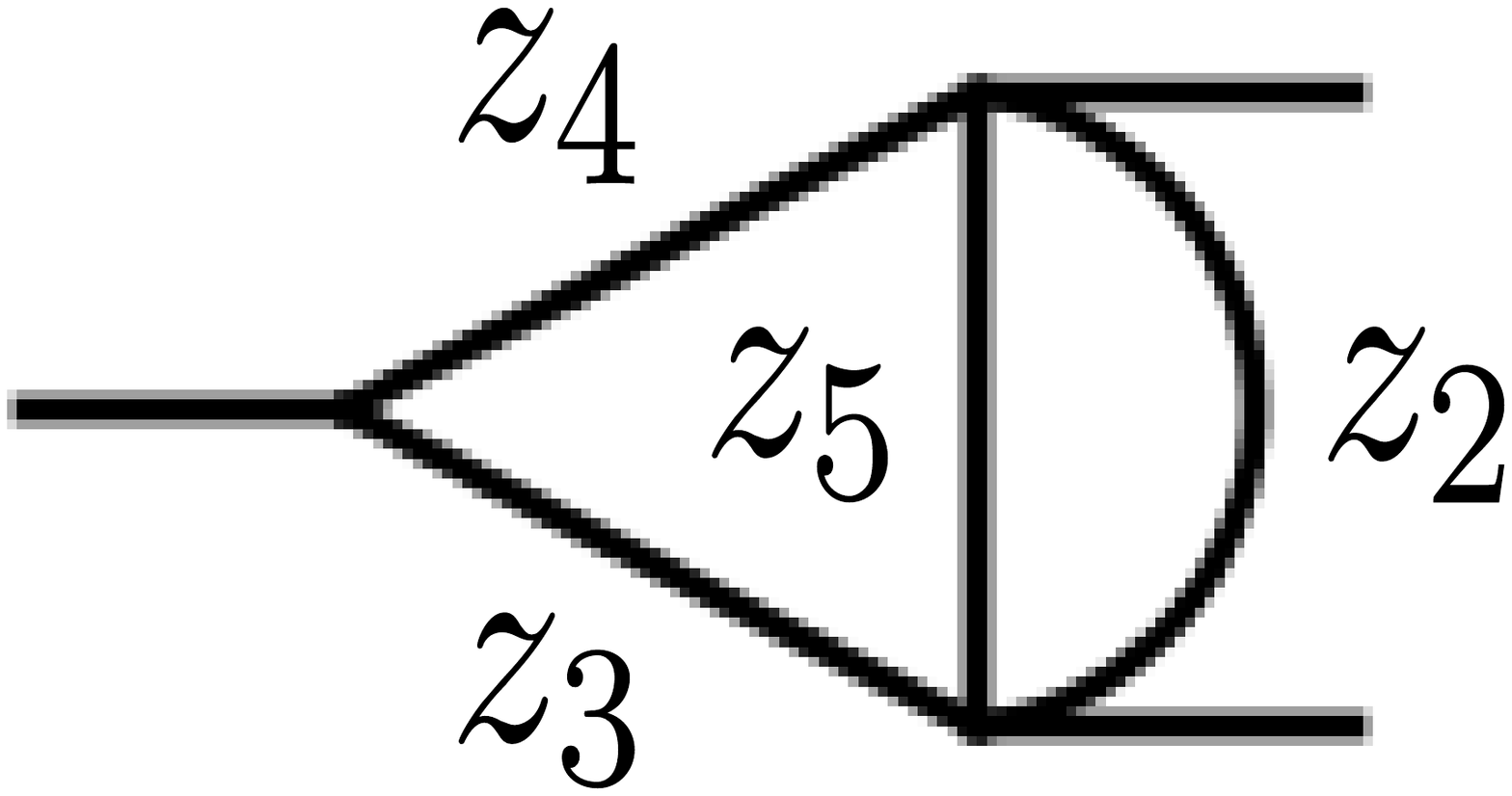}
\end{gathered} 
+ c_3 \!
\begin{gathered}
\includegraphics[scale=0.13,clip=true, trim = 0 180 0 100]{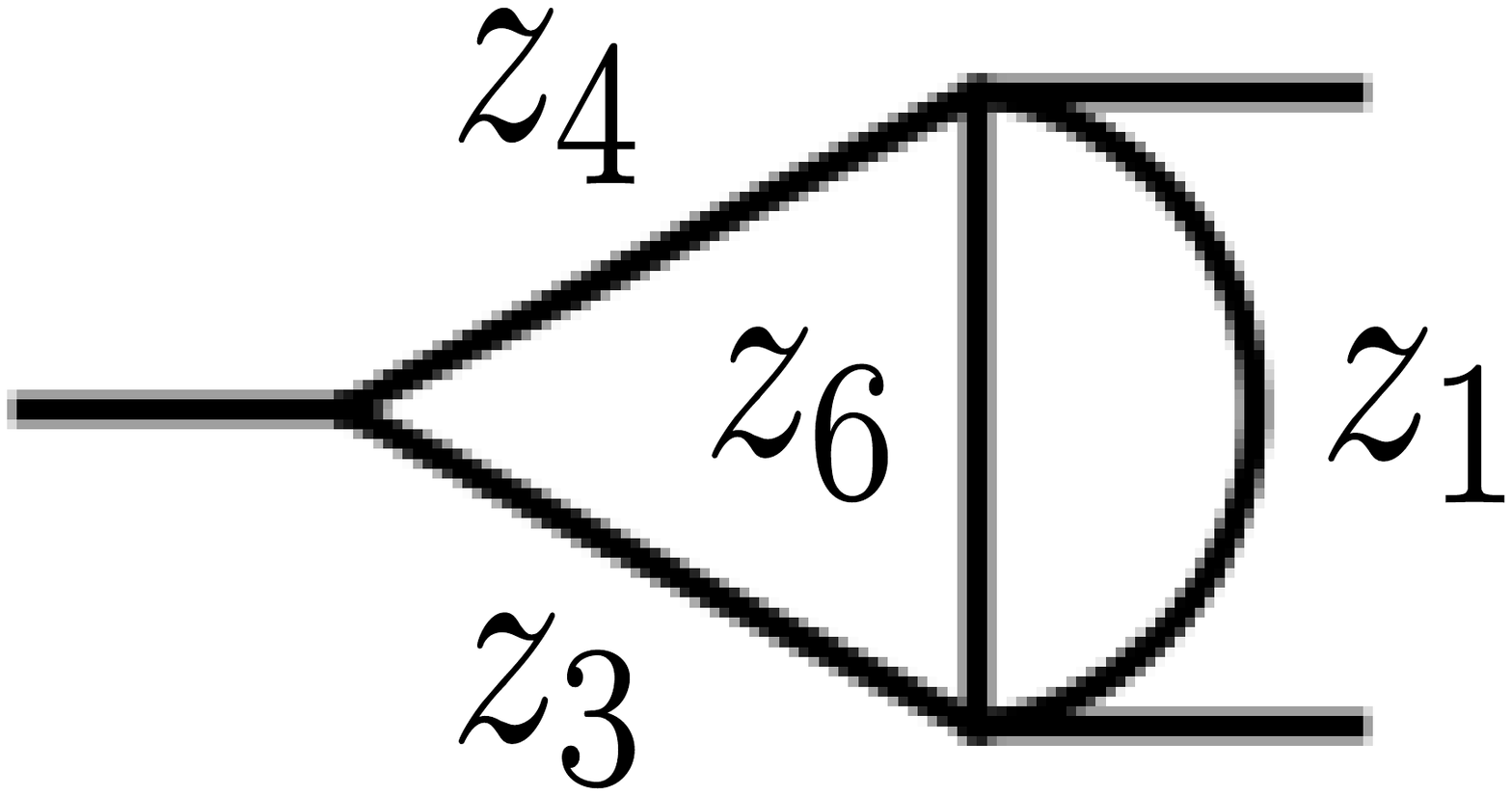}
\end{gathered} 
\nonumber \\[-6\jot]
& + c_4 \!
\begin{gathered}
\includegraphics[scale=0.12,clip=true, trim = 0 200 0 100]{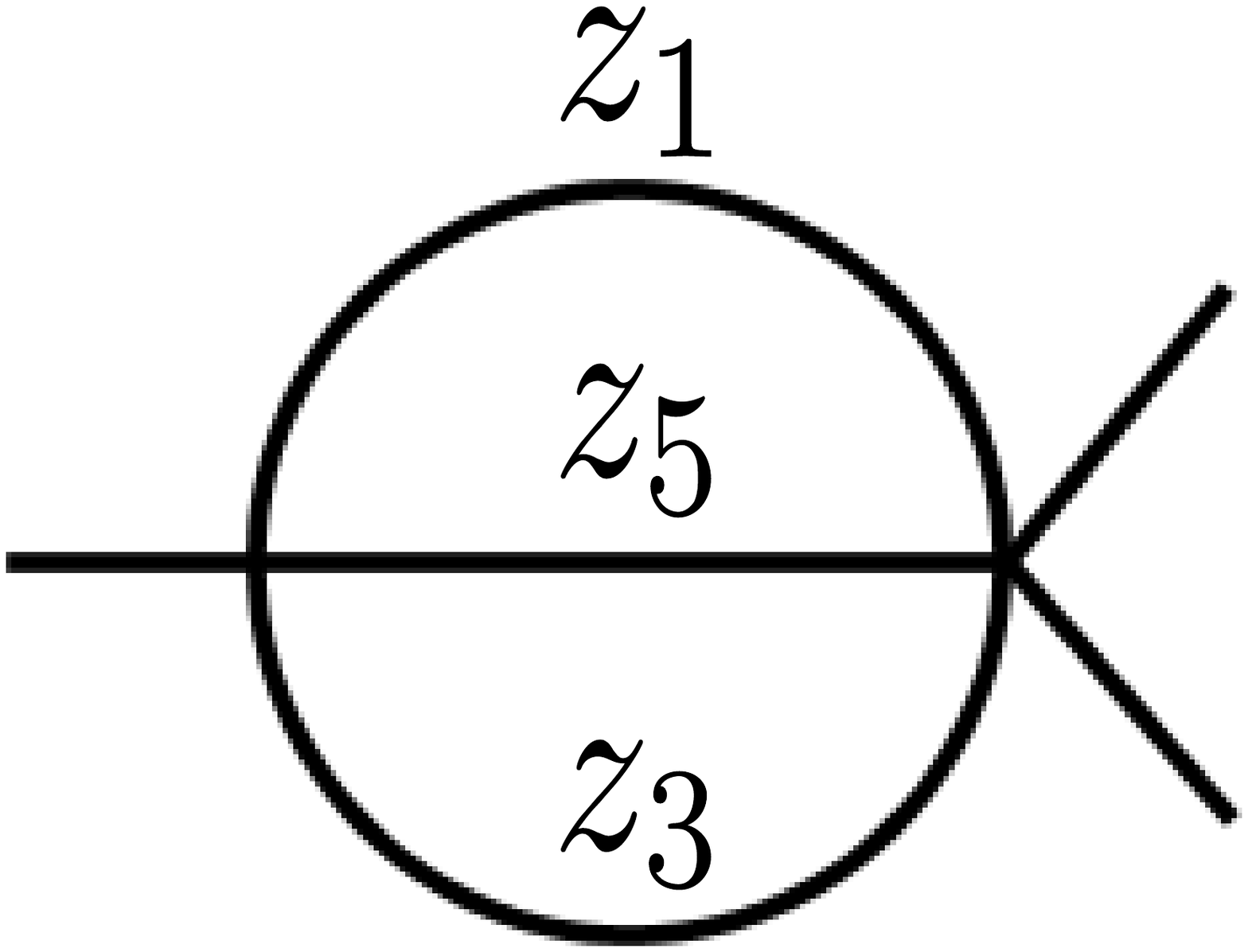}
\end{gathered}
+ c_5 \!
\begin{gathered}
\includegraphics[scale=0.12,clip=true, trim = 0 200 0 100]{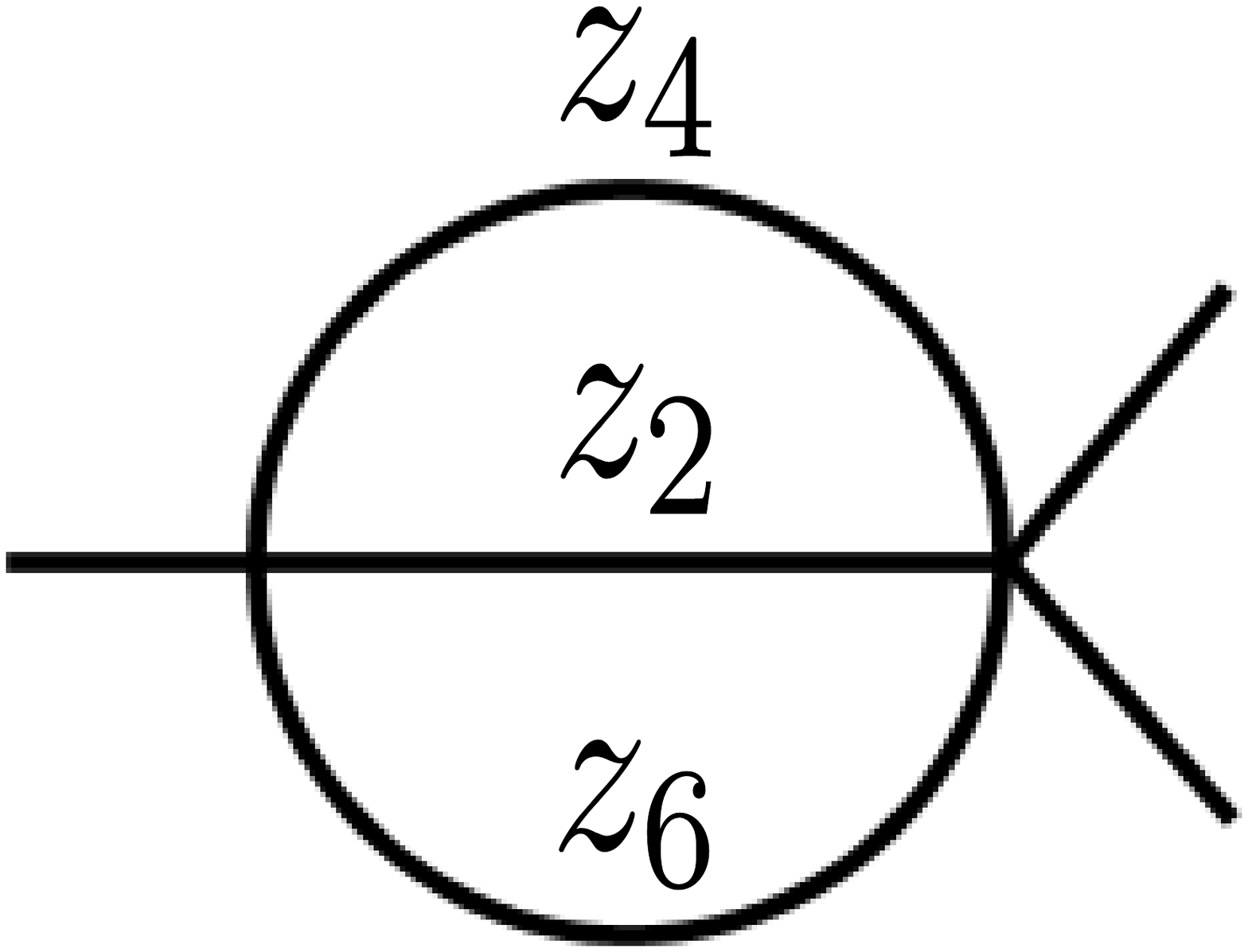}
\end{gathered} \ .
\end{align}

\clearpage

\subsection*{Massless double-box on a triple cut}

\begin{table}[!h]
\centering
\begin{tabular}{|c|c|}
\hline $\;\;$ Integral family & Denominators \\ \hline 
\hline
\hline 
\shortstack{
\includegraphics[width=0.22\textwidth,clip=true, trim = 0 275 0 200]{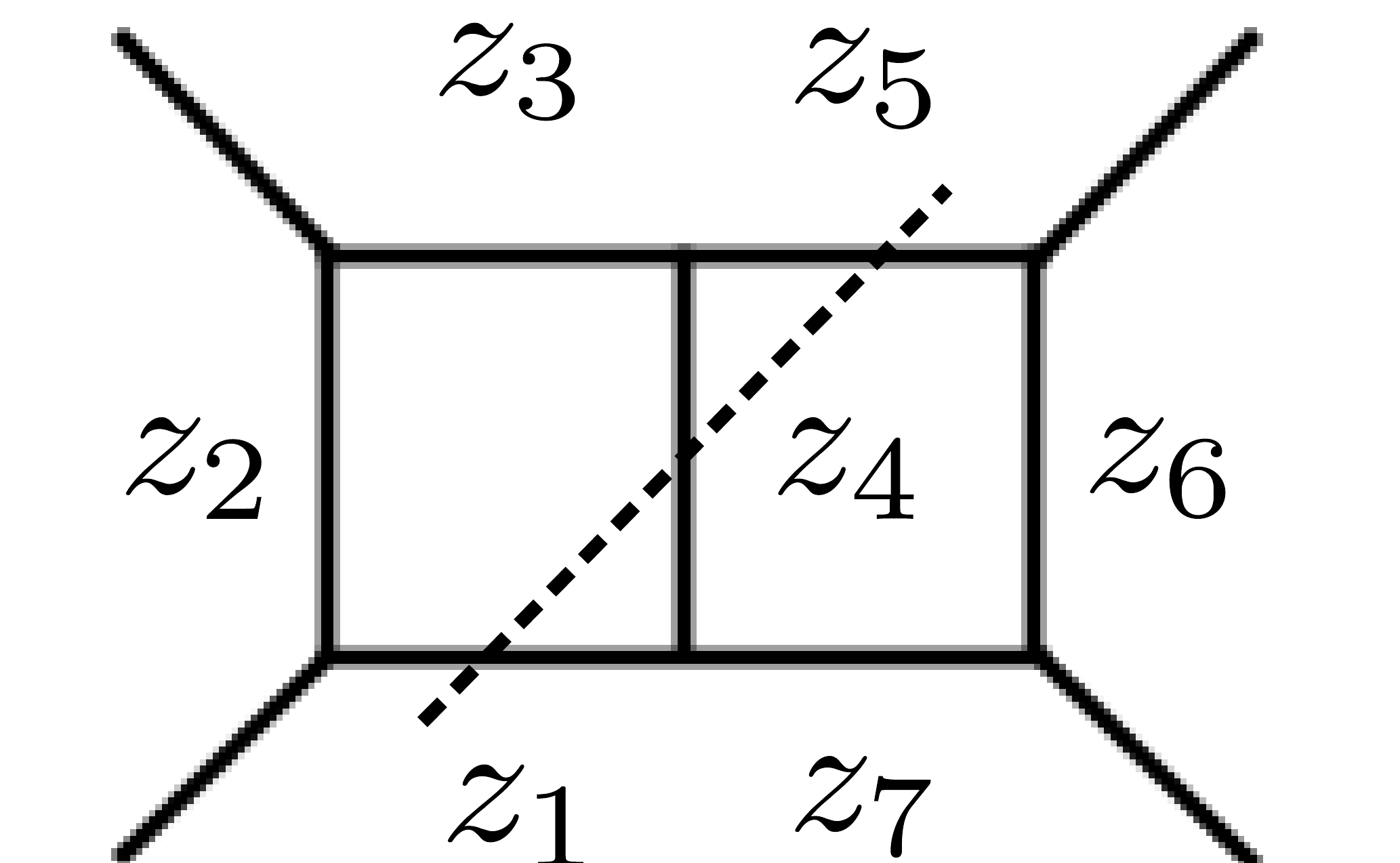}
  \\ \, \\ \, \\ \, \\ $\; \; \; \; \, s=(p_1+p_2)^2 \quad t=(p_2+p_3)^2 \, \; \; \; \;$ \\  \, \\ \, }
& \shortstack{\\  \; $z_1=k_1^2 \qquad z_2=(k_1 - p_1)^2 $ \;  \\
  $z_3=(k_1 - p_1 - p_2)^2  $ \\
  $z_4=(k_1 - k_2)^2$\\
  $z_5=(k_2 - p_1 - p_2)^2  $ \\
  $z_6=(k_2 - p_1 - p_2 - p_3)^2  $ \\
  $ z_7=k_2^2 \qquad z_8=(k_2 - p_1)^2$ \\
  $z_9=(k_1 - p_1 - p_2 - p_3)^2$ } \\
   \hline
\end{tabular}
\\[2mm]
\begin{tabular}{|c|c|c|}
\hline $\tau$ & $\nu$ & $e$ \\ \hline 
\hline
\hline 
 \multirow{6}{*}{
 \shortstack{%\{ 
 \vspace{2ex} \\
\\
$z_1=0 $ \\
$z_4=0 $ \\
$z_5=0 $ 
}
}
& $\nu_{\{8\}}=1$
&    $e^{(8)}= \left\{ 1 \right\}$ \\
& $\nu_{\{87\}}=2$
&  $e^{(87)}= \left\{1,\frac{1}{z_7} \right\}$ \\
& $\nu_{\{876\}}=2$ 
&  $e^{(876)}=\left\{\frac{1}{z_6}, \frac{1}{z_7} \right\}$ \\
& $\nu_{\{8762\}}=4$ 
&  $e^{(8762)}=\left\{\frac{1}{z_2}, \frac{1}{z_6}, \frac{1}{z_7}, \frac{1}{z_2 z_6} \right\}$ \\
& $\nu_{\{87629\}}=5$ 
&  $e^{(87629)}=\left\{1, \frac{1}{z_2}, \frac{1}{z_6}, \frac{1}{z_7}, \frac{1}{z_2 z_6} \right\}$ \\
& $\nu_{\{876293\}}=4$ 
&  $e^{(876293)}=\left\{1, \frac{1}{z_2 z_6}, \frac{1}{z_2 z_3 z_6 z_7}, \frac{z_8}{z_2 z_3 z_6 z_7} \right\}$ \\
\hline

\end{tabular}

\end{table}

\begin{align}
\begin{gathered}
\includegraphics[scale=0.13,clip=true, trim = 0 200 0 100]{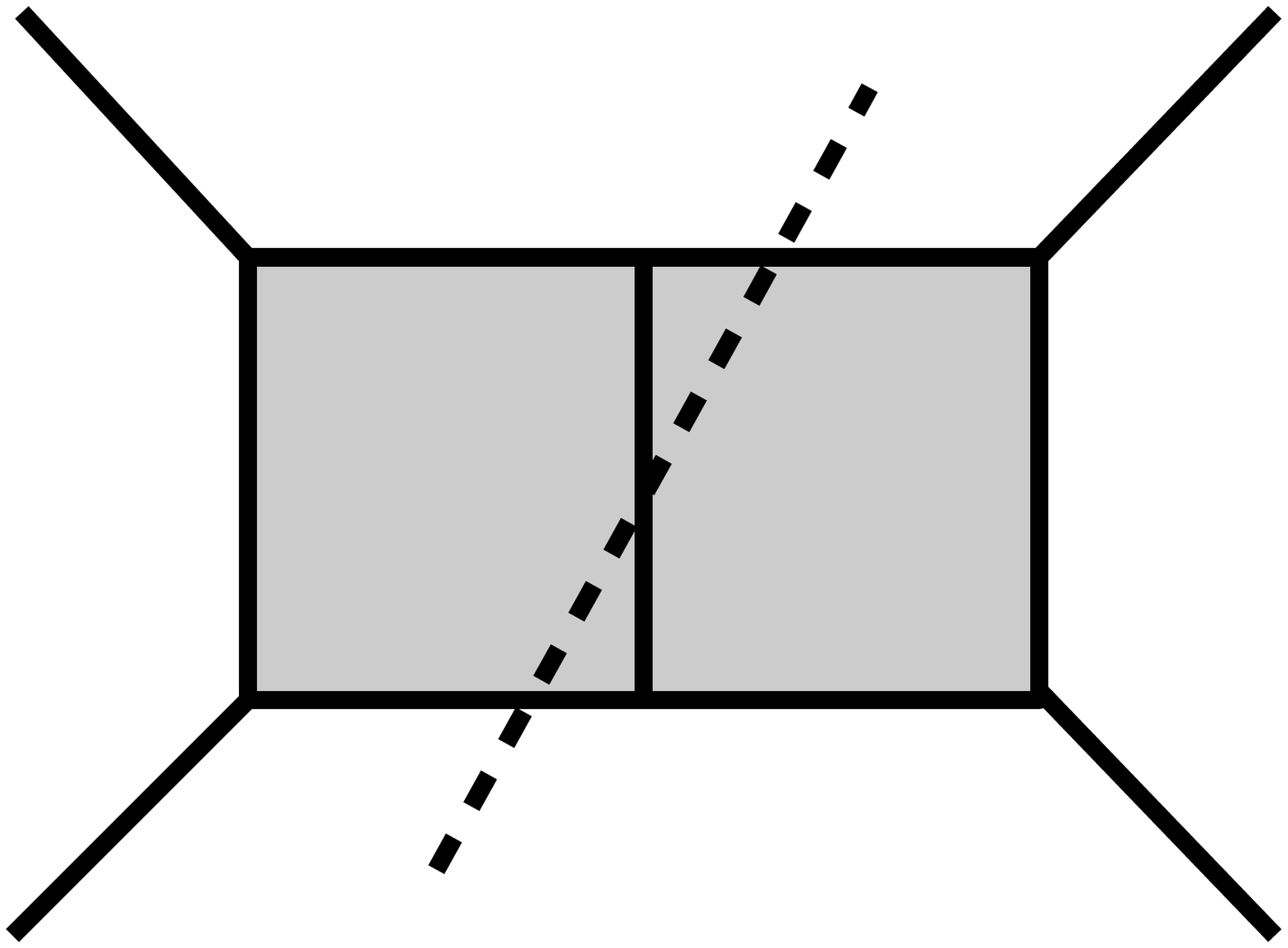}
\end{gathered} 
\! =& \;\; c_1 \!
\begin{gathered}
\includegraphics[scale=0.13,clip=true, trim = 0 200 0 100]{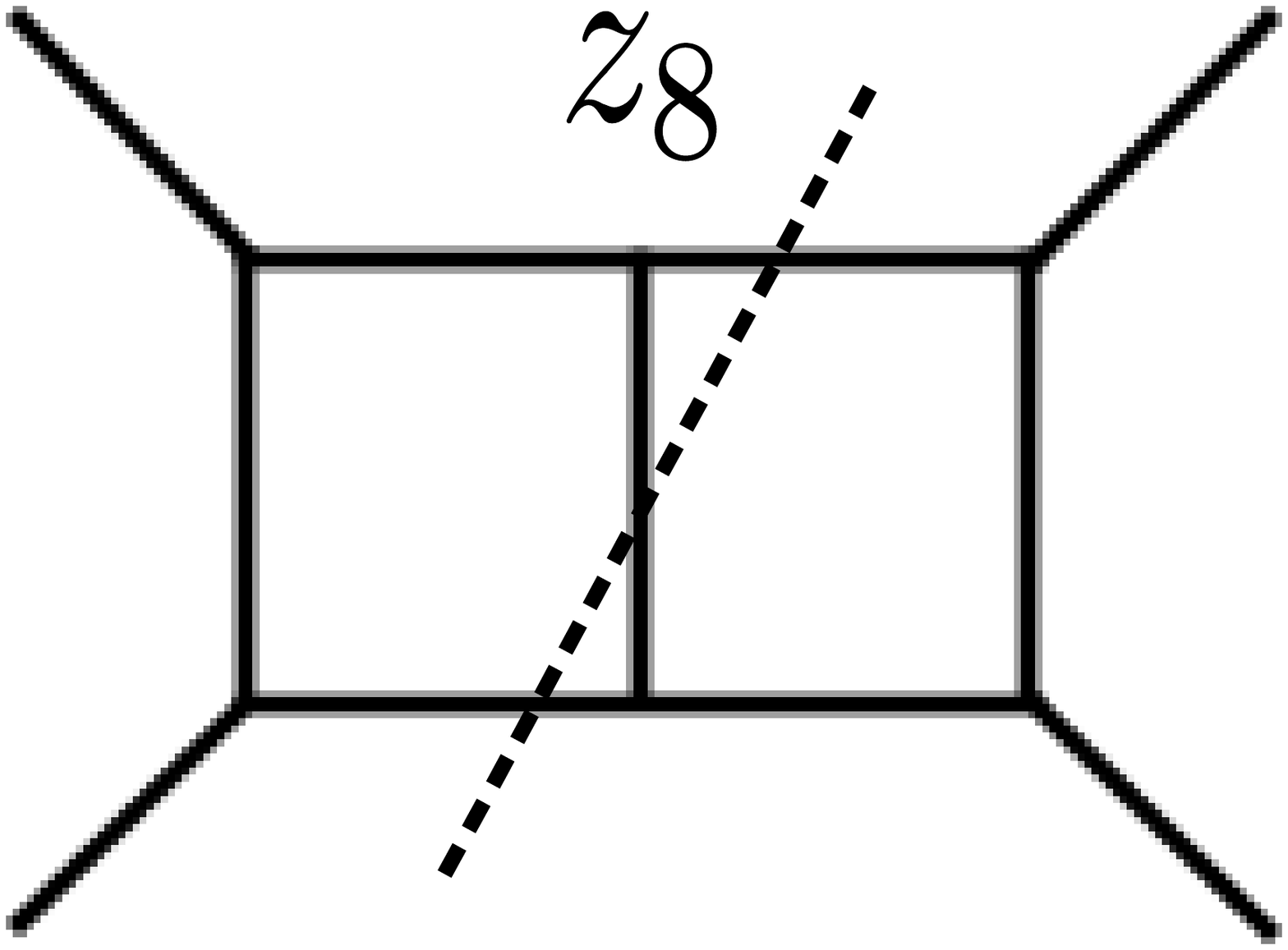}
\end{gathered}
\! + c_2 \!
\begin{gathered}
\includegraphics[scale=0.13,clip=true, trim = 0 200 0 100]{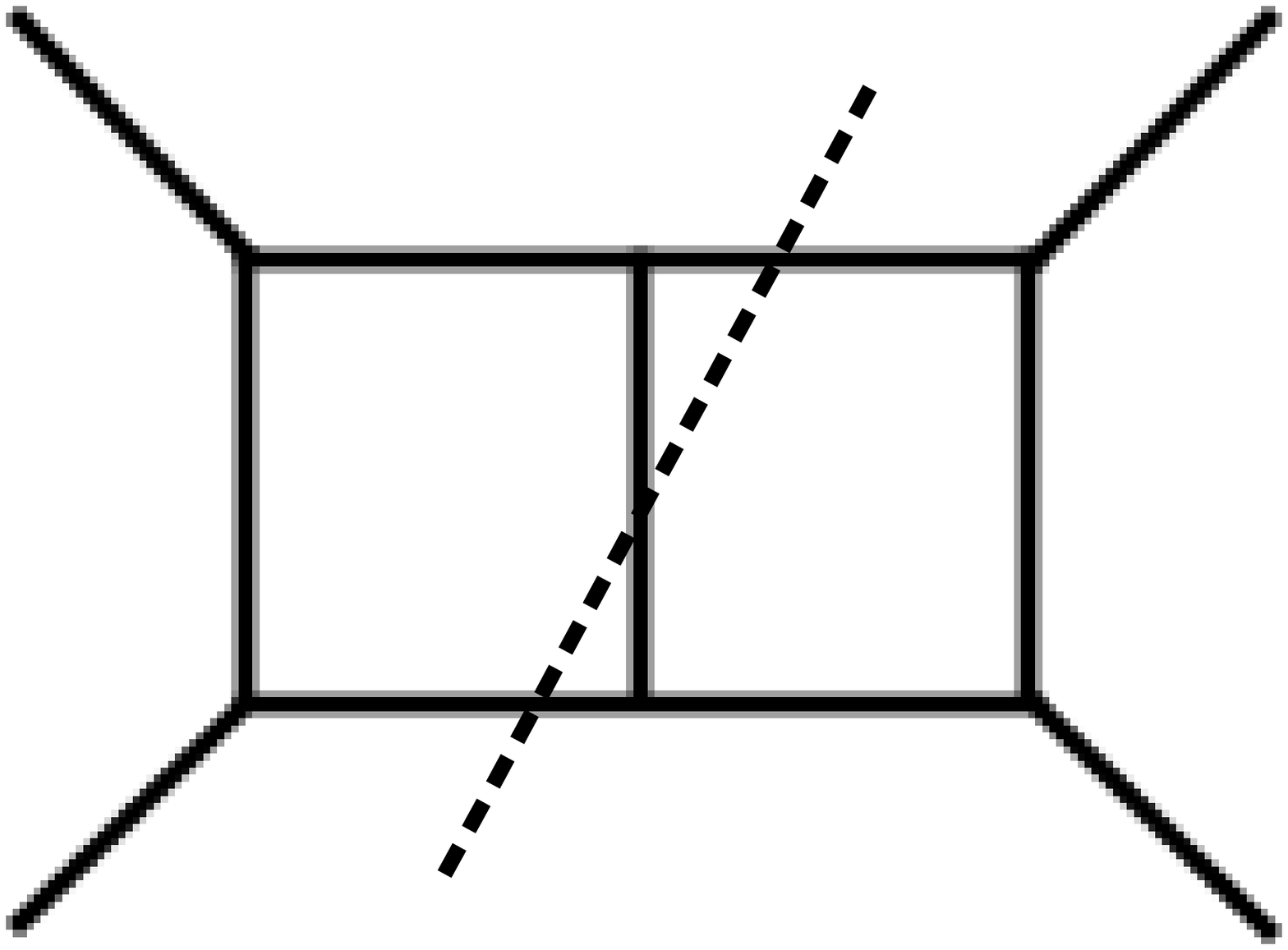}
\end{gathered} \nonumber\\
\! &\;+ c_3 \!
\begin{gathered}
\includegraphics[scale=0.13,clip=true, trim = 0 200 150 100]{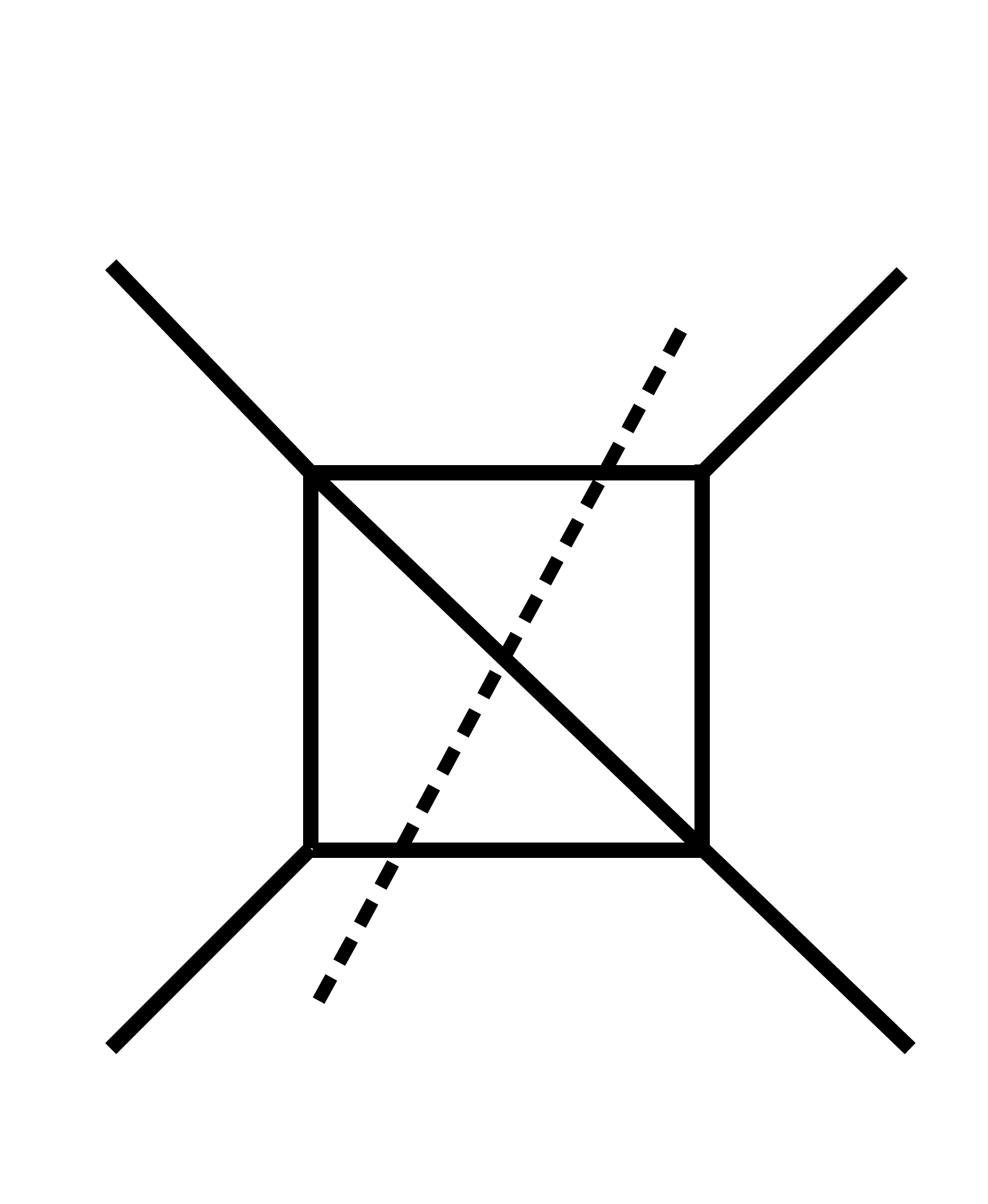}
\end{gathered}
\! + c_4 \!
\begin{gathered}
\includegraphics[scale=0.13,clip=true, trim = 50 200 50 100]{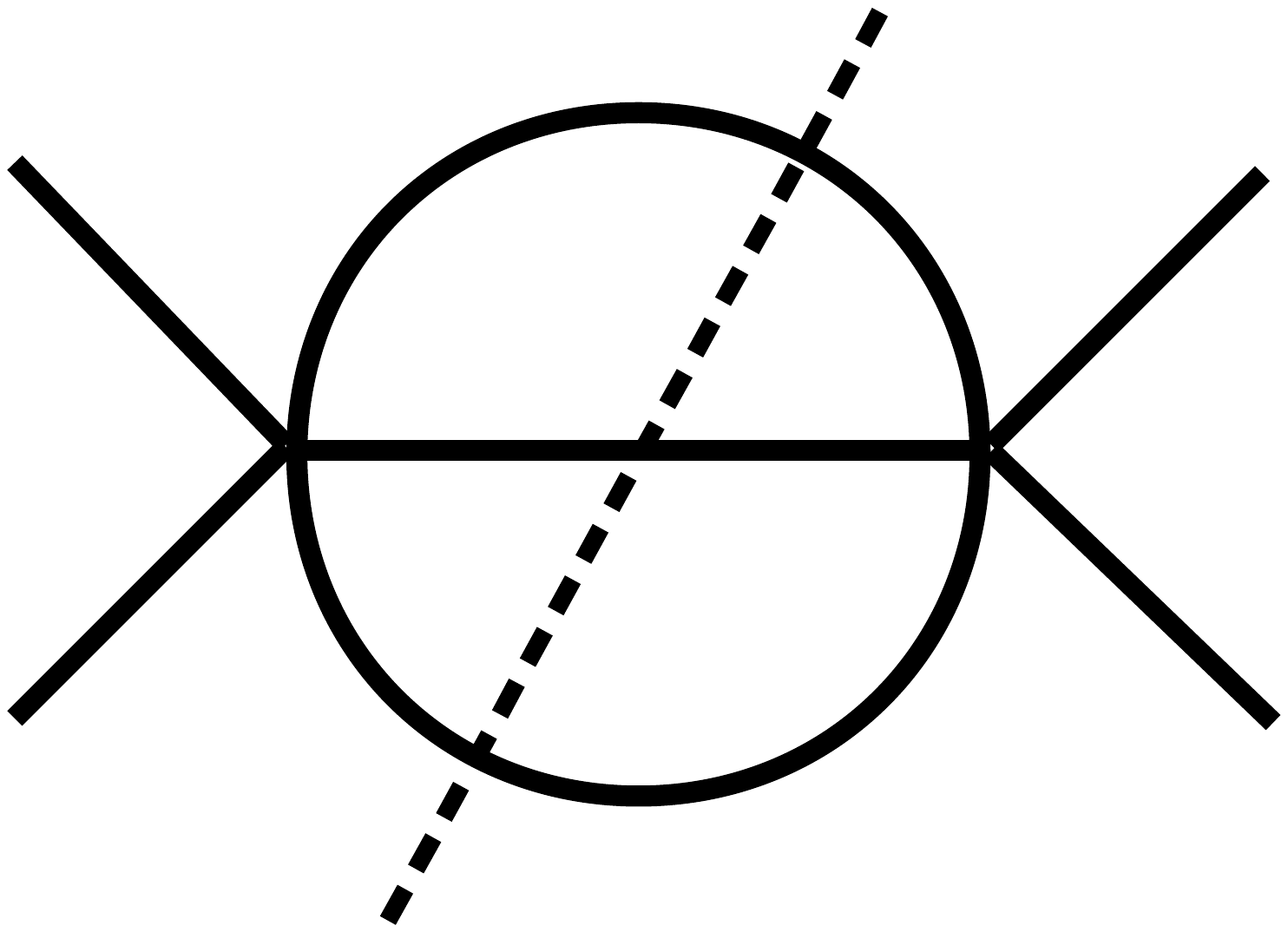}
\end{gathered} \ .
\label{eq:dboxreduction}
\end{align}

\clearpage

\section{Conclusions}
\label{sec:conclusions}

In this work, we elaborated on the vector space structure of Feynman integrals, and on the existence of what amounts to a scalar product among them first presented in Ref. \cite{Frellesvig:2019uqt}, showing a detailed description of their systematic decomposition in terms of Master Integrals.
In particular, we described the evaluation of multivariate intersection numbers for twisted cocycles, which are the key ingredient of the master decomposition formula eq. \eqref{eq:masterdeco}, in terms of a recursive algorithm boiling the computations down to univariate intersection numbers.
We applied the master decomposition formula to derive integral relations and differential equations for a number of Feynman integrals. As shown in previous works \cite{Mastrolia:2018uzb, Frellesvig:2019kgj}, they can also be used for deriving dimensional recurrence relations (finite-difference equations) for Feynman integrals. 
We discussed algebraic properties of integrals and dual integrals as well as systems of differential equations they obey.

We provided three different strategies for Feynman integral reduction, which we dubbed the {\it straight decomposition}, the {\it bottom-up decomposition}, and the {\it top-down decomposition}, which show possible combinations of the intersection-theory concepts together with unitarity-based methods and integrand decomposition.

The recursive computation of multivariate intersection numbers requires regulated integrals, not plagued by spurious irregular behavior which might emerge at the intermediate steps of the evaluation. 
For this purpose, we employed the analytic regularization procedure. On the other hand, using the richer mathematical structure of the {\it relative twisted cohomology}, the use of regulators might be avoided, thereby offering a very interesting new direction for future studies and applications to physics.

Let us conclude by observing that the decomposition formula, or better the corresponding formula for the identity resolution, in terms of multivariate intersection numbers, is applicable to generic parametric representations of Feynman integrals, including those not considered here. More generally, it 
can be used to derive linear and quadratic relations for Aomoto-Gel'fand type of integrals (and their duals), which are of broad interest and have applications in different contexts in physics as well as mathematics.

\acknowledgments 
P.M. would like to thank Nima Arkani-Hamed, Sergio Cacciatori, Thibault Damour, Maxim Kontsevich, and Ugo Moschella for stimulating discussions. 
We also thank Giulio Salvatori for interesting comments.
P.M. and S.M. would like to acknowledge the hospitality of the Center of Mathematical Sciences and Applications (CMSA), Harvard University, USA.
H.F., M.K.M. and P.M. would like to acknowledge the hospitality of the Institut des Hautes \'Etudes Scientifique (IHES), France.
We would like to thank all the participants of the workshop “MathemAmplitudes 2019: Intersection Theory and Feynman Integrals” for stimulating discussions as well as for a very lively scientific environment.
The work of H.F. is part of the HiProLoop project funded by the European Union’s Horizon 2020 research and innovation programme under the Marie Sk{\l}odowska-Curie grant agreement 747178.
The work of M.K.M., and P.M. is supported by the Supporting TAlent in ReSearch at Padova University (UniPD STARS Grant 2017 ``Diagrammalgebra''). 
The work of M.K.M is supported by Fellini - Fellowship for Innovation at INFN funded by the European Union’s Horizon 2020 research and innovation programme under the Marie Sk{\l}odowska-Curie grant agreement No 754496.
S.M. gratefully acknowledges the funding provided by Carl P. Feinberg. F.G. is supported by Fondazione Cassa di Risparmio di Padova e Rovigo (CARIPARO).
The figures were drawn with Jaxodraw~\cite{Binosi:2008ig} based on Axodraw~\cite{Axodraw:1994}.
CloudVeneto is acknowledged for the use of computing and storage facilities. \\

\appendix
\section{Master Decomposition Formula}
\label{sec:App:masterdecoproofs}
The reduction of an integral is achieved by expressing the corresponding twisted cocycle, say $\langle \varphi_L |$, as a linear combination of a set of $\nu$ basic elements, $\langle e_i|$, referred to as the {\it master forms}.
For this purpose,
we introduce a {\it dual} space of twisted cocycles, generated by the basis $| h_i \rangle$ for $i=1,2,\ldots,\nu$, and define the metric matrix ${\bf C}_{ij} \equiv \langle e_i| h_, \rangle$.
% as
% \begin{eqnarray}
% \mathbf{C}_{ij} = \langle e_i | h_j \rangle \qquad \text{for}\qquad i,j=1,2,\ldots, \nu \ . 
% \end{eqnarray}
Now, we can build the following $(\nu{+}1)\times (\nu{+}1)$ matrix ${\bf M}$, defined as,
\be
\mathbf{M} = \left(
\begin{array}{ccccc}
\langle \varphi_L | \varphi_R \rangle & \langle \varphi_L |  h_1 \rangle & 
\langle \varphi_L |  h_2 \rangle & \ldots &
\langle \varphi_L |  h_{\nu} \rangle \\
\langle  e_1 | \varphi_R \rangle & \langle  e_1 |  h_1 \rangle & 
\langle  e_1 |  h_2 \rangle & \ldots &
\langle  e_1 |  h_{\nu} \rangle \\
\langle  e_2 | \varphi_R \rangle &\langle  e_2 |  h_1 \rangle & 
\langle  e_2 |  h_2 \rangle & \ldots & \langle  e_2 |  h_{\nu} \rangle \\
\vdots & \vdots &  \vdots & \ddots & \vdots \\
\langle  e_{\nu} | \varphi_R \rangle & \langle  e_{\nu} |  h_1 \rangle & 
\langle  e_{\nu} |  h_2 \rangle & \ldots & \langle  e_{\nu} |  h_{\nu} \rangle 
\end{array}
\right)
\equiv
\left(\begin{array}{cc}
	\langle \varphi_L | \varphi_R \rangle & \mathbf{A}^\intercal\\
	\mathbf{B} & \mathbf{C} 
\end{array}\right).
\ee
We label the columns of the matrix by $|\varphi_R\rangle, |h_1\rangle, |h_2\rangle, \ldots, |h_{\nu} \rangle$ for an arbitrary $|\varphi_R\rangle$, and the rows by $\langle \varphi_L|, \langle e_1 |, \langle e_2 |, \ldots, \langle e_{\nu} |$. 
The entries of this matrix are given by the pairing (bilinear) between the corresponding rows and columns. 
In the second equality, we express the matrix $\mathbf{M}$ as 
a $\nu \times \nu$ submatrix $\mathbf{C}$, 
a column vector $\mathbf{B}$ and a row vector $\mathbf{A}^\intercal$,
respectively with elements $\mathbf{B}_i = \langle e_i | \varphi_R \rangle$ and 
$\mathbf{A}_i = \langle \varphi_L | h_i \rangle$ (for $i = 1,2, \ldots, \nu$).
Now, as the space of the twisted cocycles is $\nu$-dimensional and each entry of the matrix $\mathbf{M}$ is a bilinear, implies that the determinant of this matrix vanishes.

% The columns of the matrix $\mathbf{M}$ are labelled by $|\varphi_R\rangle, |h_1\rangle, |h_2\rangle, \ldots, |h_{\nu} \rangle$ for an arbitrary $|\varphi_R\rangle$, while the rows are labelled by $\langle \varphi_L|, \langle e_1 |, \langle e_2 |, \ldots, \langle e_{\nu} |$. 
% Each entry is given by a pairing (bilinear) of the corresponding row and column. In the second equality, we expose the structure of $\mathbf{M}$ as a $\nu \times \nu$ submatrix $\mathbf{C}$, 
% a column vector $\mathbf{B}$ and a row vector $\mathbf{A}^\intercal$,
% respectively with elements $\mathbf{B}_i = \langle e_i | \varphi_R \rangle$ and 
% $\mathbf{A}_i = \langle \varphi | h_i \rangle$ (for $i = 1,2, \ldots, \nu$).
% The fact that the $\nu{+}1$ cocycles labelling the rows and columns are necessarily linearly dependent (since the basis is $\nu$-dimensional) and that each entry of $\mathbf{M}$ is a bilinear, implies that the determinant of this matrix vanishes. 
Using the well-known identity for the determinant of a block matrix, we find:
\be
\det \mathbf{M} = \det \mathbf{C}\, \bigg( \langle \varphi_L | \varphi_R \rangle - \mathbf{A}^\intercal\, \mathbf{C}^{-1}\, \mathbf{B} \bigg) = 0.
\ee
Since $\det \mathbf{C}$ is non-zero by definition, we conclude that:
\begin{align}
\langle \varphi_L | \varphi_R \rangle &= \mathbf{A}^\intercal\, \mathbf{C}^{-1}\, \mathbf{B}\nonumber\\
&= \sum_{i,j=1}^{\nu} \langle \varphi_L | h_j \rangle \, (\mathbf{C}^{-1})_{ji}\, \langle e_i | \varphi_R \rangle \ .
\label{eq:phiLphiR_masterdeco}
\end{align}
This equation is very important. It can be exploited in three ways:
\begin{itemize}
    \item Because of the arbitrariness of both $\langle \varphi_L |$ and $| \varphi_R \rangle$, 
    the {\it r.h.s.} of the above equation implies that 
$\sum_{i,j=1}^{\nu} | h_j \rangle \, (\mathbf{C}^{-1})_{ji}\, \langle e_i | $
acts like the identity operator $\mathbb{I}_c$ (in the cohomology space) when contracted with left and right forms,
\begin{eqnarray}
\sum_{i,j=1}^{\nu} | h_j \rangle \, (\mathbf{C}^{-1})_{ji}\, \langle e_i | \equiv {\mathbb I}_c \ .
\end{eqnarray}
    \item Because of the arbitrariness of $|\varphi_R \rangle$, the equation gives the decomposition of $\langle \varphi_L |$, as
    \be\label{basis-projection}
\langle \varphi_L | = \sum_{i=1}^{\nu} \underbrace{\sum_{j=1}^{\nu} \langle \varphi_L | h_j \rangle \, \left( \mathbf{C}^{-1} \right)_{ji}\,}_{c_i} \langle e_i | \ .
\ee
\item Because of the arbitrariness of $ \langle \varphi_L|$, the equation gives the decomposition of dual forms $| \varphi_R \rangle$, as
\begin{eqnarray}
|\varphi_R \rangle = \sum_{i=1}^{\nu} \underbrace{\sum_{j=1}^{\nu} \left( \mathbf{C}^{-1} \right)_{ij} \langle e_j | \varphi_R \rangle}_{\widetilde{c}_i} \, | h_i \rangle \ .
\end{eqnarray}
\end{itemize}

With a similar approach, starting with the intersection number $[{\cal C}_L | {\cal C}_R]$ of integration contours (homology classes), 
one can derive a formula analogous to eq. (\ref{eq:phiLphiR_masterdeco}), 
\begin{align}
[ {\cal C}_L | {\cal C}_R ] &= 
 \sum_{i,j=1}^{\nu} [ {\cal C}_L | \gamma_j ]\, (\mathbf{H}^{-1})_{ji}\, [ \delta_i | {\cal C}_R ] \ . 
\label{eq:CLCR_masterdeco}
\end{align}
where $|\gamma_i]$ and $[\delta_i|$ are bases of the homology space and its dual space, respectively, and
$\mathbf{H}_{ij} \equiv [\delta_i|\gamma_j]$ is the intersection matrix of the bases elements.
As done earlier for the cohomologies, eq. (\ref{eq:CLCR_masterdeco}) can be exploited to derive the identity operator ${\mathbb I}_h$ in the homology space, 
as well as the decompositions of contours $[ {\cal C}_L |$ and dual contours $| {\cal C}_R ]$.

\section{Derivation of the connection $\mathbf\Omega$ for $n$-form intersection numbers}
\label{appendix:Omega}

Let us recall how the covariant derivative emerges in the $1$-form case, with $u=u(z_1)$.
We consider the vanishing surface term,
\begin{eqnarray}
0 &=& \int_{{\cal C}_R} d_{z_1}( \xi_L(z_1) u) = 
\int_{{\cal C}_R} (d_{z_1} \xi_L + d_{z_1} \log u \wedge \xi_L) u \nonumber \\
&\equiv& 
\int_{{\cal C}_R} \nabla_{{\bf \Omega}^{(1)}} \, \xi_L\; u \, 
\equiv \langle\nabla_{{\bf \Omega}^{(1)}} \, \xi_L|{\cal C}_R]\; ,
\end{eqnarray}
where we defined the covariant derivative, $\nabla_{{\bf \Omega}^{(1)}}\ \equiv d_{z_1} + {\bf \Omega}^{(1)}$,
using,
\begin{eqnarray}
d_{z_1}u = {\bf \Omega}^{(1)} \, u \ ,
\end{eqnarray}
and ${\bf \Omega}^{(1)} \equiv \omega_1 = d_{z_1}{\rm log} \, u$ \ . 

Let us extend the above derivation to the case of $n$-forms with $u = u(z_1,\ldots,z_n)$.
Let us consider a multivariate integral $I$ over $n$ variables $z_1, \ldots, z_n$,
\begin{eqnarray}
\langle \varphi_L^{(\bf{n})} |{\cal C}_R^{({\bf n})}] 
= \int_{{\cal C}_R^{({\bf n})}} \varphi_L^{(\bf{n})}(z_1,\ldots, z_n) u 
&=& 
\sum_{i=1}^{\nu_{\bf n-1}} \int_{{\cal C}_R^{({n})}} \varphi_{L,i}^{(n)}(z_n) \, 
\int_{{\cal C}_R^{({\bf n-1})}} e_i^{(\bf{n-1})}(z_1,\ldots, z_n)\, u  \nonumber \\
&=& 
\sum_{i=1}^{\nu_{\bf n-1}} \int_{{\cal C}_R^{({n})}} \varphi_{L,i}^{(n)}(z_n) \,  
\langle e_i^{(\bf{n-1})} | {\cal C}_R^{({\bf n-1})}] \ ,
\end{eqnarray}
where we defined 
\begin{eqnarray}
% u_i(z_2) 
\langle e_i^{(\bf{n-1})} | {\cal C}_R^{({\bf n-1})}] \equiv 
\int_{{\cal C}_R^{({\bf n-1})}} e_i^{(\bf{n-1})}(z_1,\ldots, z_n) \, u \, .
\end{eqnarray}
It it is crucial to stress that $\langle e_i^{(\bf{n-1})} | {\cal C}_R^{({\bf n-1})}]$ now plays the same role as $u$ in the univariate case.
There could exist many forms $\varphi_{L,i}^{(n)}$ that upon integration give the same result.
Let us consider the vanishing surface integral in $z_n$ of the $z_n$-derivative of $ \langle e_i^{(\bf{n-1})} | {\cal C}_R^{({\bf n-1})}] $ times an arbitrary function ($0$-form) $\xi_i(z_n)$,
\begin{eqnarray}
0 
&=&
\int_{{\cal C}_R^{({n})}} d_{z_n} \Big( \xi_i(z_n) \, \langle e_i^{(\bf{n-1})} | {\cal C}_R^{({\bf n-1})}] \Big) \ .
\label{eq:2var-equiv-class}
\end{eqnarray}
Let us notice that the integral $\langle e_i^{(\bf{n-1})} | {\cal C}_R^{({\bf n-1})}]$ %$u_i(z_2)$% 
satisfies the following differential equation in $z_n$ following Sec.~\ref{sec:DEofFormsAndDualForms}:
\begin{eqnarray}
d_{z_n} \langle e_i^{(\bf{n-1})} | {\cal C}_R^{({\bf n-1})}] 
%u_i(z_2) 
= \mathbf{\Omega}^{(n)}_{ij} 
\langle e_j^{(\bf{n-1})} | {\cal C}_R^{({\bf n-1})}] 
%u_j (z_2) 
\, ,
\label{eq:DE-of-ui}
\end{eqnarray}
where $\mathbf{\Omega}^{(n)}$ is a $\nu_{\bf n-1} \times \nu_{\bf n-1}$ matrix. 
Inserting this into eq.~(\ref{eq:2var-equiv-class}), we obtain:
\begin{eqnarray}
0 &=& 
\int_{{\cal C}_R^{({n})}} 
\bigg(
\Big( \delta_{ij} \, d_{z_n} + \mathbf{\Omega}^{(n)}_{ij}  \Big) 
\xi_i(z_n) 
\bigg) \ 
\langle e_j^{(\bf{n-1})} | {\cal C}_R^{({\bf n-1})}] 
 \nonumber \\
 &=&
 \int_{{\cal C}_R^{({n})}} \! 
 \Big( 
 \left(\nabla_{\mathbf{\Omega}^{(n)}}\right)_{ij} \, 
 \xi_i(z_n) \Big)
 \ \langle e_j^{(\bf{n-1})} | {\cal C}_R^{({\bf n-1})}] \ , 
\end{eqnarray}
where the final equation defines the connection at the $n$-th integration step (after $({\bf n-1})$-nested integrations, 
on the variables $z_1, \ldots, z_{n-1}$),
\begin{eqnarray}
\nabla_{\mathbf{\Omega}^{(n)}} \equiv {\mathbb I} \, d_{z_n} + \mathbf{\Omega}^{(n)} \ .
\end{eqnarray}
The matrix $\mathbf{\Omega}^{(n)} $ can be obtained as described in Sec.~\ref{sec:DEofFormsAndDualForms},
\begin{eqnarray}
d_{z_n} \langle e_i^{(\bf{n-1})} | {\cal C}_R^{({\bf n-1})}] 
&=& d_{z_n} \int_{{\cal C}_R^{({\bf n-1})}} \! e_i^{(\bf{n-1})}(z_1,\ldots,z_n) \, u  \nonumber \\
&=& \int_{{\cal C}_R^{({\bf n-1})}} \! \left(d_{z_n} e_i^{(\bf{n-1})}(z_1,\ldots,z_n) 
                             + d_{z_n} \log u \wedge e_i^{(\bf{n-1})}(z_1,\ldots,z_n) \right) u \nonumber \\
&=& \int_{{\cal C}_R^{({\bf n-1})}} \! 
     \left(d_{z_n}  + \omega_n \wedge \right) e_i^{(\bf{n-1})}(z_1,\ldots,z_n) \, \, u \\
& = & \langle (d_{z_n} + \omega_n \wedge) e_i^{(\bf{n-1})} | {\cal C}_R^{({\bf n-1})} ] \ , \nonumber
\end{eqnarray}
where $\omega_n \equiv d_{z_n} \log \, u$. \\
The final line can be further simplified by using the master decomposition formula in eq.~\eqref{eq:masterdeco} as,
\begin{eqnarray}
\langle (d_{z_n} + \omega_n \wedge) e_i^{(\bf{n-1})} |
&=& \langle (d_{z_n} + \omega_n \wedge) e_i^{(\bf{n-1})}| h_k^{(\bf{n-1})} \rangle 
({\mathbf C}_{{\bf (n-1)}}^{-1})_{kj} \, \langle e_j^{(\bf{n-1})} | \, .
\end{eqnarray}
Using eq.~(\ref{eq:DE-of-ui}), we can identify $\mathbf{\Omega}^{(n)}$ as,
\begin{eqnarray}
\mathbf{\Omega}^{(n)}_{ij} &=& 
\langle (d_{z_n} + \omega_n \wedge) e_i^{(\bf{n-1})} | h_k^{(\bf{n-1})} \rangle 
({\mathbf C}_{{\bf (n-1)}}^{-1})_{kj} \, .
\end{eqnarray}

\paragraph{Dual formula}
With a similar derivation, starting from the vanishing surface term,
\begin{eqnarray}
0 
&=&
\int_{{\cal C}_L^{({n})}} d_{z_n} \Big( \xi_i(z_n) \, [ {\cal C}_L^{({\bf n-1})} | h_i^{(\bf{n-1})} \rangle \Big) \ , 
\end{eqnarray}
and using,
\begin{eqnarray}
d_{z_n} \, [ {\cal C}_L^{({\bf n-1})} | h_i^{(\bf{n-1})} \rangle 
= - \, [ {\cal C}_L^{({\bf n-1})} | h_j^{(\bf{n-1})} \rangle \, {\tilde {\bf \Omega}}_{ji}^{(n)} 
\end{eqnarray}
we obtain 
\begin{eqnarray}
0 
&=&
\int_{{\cal C}_L^{({n})}}
\left( \nabla_{-\mathbf{{\tilde \Omega}}^{(n)}} \right)_{ji} \, 
 \xi_i(z_n) \,
 [ {\cal C}_L^{({\bf n-1})} | h_j^{(\bf{n-1})} \rangle \ , 
\end{eqnarray}
where the dual connection is defined as,
\begin{eqnarray}
\nabla_{-\mathbf{{\tilde \Omega}}^{(n)}} \equiv {\mathbb I} \, d_{z_n} - \mathbf{{\tilde \Omega}}^{(n)} \ ,
\end{eqnarray}
with
\begin{eqnarray}
\mathbf{{\tilde\Omega}}^{(n)}_{ji} &=& 
- ({\mathbf C}_{{\bf (n-1)}}^{-1})_{jk} \, 
\langle e_k^{(\bf{n-1})} | (d_{z_n} - \omega_n \wedge) h_i^{(\bf{n-1})} \rangle 
\ .
\end{eqnarray}

%===================================================Intersection number===================
\section{Intersection numbers for the three examples}
\label{appendix}

In this appendix we provide the explicit form of intersection numbers needed for the Feynman integral decompositions performed in Sec.~\ref{sec:examples}. Since we work in analytic regularization with a parameter $\rho$ that is taken to zero at the end of the computation, it suffices to know only the leading $\rho$-orders of intersection numbers. While our algorithm computes them exactly in $\rho$, in order to save space in this appendix we list only the \emph{leading} term for each intersection number \emph{individually}. One can check that the orders given here are sufficient for reconstructing the coefficients $c_i$ to order ${\cal O}(\rho^0)$ and that their limit as $\rho \to 0$ is in fact smooth.

\subsection{One-loop massless box}
\label{appendix:Box}

\subsubsection{Straight decomposition}
Here we provide the intersection numbers, up to the leading order in $\rho$ required for the decomposition presented in Subsec.~$\ref{subsubsec:masslesbox_straight}$:
\begin{equation}
    \mathbf{C}_{ij}= \langle e_i | h_j \rangle, \quad 1 \leq i,j \leq 3
\end{equation}
with
{\allowdisplaybreaks
\begin{align}
  \langle e_1 | h_1 \rangle &=    \frac{1}{\rho ^4} + \mathcal{O} \left(\rho^{-3}\right), \\
   \langle e_1 | h_2 \rangle &= 
 -\frac{s t}{(d-7) (d-6) \rho ^2} + \mathcal{O} \left(\rho^{-1}\right), \\
 \langle e_1 | h_3 \rangle &=    \langle e_1 | h_2 \rangle, \\
  \langle e_2 | h_1 \rangle & =
 -\frac{s t}{(d-4)(d-3)\rho ^2} + \mathcal{O} \left(\rho^{-1}\right),  \\
 \langle e_2 | h_2 \rangle &=    -\frac{s^2 t (s+t)}{4 (d-7) (d-3) \rho ^2} + \mathcal{O} \left(\rho^{-1}\right), \\
 \langle e_2 | h_3 \rangle &=    -\frac{s t \left((d-4)^2 s^2+((d-10) d+28) s t+(d-6)^2 t^2\right)}{(d-7) (d-6)^2 (d-4)^2
   (d-3)} + \mathcal{O} \left(\rho \right), \\
   \langle e_3 | h_1 \rangle & = \langle e_2 | h_1 \rangle, \\
 \langle e_3 | h_2 \rangle &=    -\frac{s t \left((d-6)^2 s^2+((d-10) d+28) s t+(d-4)^2 t^2\right)}{(d-7) (d-6)^2 (d-4)^2
   (d-3)} + \mathcal{O} \left(\rho \right), \\
 \langle e_3 | h_3 \rangle &=    -\frac{s t^2 (s+t)}{4 (d-7) (d-3) \rho ^2} + \mathcal{O} \left(\rho^{-1}\right),
\end{align}
}
and 
\begin{equation}
\langle \varphi | h_{k} \rangle, \quad 1 \leq k \leq 3 
\end{equation}
with
\begin{align}
 \langle \varphi | h_1 \rangle &=  \frac{(7-d) (d-6) (d-5)}{2 \rho ^4 s^2 t} + \mathcal{O} \left(\rho^{-3}\right), \\
 \langle \varphi | h_2 \rangle &=  \frac{(5-d) t}{2 \rho ^2 s^2} + \mathcal{O} \left(\rho^{-1}\right), \\
 \langle \varphi | h_3 \rangle &=  -\frac{(d-5) ((d-6) t+2 s)}{2(d-8) \rho^2 t^2} + \mathcal{O} \left(\rho^{-1}\right).
\end{align}

\subsubsection{Bottom-up decomposition}

Here we provide the intersection numbers required for the decomposition presented in Subsec.~\ref{subsubsec:masslessbox_bottomup}, on the $\tau=\{1,3\}$ cut:
\begin{equation}
    \mathbf{C}_{ij}= \langle e_{i,\tau}| h_{j,\tau} \rangle, \quad 1 \leq i,j \leq 2
\end{equation}
with
\allowdisplaybreaks{
\begin{align}
\langle e_{1,\tau} | h_{1,\tau} \rangle &= \frac{d-5}{\rho^2 (d-5 + 2 \rho)}, \\
\langle e_{1,\tau} | h_{2,\tau} \rangle &= \frac{-(d-5) s t}{(d-7 + 2 \rho) (d-6 + 2 \rho) (d-5 + 2 \rho)}, \\
\langle e_{2,\tau} | h_{1,\tau} \rangle &= \frac{-(d-5) s t}{(d-5 + 2 \rho) (d-4 + 2 \rho) (d-3 + 2 \rho)}, \\
\langle e_{2,\tau} | h_{2,\tau} \rangle &= \frac{(d-5) s^2 t (4 \rho^2 t - (d-6 + 4 \rho) (d-4 + 4 \rho) (s+t))}{4 (d-7 + 2 \rho) (d-6 + 2 \rho) (d-5 + 2 \rho) (d-4 + 2 \rho) (d-3 + 2 \rho)},
\end{align}}
and
\begin{equation}
    \langle \varphi_{\tau}| h_{k,\tau} \rangle, \quad 1 \leq k \leq 2
\end{equation}
with
\begin{align}
\langle \varphi_{\tau} | h_{1,\tau} \rangle &= \frac{(d-5) (d-7 + 2 \rho) ((d-6 + 4 \rho) s + 2 \rho t)}{2 (\rho-1) \rho^2 s^3 t}, \\
\langle \varphi_{\tau} | h_{2,\tau} \rangle &= \frac{(d-5) t}{2 (\rho-1) s^2} \, .
\end{align}

\subsubsection{Top-down decomposition}
For consistency with the straight decomposition and the bottom-up decomposition, we also provide here the intersection numbers needed for the top-down decomposition of Subsec.~\ref{subsubsec:masslessbox_topdown}, on the $\tau=\{1,3\}$ cut. They are
\begin{align}
\langle \phi | 1 \rangle = \frac{-(d-5) (s+t)}{2 s^2} \;,\qquad \langle 1 | 1 \rangle = \frac{-s^2 t (s+t)}{4 (d-7) (d-3)}\,.
\end{align}

\subsection{One-loop QED triangle}
\label{appendix:QEDTriangle}
Here we provide the intersection numbers, up to the leading order in $\rho$, required for the system of differential equations presented in Subsec.~$\ref{subsubsec:QED_Triangle}$:
\begin{equation}
    \mathbf{C}_{ij}= \langle e_i | h_j \rangle, \quad 1 \leq i,j \leq3
\end{equation}
with
\begin{align}
\langle e_1 | h_1 \rangle =& \frac{(d-4) \left(4 m^2-s\right)^2}{2 \left(2 (d-4)^2-2\right) \rho ^2} + \mathcal{O} \left(\rho^{-1}\right), \\    
\langle e_1 | h_2 \rangle =& \frac{s \left(4 m^2-s\right) \left(4 (2 d-9) m^2-(d-4) s\right)}{4 (d-6) (d-5) (d-3) \rho } + \mathcal{O} \left(\rho ^0\right),\\
\langle e_1 | h_3 \rangle =& \langle e_1 | h_2 \rangle, \\
\langle e_2 | h_1 \rangle =& \frac{s \left(4 m^2-s\right) \left(4 (2 d-7) m^2-(d-4) s\right)}{4 (d-5) (d-3) (d-2) \rho } + \mathcal{O} \left(\rho ^0\right), \\
\langle e_2 | h_2 \rangle =& \frac{4 m^4 s \left(4 m^2-s\right)}{\left(d^2-8 d+12\right) \rho } + \mathcal{O} \left(\rho ^0\right),\\
\langle e_2 | h_3 \rangle =& \left(s \left(-64 (d{-}5) (d{-}3) (3 (d{-}8) d{+}44) m^6{+}16 ((d{-}8) d (6 (d{-}8) d{+}173) \right. \right. \nonumber \\
& \left. \left. +1236) m^4 s {-}16 (d{-}6) (d{-}4)^2 (d{-}2) m^2 s^2{+}(d{-}6) (d{-}4)^2 (d{-}2)
   s^3\right)\right) / \nonumber \\
&   \left(4 (d{-}6)^2 (d{-}5) (d{-}4) (d{-}3) (d{-}2)^2 \right) + \mathcal{O} \left(\rho ^1\right), \\
\langle e_3 |h_1 \rangle =&  \langle e_2 | h_1 \rangle,  \\
\langle e_3 | h_2 \rangle =& \langle e_2 | h_3 \rangle, \\
\langle e_3 | h_3 \rangle =& \langle e_2 | h_2 \rangle,
\end{align}
and
\begin{equation}
    \langle \varphi | h_{k} \rangle, \quad 1 \leq k \leq3
\end{equation}
with
\begin{align}
    \langle \varphi |h_1 \rangle &= \frac{4 m^2-s}{2 (d-5) \rho ^2} + \mathcal{O} \left(\rho^{-1}\right), \\
    \langle \varphi | h_2 \rangle &= \frac{s \left(4 m^2-s\right)}{2 (d-6) (d-5) \rho } + \mathcal{O} \left(\rho ^0\right), \\
    \langle \varphi | h_3 \rangle & = \langle \varphi | h_2 \rangle.
\end{align}

\subsection{Two-loop QED sunrise}
\label{appendix:QEDSunrise}
Here we provide the intersection numbers, up to the leading order in $\rho$, required for the system of differential equations presented in Subsec.~$\ref{subsubsec:QEDSunrise}$:
\begin{equation}
    \mathbf{C}_{ij}= \langle e_{i,\tau} | h_{j,\tau} \rangle, \quad 1 \leq i,j \leq 4
\end{equation}
with (we use $\gamma = \frac{d-4}{2}$)
\setmuskip{\medmuskip}{0mu}
\allowdisplaybreaks{
\begin{align}
\langle e_{1,\tau} | h_{1,\tau} \rangle =&\frac{\gamma ^2 (s (s ((s-28) s-102)+176)-128)+8 (s-1)^2}{3 (81 \gamma ^4-45 \gamma ^2+4) \rho }+\mathcal{O}(\rho ^0),
\\
\langle e_{1,\tau} | h_{2,\tau} \rangle =& 
((s-1) (\gamma ^3 (s+8) (s ((s-39) s+48)-64)-\gamma ^2 (s+6) (s \
((s-39) s+48)
\nonumber \\
&-64)-2 \gamma  (s-1) ((s-14) s+16)-12 ((s-2) s+2)))/
(9 \
(\gamma -1) (3 \gamma -2)
\nonumber \\
& (3 \gamma -1)(3 \gamma +1) (3 \gamma +2) 
\rho) + \mathcal{O}(\rho ^0),
\\
\langle e_{1,\tau} | h_{3,\tau} \rangle =& \langle e_{1} | h_{2}\rangle,
\\
\langle e_{1,\tau} | h_{4,\tau} \rangle =& (\gamma ^4 (s-4) (s (s ((s-34) s-894)-544)+256)
\nonumber \\
&-\gamma ^3 (s (s \
(s ((s-46) s-396)+3560)+2752)-768)
\nonumber \\
&+2 \gamma ^2 (s (268-(s-24) s (4 \
s+17))+32)+12 \gamma  (s (16-3 (s-8) s)-4)
\nonumber \\
&-48 s (2 s+1))/(18 (\gamma \
-1)^2 \gamma  (81 \gamma ^4-45 \gamma ^2+4))+\mathcal{O}(\rho ^1),
\\
\langle e_{2,\tau} | h_{1,\tau} \rangle =& ((s-1) (\gamma ^3 (s+8) (s ((s-39) s+48)-64)+\gamma ^2 (s+6) (s \
((s-39) s+48)
\nonumber \\
&-64) -2 \gamma  (s-1) ((s-14) s+16)+12 ((s-2) s+2)))/(9 \
(\gamma +1) (3 \gamma -2) 
\nonumber \\
&(3 \gamma -1) (3 \gamma +1) (3 \gamma +2) \rho )+ \mathcal{O}(\rho ^0),
\\
\langle e_{2,\tau} | h_{2,\tau} \rangle =& (-72 ((s-1) s (s^2+2)+1)+\gamma ^4 (s (s (s (s ((s-36) \
s-1563)+1516)
\nonumber \\
&-3168)+3840)-2048)+\gamma ^2 (1280-s (s (s (s ((s-36) \
s-915)+1108)
\nonumber \\
&-2232)+2544)))/(27 (\gamma ^2 (7-9 \gamma ^2)^2-4) \rho \
)+ \mathcal{O}(\rho ^0),
\\
\langle e_{2,\tau} | h_{3,\tau} \rangle =& (\gamma ^4 (s (s (s (s ((s-36) \
s+624)+1516)-3168)+3840)-2048)
\nonumber \\
&-\gamma ^2 (s (s (s (s ((s-36) \
s+300)+1108)-2232)+2544)-1280)
\nonumber \\
&+36 (s (s (s (s+2)-4)+4)-2))/(27 (\gamma \
^2 (7-9 \gamma ^2)^2-4) \rho )+\mathcal{O}(\rho ^0),
\\
\langle e_{2,\tau} | h_{4,\tau} \rangle =& (16 \gamma ^2 (32 \gamma  (\gamma +1) (8 \gamma ^2-5)+9)+144 \
\gamma +(\gamma -1) \gamma ^3 (\gamma +1)^2 s^6
\nonumber \\
&-6 (\gamma -1) \gamma \
^2 (\gamma +1)^2 (7 \gamma -1) s^5-3 \gamma  (\gamma +1) (\gamma  \
(\gamma  (\gamma  (445 \gamma +98)-281)
\nonumber \\
&-38)+16) s^4 +16 (\gamma +1) \
(\gamma  (\gamma  (\gamma  (\gamma  (379 \gamma -99)-277)+108)+18)
\nonumber \\
&-9) s^3 +24 (568 \gamma ^6-503 \gamma ^4+121 \gamma ^2-6) s^2-48 (\gamma \
+1) (4 \gamma -1) (4 \gamma +1)
\nonumber \\
&(\gamma  (\gamma  (14 \gamma \
-5)-8)+3) s)/(54 \gamma  (\gamma ^2-1)^2 (81 \gamma ^4-45 \gamma \
^2+4))+\mathcal{O}(\rho ^1),
\\
\langle e_{3,\tau} | h_{1,\tau} \rangle =& \langle e_{2} | h_{1} \rangle,
\\
\langle e_{3,\tau} | h_{2,\tau} \rangle =& \langle e_{2} | h_{3} \rangle,
\\
\langle e_{3,\tau} | h_{3,\tau} \rangle =& \langle e_{2} | h_{2} \rangle,
\\
\langle e_{3,\tau} | h_{4,\tau} \rangle =& \langle e_{2} | h_{4} \rangle,
\\
\langle e_{4,\tau} | h_{1,\tau} \rangle =& (\gamma ^4 (s-4) (s (s ((s-34) s-894)-544)+256)
\nonumber \\
&+\gamma ^3 (s (s \
(s ((s-46) s-396)+3560)+2752)-768)
\nonumber \\
&+2 \gamma ^2 (s (268-(s-24) s (4 \
s+17))+32)+12 \gamma  (s (3 (s-8) s-16)+4)
\nonumber \\
&-48 s (2 s+1))/(18 \gamma  (\
\gamma +1)^2 (81 \gamma ^4-45 \gamma ^2+4))+\mathcal{O}(\rho ^1),
\\
\langle e_{4,\tau} | h_{2,\tau} \rangle =& (16 (\gamma -1) \gamma  (4 \gamma -3) (4 \gamma -1) (4 \gamma \
+1) (4 \gamma +3)+(\gamma -1)^2 \gamma ^3 (\gamma +1) s^6
\nonumber \\
&-6 (\gamma \
-1)^2 \gamma ^2 (\gamma +1) (7 \gamma +1) s^5-3 (\gamma -1) \gamma  (\
\gamma  (\gamma  (\gamma  (445 \gamma -98)-281)
\nonumber \\
&+38) +16) s^4+16 \
(\gamma -1) (\gamma  (\gamma  (\gamma  (\gamma  (379 \gamma \
+99)-277)-108)+18)
\nonumber \\
&+9) s^3+24 (568 \gamma ^6-503 \gamma ^4+121 \gamma \
^2-6) s^2-48 (\gamma -1) (4 \gamma -1) (4 \gamma +1)
\nonumber \\
&
(\gamma  (\gamma \
 (14 \gamma +5)-8)-3) s)/(54 \gamma  (\gamma ^2-1)^2 (81 \gamma ^4-45 \
\gamma ^2+4))+\mathcal{O}(\rho ^1),
\\
\langle e_{4,\tau} | h_{3,\tau} \rangle =& \langle e_{4} | h_{2} \rangle,
\\
\langle e_{4,\tau} | h_{4,\tau} \rangle =& \frac{2 (4 \gamma ^2-1) s^2}{\gamma  (\gamma ^2-1)^2}+\mathcal{O}(\rho ^1),
\end{align}}
and
\begin{equation}
    \langle \varphi_{k,\tau} | h_{l,\tau} \rangle, \quad 1 \leq k,l \leq 4
\end{equation}
with
\allowdisplaybreaks{
\begin{align}
\langle \varphi_{1,\tau} | h_{1,\tau} \rangle =& \big(\gamma ^2 (s^4-14 s^3-88 s+128)+2 \gamma ^3 (s-1) (s ((s-21) \
s-24)-64)
\nonumber \\
&
+4 \gamma  (s-2) (s-1)^2+8 (s-1)\big)/\big(3 (81 \gamma ^4-45 \gamma \
^2+4) \rho  s\big)+\mathcal{O}(\rho ^0),
\\
\langle \varphi_{1,\tau} | h_{2,\tau} \rangle =& (-8 (\gamma -3)+\gamma ^2 (s ((s-20) s (s^2+8)+452)-416)+2 \
\gamma  (s-3) s^2 (3 s
\nonumber \\
&
-5)+\gamma ^4 (s (s (s (73-2 (s-26) \
s)+56)-448)+512)
\nonumber \\
&
+\gamma ^3 (s-1) (s+2) (s ((s-39) s+48)-64)+12 (s-2) \
s)/(9 \rho  ((\gamma -1) (3 \gamma 
\nonumber \\
&
-2)(3 \gamma -1) (3 \gamma +1) (3 \
\gamma +2) s))+\mathcal{O}(\rho ^0),
\\
\langle \varphi_{1,\tau} | h_{3,\tau} \rangle =& \langle \varphi_1 | h_{2} \rangle,
\\
\langle \varphi_{1,\tau} | h_{4,\tau} \rangle =& (2 \gamma ^5 (s-1) (s+8) (s ((s-39) s+48)-64)
\nonumber \\
&
+\gamma ^4 (s \
(448-s (s ((s-56) s-14)+1988))+256)
\nonumber \\
&
+\gamma ^3 (s (s (s (154-(s-20) \
s)+640)+1720)-832)
\nonumber \\
&
-2 \gamma ^2 (s (s (s (6 s-35)-126)+408)+8)-12 \
\gamma  (s (s+2) (s+4)-4)
\nonumber \\
&
+48 s)/(18 (\gamma -1)^2 \gamma  (81 \gamma \
^4-45 \gamma ^2+4) s)+\mathcal{O}(\rho ^1),
\\
\langle \varphi_{2,\tau} | h_{1,\tau} \rangle 
=&
\Big(12 (s-1) (s^2-2)+\gamma ^3 (s (s (s (2 (s-26) \
s-73)-56)+448)-512)
\nonumber \\
&+\gamma ^2 (s+2) (s (s (2 (s-27) s-45)+208)-192)+2 \
\gamma  (s-1)^2 ((s-2) s
\nonumber \\
&+16)\Big)/\Big(9 (81 \gamma ^4-45 \gamma ^2+4) \rho  \
s\Big)+\mathcal{O}(\rho ^0),
\\
\langle \varphi_{2,\tau} | h_{2,\tau} \rangle 
=&
\Big(36 (-2 s^4+s^3-2 s+2)+\gamma ^4 (s (s (s (s (2 (s-30) \
s-1323)+236)+672)
\nonumber \\
&-2304)+2048)+6 \gamma ^3 (s-1) s (s ((s-39) \
s+48)-64)+\gamma ^2 (s (s (s (s (855
\nonumber \\
&-2 (s-30) \
s)-392)-132)+1296)-1280)-6 \gamma  (s-1) s ((s-6) (s-3) s
\nonumber \\
&
-4)\Big)/\Big(27 \
(\gamma -1) (3 \gamma -2) (3 \gamma -1) (3 \gamma +1) (3 \gamma +2) 
\rho  s\Big)+\mathcal{O}(\rho ^0),
\\
\langle \varphi_{2,\tau} | h_{3,\tau} \rangle 
=& 
\Big(36 (s^4+s^3-2 s+2)+2 \gamma ^4 (s (s (s (s ((s-30) \
s+432)+118)+336)-1152)
\nonumber \\
&
+1024)+6 \gamma ^3 (s-1) s (s ((s-39) \
s+48)-64)-2 \gamma ^2 (s (s (s (s ((s-30) s+180)
\nonumber \\
&
+196)+66)-648)+640)-6 \
\gamma  (s-1) s ((s-6) (s-3) s
\nonumber \\
&
-4)\Big)/\Big(27 (\gamma -1) (3 \gamma -2) (3 \
\gamma -1) (3 \gamma +1) (3 \gamma +2) \rho  s\Big)+\mathcal{O}(\rho ^0),
\\
\langle \varphi_{2,\tau} | h_{4,\tau} \rangle 
=&
\Big(-16 \gamma  (\gamma +1) (4 \gamma -3) (4 \gamma -1) (4 \gamma \
+1) (4 \gamma +3)+2 (\gamma -1) \gamma ^3 (\gamma +1)^2 s^6
\nonumber \\
&
-18 \
(\gamma -1) \gamma ^2 (\gamma +1)^2 (4 \gamma -1) s^5-3 \gamma  \
(\gamma +1) (\gamma  (\gamma  (\gamma  (313 \gamma +264)-273)-84)
\nonumber \\
&
+20) \
s^4+8 (\gamma +1) (\gamma  (\gamma  (379 \gamma ^3-475 \gamma \
+198)+36)-18) s^3-144 (\gamma -1) \gamma  (44 \gamma ^4
\nonumber \\
&
-31 \gamma \
^2+2) s^2+48 (\gamma +1) (4 \gamma -1) (4 \gamma +1) (5 \gamma -3) (2 \
\gamma ^2
\nonumber \\
&
-1) s\Big)/\Big(54 (\gamma -1)^2 \gamma  (\gamma +1) (3 \gamma -2) (3 \
\gamma -1) (3 \gamma +1) (3 \gamma +2) s\Big)+\mathcal{O}(\rho ^1),
\\
\langle \varphi_{3,\tau} | h_{1,\tau} \rangle
=&
\Big(12 (s-1) (s^2-2)+\gamma ^3 (s (s (s (2 (s-26) \
s-73)-56)+448)-512)
\nonumber \\
&
+\gamma ^2 (s+2) (s (s (2 (s-27) s-45)+208)-192)+2 \
\gamma  (s-1)^2 ((s-2) s
\nonumber \\
&
+16)\Big)/\Big(9 (81 \gamma ^4-45 \gamma ^2+4) \rho  \
s\Big)+\mathcal{O}(\rho ^0),
\\
\langle \varphi_{3,\tau} | h_{2,\tau} \rangle
=&
\Big(36 (s^4+s^3-2 s+2)+2 \gamma ^4 (s (s (s (s ((s-30) \
s+432)+118)+336)
\nonumber \\
&
-1152)+1024)+6 \gamma ^3 (s-1) s (s ((s-39) \
s+48)-64)-2 \gamma ^2 (s (s (s (s ((s-30) s
\nonumber \\
&
+180)+196)+66)-648)+640)-6 \
\gamma  (s-1) s ((s-6) (s-3) s
\nonumber \\
&
-4)\Big)/\Big(27 (\gamma -1) (3 \gamma -2) (3 \
\gamma -1) (3 \gamma +1) (3 \gamma +2) \rho  s\Big)+\mathcal{O}(\rho ^0),
\\
\langle \varphi_{3,\tau} | h_{3,\tau} \rangle
=&
\Big(36 (-2 s^4+s^3-2 s+2)+\gamma ^4 (s (s (s (s (2 (s-30) \
s-1323)+236)+672)
\nonumber \\
&
-2304)+2048)+6 \gamma ^3 (s-1) s (s ((s-39) \
s+48)-64)+\gamma ^2 (s (s (s (s (855
\nonumber \\
&
-2 (s-30) \
s)-392)-132)+1296)-1280)-6 \gamma  (s-1) s ((s-6) (s-3) s
\nonumber \\
&
-4)\Big)/\Big(27 \
(\gamma -1) (3 \gamma -2) (3 \gamma -1) (3 \gamma +1) (3 \gamma +2) 
\rho  s\Big)+\mathcal{O}(\rho ^0),
\\
\langle \varphi_{3,\tau} | h_{4,\tau} \rangle
=&
\Big(-16 \gamma  (\gamma +1) (4 \gamma -3) (4 \gamma -1) (4 \gamma \
+1) (4 \gamma +3)+2 (\gamma -1) \gamma ^3 (\gamma +1)^2 s^6
\nonumber \\
&
-18 \
(\gamma -1) \gamma ^2 (\gamma +1)^2 (4 \gamma -1) s^5-3 \gamma  \
(\gamma +1) (\gamma  (\gamma  (\gamma  (313 \gamma +264)-273)
\nonumber \\
&
-84)+20) \
s^4+8 (\gamma +1) (\gamma  (\gamma  (379 \gamma ^3-475 \gamma \
+198)+36)-18) s^3-144 (\gamma
\nonumber \\
&
-1) \gamma  (44 \gamma ^4-31 \gamma \
^2+2) s^2+48 (\gamma +1) (4 \gamma -1) (4 \gamma +1) (5 \gamma -3) (2 \
\gamma ^2
\nonumber \\
&
-1) s\Big)/\Big(54 (\gamma -1)^2 \gamma  (\gamma +1) (3 \gamma -2) (3 \
\gamma -1) (3 \gamma +1) (3 \gamma +2) s\Big)+\mathcal{O}(\rho ^1),
\\
\langle \varphi_{4,\tau} | h_{1,\tau} \rangle 
=&
\Big((s-1) (\gamma ^3 (s+8) (s ((s-39) s+48)
-64)+\gamma ^2 (s+6) (s \
((s-39) s+48)
\nonumber \\
&
-64)-2 \gamma  (s-1) ((s-14) s+16)+12 ((s-2) s+2))\Big)/\Big(9 \
(\gamma +1) (3 \gamma 
\nonumber \\
&
-2) (3 \gamma -1) (3 \gamma +1) (3 \gamma +2) \
s\Big)+\mathcal{O}(\rho ^1),
\\
\langle \varphi_{4,\tau} | h_{2,\tau} \rangle 
=&
\Big(\gamma ^4 (s (s (s (s (2 (s-36) s-939)+3032)-6336)+7680)-4096)
\nonumber \\
&
+
\gamma ^2 (s (s (s (s (615-2 (s-36) s)-2216)+4464)-5088)+2560)
\nonumber \\
&
-36 \
((s-2) s+2)^2\Big)/\Big(54 (\gamma ^2 (7-9 \gamma ^2)^2-4) s\Big)+\mathcal{O}(\rho^1),
\\
\langle \varphi_{4,\tau} | h_{3,\tau} \rangle 
=& \langle \varphi_{4,\tau} | h_{2,\tau} \rangle,
\\
\langle \varphi_{4,\tau} | h_{4,\tau} \rangle 
=&
\Big(\rho  (16 \gamma ^2 (32 \gamma  (\gamma +1) (8 \gamma \
^2-5)+9)+144 \gamma +(\gamma -1) \gamma ^3 (\gamma +1)^2 s^6
\nonumber \\
&
-6 \
(\gamma -1) \gamma ^2 (\gamma +1)^2 (7 \gamma -1) s^5-3 \gamma  \
(\gamma +1) (\gamma  (\gamma  (\gamma  (445 \gamma +98)-281)
\nonumber \\
&
-38)+16) \
s^4+16 (\gamma +1) (\gamma  (\gamma  (\gamma  (\gamma  (379 \gamma \
-99)-277)+108)+18)-9) s^3
\nonumber \\
&
+24 (568 \gamma ^6-503 \gamma ^4+121 \gamma \
^2-6) s^2-48 (\gamma +1) (4 \gamma -1) (4 \gamma +1) (\gamma  (\gamma \
 (14 \gamma
\nonumber \\
& 
 -5)-8)+3) s)\Big)/\Big(54 \gamma  (\gamma ^2-1)^2 (81 \gamma ^4-45 \
\gamma ^2+4) s\Big)+\mathcal{O}(\rho^2).
\end{align}
}

\bibliographystyle{JHEP}
\bibliography{biblio}

\end{document}